\documentclass[11pt,a4paper]{article}
\pdfoutput=1
\usepackage{jheppub}

\usepackage{bbm}
\usepackage{slashed}
\usepackage{mathtools}
\usepackage[matrix,arrow]{xy}
\usepackage{float}
\usepackage{upgreek}
\usepackage[stable]{footmisc}

\newcommand{\qed}{\nobreak \ifvmode \relax \else
      \ifdim\lastskip<1.5em \hskip-\lastskip
      \hskip1.5em plus0em minus0.5em \fi \nobreak
      \vrule height0.75em width0.5em depth0.25em\fi}

\DeclareMathAlphabet{\mathpzc}{OT1}{pzc}{m}{it}
\def\tHooft{\mbox{'t Hooft }}

\def\NN{\mathcal{N}}

\def\AA{\mathcal{A}}

\def\EE{\mathcal{E}}

\def\LL{\mathcal{L}}
\def\AA{\mathcal{A}}
\def\SS{\mathcal{S}}

\def\MM{\mathcal{M}}
\def\GG{\mathcal{G}}

\def\HH{\mathcal{H}}
\def\ZZ{\mathcal{Z}}

\def\HH{\mathcal{H}}

\def\II{\mathcal{I}}
\def\UU{\mathcal{U}}

\def\cg{\mathpzc{g}}

\def\fMM{\overline{\underline{\MM}}}
\def\gm{\gamma_{\rm m}}

\def\uj{\underline{j}}
\def\uk{\underline{k}}

\def\Rng{ \textrm{Rng}}
\def\pd{\partial}
\def\ed{  \textrm{d}}
\def\eD{ \textrm{D}}
\def\rnk{ \, \textrm{rnk} \, }
\def\Re{ \, \textrm{Re} \, }

\def\Hom{ \, \textrm{Hom} \, }

\def\Ad{ \, \textrm{Ad}  }
\def\ad{ \, \textrm{ad} }
\def\Tr{ \, \textrm{Tr}  }
\def\tr{ \, \textrm{tr} }
\def\ind{ \, \textrm{ind}  }
\def\coker{ \, \textrm{coker} \, }
\def\diag{ \, \textrm{diag}}
\def\sgn{ \, \textrm{sgn} }
\def\vol{ \, \textrm{vol} }
\def\half{\frac{1}{2}}

\def\ie{{\it i.e.}}
\def\eg{{\it e.g.}}

\def\be{\begin{equation}}
\def\ee{\end{equation}}
\def\bea{\begin{eqnarray}}
\def\eea{\end{eqnarray}}

\title{Parameter counting for singular monopoles on $\mathbb{R}^3$}
\author[a]{Gregory W.~Moore,}
\author[b]{Andrew B.~Royston,}
\author[c]{Dieter Van den Bleeken}

\affiliation[a]{NHETC and
Department of Physics and Astronomy, Rutgers University \\
126 Frelinghuysen Rd., Piscataway NJ 08855, USA}
\affiliation[b]{George P.\ \& Cynthia Woods Mitchell Institute for Fundamental Physics and Astronomy, \\
Texas A\&M University, College Station, TX 77843, USA}
\affiliation[c]{Physics Department, Bo\u{g}azi\c{c}i University\\
 34342 Bebek / Istanbul, TURKEY}

\emailAdd{gmoore@physics.rutgers.edu}
\emailAdd{aroyston@physics.tamu.edu}
\emailAdd{dieter.van@boun.edu.tr}

\abstract{We compute the dimension of the moduli space of gauge-inequivalent solutions to the Bogomolny equation on $\mathbb{R}^3$ with prescribed singularities corresponding to the insertion of a finite number of 't Hooft defects.  We do this by generalizing the methods of C.\ Callias and E.\ Weinberg to the case of $\mathbb{R}^3$ with a finite set of points removed.  For a special class of Cartan-valued backgrounds we go further and construct an explicit basis of $\LL^2$-normalizable zero-modes.  Finally we exhibit and study a two-parameter family of spherically symmetric singular monopoles, using the dimension formula to provide a physical interpretation of these configurations.  This paper is the first in a series of three on singular monopoles, where we also explore the role they play in the contexts of intersecting D-brane systems and four-dimensional $\NN = 2$ super Yang--Mills theories.}


\begin{document}
\begin{flushright} MIFPA-14-13 \end{flushright}
\maketitle

\section{Introduction and summary of results}

In this paper we consider Yang--Mills--Higgs theory with compact simple gauge group $G$ and Higgs field $\Phi$ in the adjoint representation.  This theory famously has magnetic monopoles: smooth, finite-energy, localized solutions to the classical equations of motion, discovered independently by 't Hooft \cite{'tHooft:1974qc} and Polyakov \cite{Polyakov:1974ek}.  In the Bogomolny--Prasad--Sommerfield limit \cite{Bogomolny:1975de,Prasad:1975kr}, the energy functional is minimized by time-independent field configurations satisfying the first-order differential equations $F = \star \eD \Phi$ on an oriented three-dimensional Euclidean space.  The requirement of finite energy imposes asymptotic boundary conditions on the fields that are specified by two quantities, the asymptotic Higgs field, $\Phi_{\infty}$, and the magnetic charge, $\gamma_{\rm m}$:
\begin{equation}\label{largerbc}
\Phi = \Phi_\infty - \frac{1}{2r} \gm + O(r^{-(1 + \updelta)})~, \quad F = \half \gm \sin{\theta} \ed\theta \ed\phi + O(r^{-(2+\updelta)})~, \quad \textrm{as $r \to \infty$}~,
\end{equation}
for any $\updelta > 0$.\footnote{Here $(r,\theta,\phi)$ are standard spherical coordinates on $\mathbbm{R}^3$, with orientation $\ed x \wedge \ed y \wedge \ed z = r^2\sin{\theta} \ \ed r \wedge \ed \theta \wedge \ed \phi$.  The behavior of $\ed\theta$ and $\ed\phi$ is $O(r^{-1})$ when expressed in terms of an orthonormal basis of one-forms.  Hence the leading terms of $F$ in \eqref{largerbc} and \eqref{tHooftpole} are $O(r^{-2})$ and $O(r_{n}^{-2})$ respectively.}  (See the discussion around equations \eqref{finen} and \eqref{asymptoticbcs} below for details.) For a given pair $(\gamma_{\rm m}; \Phi_\infty)$ there is a space of gauge inequivalent solutions $\MM(\gamma_{\rm m}; \Phi_\infty)$.  This space is endowed with a natural metric descending from the kinetic terms of the energy functional and, in favorable circumstances which we will review below, it is a finite-dimensional Riemannian manifold possessing a number of remarkable properties including hyperk\"ahlerity and various isometries.  The study of the Bogomolny equation and its associated monopole moduli spaces has had a profound impact on both mathematics and physics.  Foundational work on the subject includes \cite{Weinberg:1979ma,Weinberg:1979zt,Manton:1981mp,Nahm:1981nb,Taubes:1981gw,Hitchin:1982gh,Taubes:1983bm,Donaldson:1985id,Atiyah:1985dv}.  Classic texts are \cite{MR614447,MR934202}; modern reviews with extensive references include \cite{Harvey:1996ur,MR2068924,Tong:2005un,Weinberg:2006rq}.

We study solutions to the Bogomolny equation which are smooth on $\mathbb{R}^3 \setminus \{ \vec{x}_n \}_{n=1}^{N_t}$.  At the points $\vec{x}_n$ the fields are required to have a specific singularity structure which in physical language corresponds to the insertion of an 't Hooft line defect \cite{'tHooft:1977hy,Kapustin:2005py}.  In the vicinity of a line defect at $\vec{x}_n$ with charge $P_n$ we have\footnote{Our defect boundary conditions differ in an important way from those considered previously, in that we allow for subleading behavior that is still singular as $r \to 0$.  We will see in explicit examples that this behavior occurs for the components of the fields along root directions $E_\alpha$ in the Lie algebra when the root $\alpha$ and the \tHooft charge $P$ have pairing $\langle \alpha, P \rangle = \pm 1$.  This behavior can also be extracted directly from the explicit $G = SO(3)$ solution of \cite{Cherkis:2007jm,Cherkis:2007qa}.}
\begin{equation}\label{tHooftpole}
\Phi = - \frac{1}{2r_{n}} P_{n} + O(r_{n}^{-1/2})~, \qquad F = \half P_n \sin{\theta}_{n} \ed \theta_{n} \ed \phi_{n} + O(r_{n}^{-3/2})~, \quad \textrm{as $r_n \to 0$}~,
\end{equation}
where $(r_{n} = |\vec{x} - \vec{x}_n|, \theta_{n},\phi_{n})$ are standard spherical coordinates centered on the defect, and $P_n$ is a covariantly constant section of the adjoint bundle restricted to the infinitesimal two-sphere surrounding the defect.  The boundary condition \eqref{tHooftpole} is consistent with the Bogomolny equation $F = \star \eD \Phi$ in that truncating the fields to their leading order behavior yields a field configuration that solves the equation.\footnote{One can modify these boundary conditions by changing the sign of the pole term in $\Phi$, in which case they will be consistent with the equation $F = - \star \eD \Phi$.  This $\mathbb{Z}_2$ choice is part of the data of the defect and dictates which form of the Bogomolny equation one is considering.}  By making local gauge transformations in the northern and southern patches of the sphere we can take $P_{n}$ to be a constant, valued in a Cartan subalgebra.  Single-valuedness of the transition function on the overlap implies $\exp{(2\pi P_{n})} = 1_G$, the identity element in $G$.  Thus we may think of the 't Hooft  defect as a Dirac monopole embedded into the gauge group $G$, where $P_{n}$ determines the embedding $U(1) \hookrightarrow T \subset G$ of $U(1)$ into a Cartan torus of $G$.

Given the data of a set of defects, $ (\vec{x}_{n}, P_{n} )_{n = 1}^{N_t}$, together with the asymptotic Higgs field and magnetic charge, one can define $\fMM\left( (\vec{x}_{n}, P_{n} )_{n = 1}^{N_t} ; \gamma_{\rm m} ; \Phi_{\infty} \right)$, the moduli space of gauge-inequivalent solutions to the Bogomolny equation with singularities \eqref{tHooftpole} at the $\vec{x}_n$ and asymptotic boundary conditions determined by $(\gamma_{\rm m}; \Phi_\infty)$.  This space comes equipped with a natural metric and around generic points it is a smooth Riemannian manifold.  The main result of this paper is a formula for the dimension of $\fMM$.  We state and discuss this result in section \ref{sec:resultsummary} below.  First, however, we provide some context and motivation for it.

There has been a great deal of work on singular monopoles and their moduli spaces originating with Kronheimer \cite{Kronheimer}.  He exhibited an intriguing correspondence between singular $SU(2)$ monopoles on $\mathbb{R}^3$ and $SU(2)$ instanton configurations on the (multi-centered) Taub-NUT manifold, invariant under a certain $U(1)$ action.  He then went on to set up a minitwistor approach to singular monopole moduli space along the lines of Hitchin's work \cite{Hitchin:1982gh}.  The connection to $U(1)$-invariant instantons on Taub-NUT is analogous to the relation between smooth monopoles and $U(1)$-invariant instantons on $\mathbb{R}^3 \times S^1$.  In the Taub-NUT case the singularities of the monopole configuration on the $\mathbb{R}^3$ base are neatly encoded by the shrinking of the circle fiber at the nuts.

Singular $SU(2)$ monopoles on arbitrary compact Riemannian three-manifolds were considered by Pauly \cite{MR1624279}, who computed the dimension of the moduli space by exploiting the relation with $U(1)$-invariant instantons and applying the Atiyah--Singer fixed point theorem to the appropriate zero-mode operator.  Global smooth solutions to the Bogomolny equation on compact three-folds are rather trivial---the connection must be flat and the Higgs field covariantly constant \cite{MR1624279}---so it is natural to consider singular monopoles on such spaces.  Equivariant index techniques have not been applied to compute the dimension of singular monopole moduli spaces in the case of $\mathbb{R}^3$, presumably due to the difficulties in working with an equivariant Atiyah--Patodi--Singer index theorem for manifolds with boundary.\footnote{See \cite{MR511246} for the formulation of such a theorem.}

Singular monopoles and their moduli spaces have also made various appearances in the physics literature.  In configurations of D1-branes stretched between D3-branes, the endpoint of the D1-brane induces a magnetic monopole configuration in the low energy D3-brane worldvolume theory.   Finite length D1-branes lead to smooth monopole configurations while semi-infinite D1-branes ending on D3-branes give singular monopoles, as was first pointed out in the T-dual context of Hanany and Witten \cite{Hanany:1996ie}.  As shown by Diaconescu, the D-brane picture provides an explicit geometric realization of the Nahm, or ADHM--N, construction of magnetic monopoles \cite{Diaconescu:1996rk}.  Cherkis and Kapustin described singular monopoles in terms of solutions to the Nahm equation on a semi-infinite interval, and went on to construct explicit moduli spaces in several examples for the $G = SU(2)$ theory \cite{Cherkis:1997aa,Cherkis:1998xca,Cherkis:1998hi}.

More recently, Cherkis has developed the bow formalism \cite{Cherkis:2008ip,Cherkis:2010bn} for constructing instanton configurations on Taub-NUT space.  It is a synthesis of the Nahm transform and the quiver techniques of Kronheimer--Nakajima \cite{MR1075769} for studying instantons on ALE spaces.  The moduli space of bow data is argued to be isometric to the instanton moduli space in \cite{Cherkis:2010bn}, and this leads to a presentation of the moduli spaces in terms of finite dimensional hyperk\"ahler quotients.  A special subclass of bows, referred to as \emph{Cheshire} bows, represents $U(1)$-invariant instantons and hence, by \cite{Kronheimer}, singular monopole configurations.  The Cheshire bow formalism has been used to produce explicit solutions for one $SU(2)$ 't Hooft--Polyakov monopole in the presence of an arbitrary number of minimal \tHooft defect singularities \cite{Blair:2010vh}.  Cheshire bows have not yet been used to study singular monopoles in higher rank gauge groups or with arbitrary defect charges $P_n$.  Comparison of the bow formalism with our results is an interesting problem.

Moduli spaces of singular monopoles on compact manifolds of the type $I \times C$ with $I$ an interval and $C$ a Riemann surface play an important role in the work of Kapustin and Witten on the geometric Langlands program \cite{Kapustin:2006pk}.  One of many results obtained in that paper is a generalization of Pauly's formula \cite{MR1624279} for the dimension of the moduli space to arbitrary compact, simple $G$.  We discuss the relation of this formula to ours in section \ref{sec:resultsummary} below.

In the remainder of this section we briefly summarize our main result and provide a physical interpretation of it.  We then lay out a brief outline of results to appear in two subsequent papers.  In section \ref{sec:monopolereview} we review some monopole basics, give a precise definition of the moduli space $\fMM \left( (\vec{x}_{n}, P_{n} )_{n = 1}^{N_t}; \gamma_{\rm m} ; \Phi_{\infty} \right)$, and set up the deformation problem.  In section \ref{sec:index} we recall Weinberg's original computation of the formal\footnote{Formal in the sense that one assumes the existence of the background solution about which the linearized deformation analysis takes place.  One needs an existence theorem as in \cite{Taubes:1981gw} to show that the moduli space is non-empty.  Then its dimension is given by the formal dimension.} dimension in the smooth case \cite{Weinberg:1979ma,Weinberg:1979zt}, which makes use of the Callias index theorem for Dirac operators on open Euclidian space \cite{Callias:1977kg}.  We then extend the analysis to the singular case.  This involves the use of an explicit basis of eigenfunctions of the Dirac operator coupled to the leading order gauge and Higgs field of the 't Hooft defect \eqref{tHooftpole}.  The construction of this basis is a slight generalization of the calculation in \cite{HarishChandra:1948zz}, and is summarized in appendix \ref{app:Dirac}.  Note that the gauge and Higgs field configuration with $F = \half P \sin{\theta} \ed \theta \ed \phi$ and $\Phi = \Phi_{\infty} - \frac{1}{2r} P$ is an exact solution to the Bogomolny equation with an 't Hooft defect of charge $P$ at $\vec{x}_0 = 0$ and asymptotic data $(\gamma_{\rm m} = P; \Phi_\infty)$.  In section \ref{sec:cartanzms} we verify our dimension formula by constructing the explicit basis of $\LL^2$-normalizable zero-modes about this background.  In section \ref{sec:cfamily} we exhibit and study a two-parameter family of singular monopoles in $\mathfrak{su}(2)$ gauge theory, and argue that this family parameterizes a surface inside an eight-dimensional moduli space.  We describe several directions for further study in section \ref{sec:further}.

\subsection{Dimension formula}\label{sec:resultsummary}

We restrict to the case of maximal symmetry breaking, \ie\ regular values of $\Phi_{\infty}$, where the group of global gauge transformations leaving $\Phi_\infty$ invariant is a Cartan torus $T \subset G$.  We can choose a gauge where both $\Phi_{\infty},\gamma_{\rm m}$ are constant over the asymptotic two-sphere and valued in the Cartan subalgebra of the Lie algebra, $\mathfrak{t} \subset \mathfrak{g}$.  Let $\Delta$ be the corresponding root system for the Lie algebra.  Then our result for the dimension of $\fMM\left( (\vec{x}_{n}, P_{n} )_{n = 1}^{N_t} ; \gamma_{\rm m} ; \Phi_{\infty} \right)$, when non-empty, is
\begin{equation}\label{dim1}
\dim_{\mathbb{R}} \fMM = \sum_{\alpha \in \Delta} \left( \frac{ \langle \alpha, \Phi_{\infty} \rangle \langle \alpha, \gamma_{\rm m} \rangle }{ | \langle \alpha, \Phi_\infty \rangle |} + \sum_{n}  | \langle \alpha, P_{n} \rangle | \right) ~.
\end{equation}
Here $\langle~,~\rangle : \mathfrak{t}^\ast \otimes \mathfrak{t} \to \mathbb{R}$ denotes the canonical pairing between $\mathfrak{t}$ and its vector space dual.  By making local gauge transformations, \ie\ ones that go to the identity at infinity but are nontrivial at the singularity, one can conjugate any 't Hooft charge $P_{n}$ by a Weyl transformation.  Thus it is only the Weyl orbit of an 't Hooft charge that is gauge invariant.   Similarly global gauge transformations can be used to implement Weyl transformations on the asymptotic data $(\gamma_{\rm m}; \Phi_{\infty} )$.  Formula \eqref{dim1} is manifestly invariant under such transformations.  This formula follows from a more general result derived in the text (see \eqref{indLrho}) once one restricts to the adjoint representation of the Lie algebra.

A key step in the derivation of this formula is an expression of the index as the integral of a local index density which is a total derivative on the Riemannian three-manifold $\mathbb{R}^3 \setminus \{ \vec{x}_n \}_{n=1}^{N_t}$. (Equation \eqref{indexdensity} below.)  Indeed the expression as a total derivative generalizes to an arbitrary Riemannian metric and hence can be extended to general three-manifolds.  Each term in the parentheses in equation \eqref{dim1} above  originates from a different boundary contribution: the term involving the $n^{\rm th}$ 't Hooft charge $P_{n}$ is the boundary contribution from an infinitesimal two-sphere surrounding $\vec{x}_n$ and the first term involving the asymptotic data is the boundary contribution from the two-sphere at infinity.  If we drop the term involving the asymptotic data then the local contributions are equivalent to those derived in \cite{MR1624279,Kapustin:2006pk} for compact three-manifolds.  Meanwhile in the absence of 't Hooft charges this formula reduces to the classic result of \cite{Weinberg:1979zt}. Given the local nature of the contributions it is quite natural that we simply take the sum of these previous results. Thus, one may view our computation as an alternative derivation of the local contributions near the 't Hooft charges derived by different means in \cite{MR1624279,Kapustin:2006pk}.  As we have said the expression could be generalized to a much larger class of Riemannian three manifolds with boundary and it would be interesting to evaluate the contributions arising in various hyperbolic geometries, in particular making connections with hyperbolic monopoles \cite{MR893593}, but we will not address that in this paper.

Another property, which is not obvious in the form \eqref{dim1}, is that the dimension is always an integer divisible by four.  This is important since it is expected that $\fMM$, with the natural metric induced from the flat metric on field configuration space, is a hyperk\"ahler manifold.  Hyperk\"ahlerity is expected since $\fMM$ can be formally constructed as an infinite-dimensional hyperk\"ahler quotient.  Also, the results of \cite{Cherkis:2010bn} imply that the metric is hyperk\"ahler for the class of examples to which the Cheshire bow construction can be applied.  Note it is crucial that we are considering singular monopoles on $\mathbb{R}^3$ for this property.\footnote{More specifically, one requires that the auxiliary four-manifold on which the corresponding $U(1)$-invariant instanton configuration is constructed should be hyperk\"ahler.  Both $\mathbb{R}^3 \times S^1$ for the smooth case and Taub-NUT for the singular case have this property.}  In particular, without the contribution from the asymptotic boundary, \eqref{dim1} will not in general be an integer multiple of four.

In order to show that \eqref{dim1} is an integer multiple of four we must discuss the lattices the charges $\gamma_{\rm m}$, $P_{n}$ live in.  First, the 't Hooft charges sit in the lattice $\Hom(U(1),T) \cong \{ H \in \mathfrak{t}~ |~ \exp(2\pi H) = 1_G \}$, which is known as the co-character lattice, $\Lambda_G$, or equivalently the character lattice of the GNO or Langlands dual group, $\Lambda_{{}^LG}^\vee$, \cite{Goddard:1976qe}.  For smooth monopoles, topological considerations imply that the asymptotic magnetic charge sits in the co-root lattice, $\Lambda_{\rm cr}$.  This is in general a sub-lattice of $\Lambda_G$, and we have $\Lambda_G /\Lambda_{\rm cr}  \cong \pi_1(G)$ --- the two agree only if $G$ is simply-connected.\footnote{See appendix \ref{app:Lie} for more details.}  However when there are 't Hooft defects present the same arguments can be used to show that the possible values of $\gamma_{\rm m}$ are shifted by the 't Hooft charges and thus sit in the shifted lattice $\Lambda_{\rm cr} + (\sum_{n} P_{n})$.  This is the same set as the co-root lattice if and only if $\sum_n P_{n}$ is in the co-root lattice; in general it is a torsor for the co-root lattice.  Note that if $P_{n}, P_{n}'$ are related by a Weyl transformation, then $P_{n} - P_{n}' \in \Lambda_{\rm cr}$.

Now, given $\Phi_{\infty}$, we can define a system of positive roots, $\Delta^+$, by the condition $\alpha \in \Delta^+ \iff \langle \alpha, \Phi_\infty \rangle > 0$.  (Here we are using the maximal symmetry breaking assumption).  The positive roots determine a unique set of simple roots.  Formula \eqref{dim1} is invariant under $\alpha \to -\alpha$ so we can write it as twice the sum over the positive roots, and when we do this the $\Phi_\infty$ factors cancel out because of our choice of root system.  For each $P_{n}$ let $P_{n}^-$ be the unique element in the Weyl orbit of $P_{n}$ which lies in the closure of the anti-fundamental Weyl chamber, such that $\langle \alpha, P_{n}^- \rangle \leq 0$ for all $\alpha \in \Delta^+$, and define the \emph{relative magnetic charge} $\tilde{\gamma}_{\rm m} \in \Lambda_{\rm cr}$ by $\tilde{\gamma}_{\rm m} := \gm - \sum_n P_{n}^-$.  Finally let $\{ H_I \}_{I=1}^{\rnk{\mathfrak{g}}}$ denote the basis of simple co-roots, and write $\tilde{\gamma}_{\rm m} = \sum_I \tilde{m}^I H_I$, where the $\tilde{m}^I$ are integers.  Then, noting that we are free to replace $P_n$ by $P_{n}^-$ in \eqref{dim1}, we have
\begin{align}\label{dim2}
\dim_{\mathbb{R}} \fMM =&~ 2 \sum_{\alpha \in \Delta^+} \left( \langle \alpha, \gamma_{\rm m} \rangle + \sum_n | \langle \alpha, P_{n}^- \rangle |  \right) = 2 \sum_{\alpha \in \Delta^+} \left( \langle \alpha, \gamma_{\rm m} \rangle - \sum_n \langle \alpha, P_{n}^- \rangle  \right) \cr
 =&~ 2 \sum_{\alpha \in \Delta^+} \langle \alpha, \tilde{\gamma}_{\rm m} \rangle = 4 \langle \varrho, \tilde{\gamma}_{\rm m} \rangle =  4 \sum_{I=1}^{\rnk{\mathfrak{g}}} \tilde{m}^I \langle \rho, H_I \rangle = 4 \sum_{I=1}^{\rnk{\mathfrak{g}}} \tilde{m}^I ~.
\end{align}
where we recalled that the Weyl element $\varrho := \half \sum_{\alpha \in \Delta^+} \alpha$, which is also equal to the sum over all fundamental weights, satisfies $\langle \varrho, H_I \rangle = 1$, for all $I$.

This formula is reminiscent of the one for smooth monopoles \cite{Weinberg:1979zt} in the maximal symmetry breaking case and reduces to it in the absence of 't Hooft defects since then $\tilde{\gamma}_{\rm m} \to \gamma_{\rm m}$.  In that case there is a natural physical interpretation of the result due to Weinberg.  There are $\rnk(\mathfrak{g})$ species of ``fundamental'' monopoles---one for each simple root of the Lie algebra---and a configuration with total charge $\gamma_{\rm m} = \sum_I m^I H_I$ can be thought of as containing $m^I$ monopoles of species $I$ for each $I = 1,\ldots, \rnk{\mathfrak{g}}$.  Each fundamental monopole has four moduli associated with it: three for its position and one for a $U(1)$ phase parameter whose conjugate momentum corresponds to electric charge.  From this point of view one intuitively expects to have solutions to the first order Bogomolny equation $F = \star \eD \Phi$, only when all of the $m^I$ are non-negative.  Configurations with only anti-monopoles ($m^I \leq 0$) would solve $F = - \star \eD \Phi$.  In the case of smooth monopoles this can be rigorously demonstrated, and in fact the statement has a generalization to arbitrary symmetry breaking \cite{Murray:2003mm}.  It was furthermore demonstrated in \cite{Taubes:1981gw} that for any collection of non-negative $\{ m^I \}$ such that $\sum_I m^I > 0$, solutions exist.

We would like to put forward the same interpretation here, and suggest that the configuration with 't Hooft charges $P_{n}$ and asymptotic charge $\gamma_m = \tilde{\gamma}_m + \sum_n P_{n}^-$ can be thought of as $\tilde{m}^I$ smooth, mobile monopoles of species $I$ in the presence of the fixed line defects.  In particular we conjecture that $\fMM$ is non-empty if and only if all $\tilde{m}^I$ are non-negative.  (We will explicitly show that that $\fMM$ is a point in the case when all $\tilde{m}^I = 0$).  We will not prove this conjecture though we show in \cite{MRVdimP2} that it is strongly motivated from intersecting brane configurations.  One could perhaps use the gluing techniques of \cite{MR614447,Taubes:1981gw}, additionally gluing in the appropriate singular field configuration \eqref{tHooftpole} in the vicinity of the 't Hooft defects, to prove existence.  Another conjecture we state here is that the moduli spaces $\fMM$ are connected.  This seems physically reasonable from the picture of fundamental monopoles moving around in the presence of defects, but our analysis of the dimension is local and does not shed light on this issue.

One may wonder what is the physical reason for selecting $P^-$ as the natural representative of the Weyl orbit of $P$ to use in defining the relative magnetic charge $\tilde{\gamma}_{\rm m}$.  Our intuition from the dimension formula \eqref{dim1} in the case without \tHooft defects is that it is the sign of the components of $\gm$ (with respect to some basis of $\mathfrak{t}$) relative to the sign of the components of $\Phi_{\infty}$ (with respect to the same basis) that is physically relevant.  In the case of an \tHooft defect it should be the sign of the components of $P$ relative to sign of the components of the \emph{local} Higgs field that is relevant.  The local Higgs field has a simple pole with residue $-P$.  In order to compare the asymptotic magnetic charge with the \tHooft charge, by making local gauge transformations we should conjugate $-P$ (the local Higgs field) to the closure of the fundamental Weyl chamber defined by $\Phi_\infty$ so that they define the same ``polarization'' of $\mathfrak{t}$, \ie\ the same splitting into positive and negative half-spaces.  Equivalently $P$ should be conjugated to the closure of the anti-fundamental Weyl chamber.\footnote{If we study solutions to $F = - \star \eD \Phi$ instead of $F = \star \eD \Phi$, the definition of $\tilde{\gamma}_{\rm m}$ and the dimension formula will be modified by some signs.  This is discussed in the main text.}

\subsection{Preview of subsequent papers}

This is the first paper in a series of three exploring singular monopoles and the role they play in certain four-dimensional quantum field theories with $\NN = 2$ supersymmetry.

In the second paper of the series \cite{MRVdimP2} we review and expand on the embedding of singular monopole configurations into systems of intersecting D-branes in string theory. The brane realization of monopoles indicates that one should be able to construct singular monopole configurations for gauge group $G$ by taking limits of smooth monopole configurations for gauge group $G'$, with $\rnk{G'} > \rnk{G}$, in which the masses of a subset of the smooth monopoles become infinite. We provide a detailed and precise implementation of this idea in a class of examples. We then demonstrate how our dimension formula \eqref{dim2} agrees with expectations for the dimension of $\fMM$ based on identifying motion on moduli space with motion of branes. This in turn provides strong evidence for the conjecture given above stating the precise conditions on the data $\left( (\vec{x}_{n}, P_{n} )_{n = 1}^{N_t}; \gamma_{\rm m} ; \Phi_{\infty} \right)$ such that solutions to the Bogomolny equation exist. We find that when \tHooft defects are present it is important to take into account the effects of brane bending. This leads us to a physical picture of monopole bubbling and a new, distinct process that we dub ``monopole extraction''. Finally we show how the brane systems can be utilized to understand certain wall-crossing properties of the index formula derived here. See \eg\ \eqref{eq:IndexJump}.

The original motivation for our work comes from the role played by 't Hooft defects, and more general line defects, in four-dimensional gauge theories on $\mathbb{R}^{1,3}$ with $\NN = 2$ supersymmetry. The insertion of an 't Hooft line defect with worldline $\mathbb{R}_t \times \{ \vec{x}_0 \} \subset \mathbb{R}^{1,3}$ modifies the theory in such a way as to preserve half of the original supersymmetry. One can inquire about the existence of BPS states in the modified theory. These were dubbed ``framed BPS states'' in \cite{Gaiotto:2010be},\footnote{This is also the origin of our ``overbar-underbar'' notation $\fMM$ for the moduli spaces considered here.} where an analysis of their properties led to new insights in both mathematics and physics, including a physical derivation of the Kontsevich--Soibelman wall-crossing formula \cite{2008arXiv0811.2435K}, connections with integrable systems, and with moduli spaces of flat connections on Riemann surfaces.

In the third paper of this series \cite{MRVmain} we develop the semiclassical description of framed BPS states. This involves a supersymmetric quantum mechanics on the moduli space of singular monopoles, in which framed BPS states are represented by zero-modes of a Dirac-type operator on $\fMM$. We describe the action of $SU(2)$ $R$-symmetry geometrically and express the protected spin characters introduced in \cite{Gaiotto:2010be}   as weighted traces over the kernels of these Dirac operators. We review and elaborate on some positivity conjectures that have been made for protected spin characters, and proven for 't Hooft defects in pure $SU(N)$ $\NN=2$ gauge
theories \cite{Chuang:2013wt}. We  translate these positivity theorems into a statement about the kernel of the Dirac operator; for example, one form of the theorem is equivalent to the statement that the kernel is chiral. We also use the semiclassical construction to prove a simple vanishing theorem, allowing us to determine the exact spectrum of a class of theories on a special locus in the weak coupling regime. We study some examples in detail, verifying Denef's bound state radius formula and the existence of higher spin states at weak coupling using explicit spinor zero-modes on moduli space. Finally, we explain how some explicit computations of certain line defect vacuum expectation values in \cite{Gaiotto:2010be} can be translated into nontrivial predictions for dimensions of spinor zero-modes on some moduli spaces of singular monopoles.

Recently, the moduli space dynamics of vortices in the presence of defects \cite{Tong:2013iqa} and of monopoles in the presence of Wilson lines \cite{Tong:2014yla} have been studied. These works are similar in spirit to \cite{MRVmain}.

\section{Monopole basics}\label{sec:monopolereview}

Our main goal in this section is to define the moduli space and to set up the linearized deformation problem that determines its formal dimension.  This is mostly a straightforward extension of standard constructs for smooth monopoles to the case with defects.  We will motivate the definition of the moduli space from a physical point of view, using the discussion to set up our notation and conventions.  One aspect we explain, that does not seem to have been appreciated in the previous literature on \tHooft defects, is that having a well-defined variational principle for the Yang--Mills--Higgs action functional with line defect boundary conditions requires the addition of boundary terms to the action that are localized at the defects.  These boundary terms have the added benefit of rendering the energy of singular monopole configurations finite.  This allows one to derive a BPS-type bound on the energy.  This bound agrees with the classical limit of the BPS bound obtained in \cite{Gaiotto:2010be} for framed BPS states in gauge theories with $\NN = 2$ supersymmetry.

\subsection{Boundary terms, finite energy, and boundary conditions}

Let us begin with Yang--Mills--Higgs theory on flat Minkowski space.  In the presence of \tHooft defects the theory will be defined on $M = \mathbb{R}_t \times \UU$, where $\UU := \mathbb{R}^3 \setminus \{ \vec{x}_{n} \}_{n = 1}^{N_t}$.  It consists of a gauge field and adjoint-valued Higgs field, $(A_\mu, \Phi)$, and we take simple and compact gauge group $G$.  We work in geometric conventions where generators of the Lie algebra, $\mathfrak{g}$, are represented by anti-Hermitian matrices, the field strength is $F_{\mu\nu} = 2 \pd_{[\mu} A_{\nu]} + [A_\mu, A_\nu]$, and the covariant derivative $D_\mu \Phi = \pd_\mu \Phi + [A_\mu, \Phi]$.  The Hamiltonian, or energy functional, for the system in the BPS limit of vanishing scalar potential is
\begin{align}\label{Ham}
E =&~ K + V~, \cr
K =&~ \frac{1}{g^2} \int_\UU \ed^3 x \Tr \left\{ E_i E^i + D_0 \Phi D_0 \Phi \right\}~, \cr
V =&~ \frac{1}{g^2} \int_\UU \ed^3 x \Tr \left\{ B_i B^i + D_i \Phi D^i \Phi \right\} + V_{\rm bndry}~,
\end{align}
where $g$ is the Yang--Mills coupling, and $E_i = F_{i0}$ and $B_i = \half \epsilon_{ijk} F^{jk}$ are the non-Abelian electric and magnetic field.  The indices $\mu,\nu = 0,1,2,3,$ while $i,j = 1,2,3$.  We also use form notation $F = \half F_{\mu\nu} \ed x^\mu \ed x^\nu$, $\eD \Phi = D_\mu \Phi \ed x^\mu$ when convenient.  We use ``$\Tr$'' to denote a positive-definite bi-invariant form on $\mathfrak{g}$.\footnote{A standard choice is $\Tr = - (2 h^\vee)^{-1} \tr_{\rm adj}$ in terms of the Cartan-Killing form and the dual Coxeter number.  With this choice we have $\Tr = -\tr_{\bf N}$, minus the trace in the fundamental representation, for $G = SU(N)$.  Further details on our Lie algebra conventions can be found in appendix \ref{app:Lie}.}  The canonical variables are $(A_i,\Phi)$ with conjugate momenta $\pi_i = E_i$ and $\pi_\Phi = D_0 \Phi$.  In order to give Lorentz covariant dynamics they should be subjected to the Gauss Law constraint, $D^i E_i - \half [\Phi, D_0 \Phi] = 0$, which arises as the $A_0$ equation of motion in the Lagrangian formulation.  $V_{\rm bndry}$ denotes boundary terms associated with the defects.  Their presence is required in order to have a well-defined variational principle and furthermore leads to finite energies for singular monopole configurations, as we will see below.

Our interest here is in static configurations.  We fix the temporal dependence of the gauge freedom by working in $A_0$ = 0 gauge.  Bogomolny observed that the potential can be written in the form
\begin{equation}\label{Vibp}
V = \frac{1}{g^2} \int_\UU \ed^3 x \Tr \left| B_i \mp D_i \Phi \right|^2 \pm \frac{2}{g^2} \int_{\pd \UU} \Tr \{ \Phi F \} + V_{\rm bndry}~.
\end{equation}
For static, finite energy configurations this implies the bound on the energy
\begin{equation}\label{bbound}
E \geq \pm \frac{2}{g^2} \int_{\pd \UU} \Tr \left\{ \Phi F \right\} + V_{\rm bndry}~,
\end{equation}
which is saturated by field configurations solving the first order equations $B_i = \pm D_i \Phi$.  Such configurations are necessarily solutions to the second order equations of motion.  In the absence of defects the sign in the above two formulae should be chosen such that the bound \eqref{bbound} is maximal; the choice will depend on the asymptotic form of $\Phi,F$.  When defects are present this sign will instead be dictated by the defect boundary conditions.  In the following we use $\sigma = \pm$ to encode this sign.

In the absence 't Hooft defects, such that $\UU = \mathbb{R}^3$ and $V_{\rm bndry} = 0$, finite energy follows from the large $r$ boundary conditions
\begin{align}\label{finen}
\Phi =&~ \tilde{\Phi}(\hat{r}) - \sigma \frac{M(\hat{r})}{2 r} + O(r^{-(1+\updelta)})~, \qquad B = \frac{M(\hat{r})}{2 r^2} \hat{r} + O(r^{-(2+\updelta)})~.
\end{align}
Here $\updelta > 0$, $\tilde{\Phi},M$ are commuting, covariantly constant sections of the adjoint bundle over $S_{\infty}^2$,
\begin{equation}
D_i \tilde{\Phi} |_{S_{\infty}^2} = D_i M |_{S_{\infty}^2} = 0~, \qquad [\tilde{\Phi},M] = 0~,
\end{equation}
and the Bogomolny equation $B_i =  \sigma D_i \Phi$ has been used to relate the $r^{-1}$ and $r^{-2}$ terms in the Higgs and magnetic field.  It has been proven that these are also necessary conditions for finite energy when the gauge group is $SU(2)$ \cite{MR614447}, and this is expected to be true in general \cite{MR1625475,Murray:2003mm}.  (See also the discussion in \cite{Kampmeijer:2008wz}.)

When \tHooft defects are present there are certain boundary terms that should be included in the energy functional.  This follows from demanding consistency of the defect boundary conditions \eqref{tHooftpole} with a variational principle, as we now demonstrate.  The Hamiltonian with Gauss constraint can be derived from the action
\begin{align}\label{action}
S_{\rm YMH} =&~ -\frac{1}{g_{0}^2} \int \ed^4 x \Tr \left\{ \half F_{\mu\nu} F^{\mu\nu} + D_\mu \Phi D^\mu \Phi \right\} - \int \ed t V_{\rm bndry} \cr
= &~ S_{\rm bulk} + S_{\rm bndry}~,
\end{align}
where $S_{\rm bndry}$ is minus the time integral of the boundary potential.  Variation of the bulk term yields
\begin{align}
\delta S_{\rm bulk} =&~ \frac{2}{g^2} \int_M \ed^4 x \Tr \left\{ \left( D^\mu F_{\mu\nu} - [\Phi, D_\mu \Phi] \right) \delta A^\nu + 2 (D^\mu D_\mu \Phi) \delta \Phi \right\} + \cr
&~ - \frac{2}{g^2} \int_{\pd M} \ed^3 x \sqrt{\gamma_\pd} n^\mu \Tr \left\{ F_{\mu\nu} \delta A^\nu + (D_\mu \Phi) \delta \Phi \right\}~,
\end{align}
where the second line can be decomposed into a sum of integrals over each boundary component of $M$, $\ed^3x \sqrt{\gamma_\pd}$ is the induced volume form on the boundary, and $n^\mu$ is the unit normal vector.  Boundary conditions on the fields follow from the boundary terms in the variation, since a solution to the equations of motion should extremize the action: $\delta S = 0$.  The temporal boundary terms at $t = \pm \infty$ are zero for the class of static field configurations that we consider.

In the presence of defects there are spatial boundary terms associated with the boundary components $\mathbbm{R}_t \times S_{\varepsilon_n}^2$, where $S_{\varepsilon_n}^2$ is an infinitesimal two-sphere of radius $\varepsilon_n$ surrounding $\vec{x}_n$.  Let $\vec{r}_{n} = \vec{x} - \vec{x}_n$.  On a static solution to the equations of motion we then have
\begin{align}
\delta S_{\rm bulk} =&~ \frac{2}{g_{0}^2} \int \ed t \left( \lim_{r \to \infty} \int_{S_{\infty}^2} \ed\Omega r^2 \hat{r}^i -\sum_n \lim_{\varepsilon_n \to 0} \int_{S_{\varepsilon_n}^2} \ed \Omega_n \varepsilon_{n}^2 \hat{r}_{n}^i \right)  \times \cr
& \qquad \qquad \qquad \qquad \qquad \qquad \qquad \qquad  \times \Tr \left\{ \epsilon_{ijk} B^k \delta A^j + (D_i \Phi) \delta \Phi \right\}~, \qquad
\end{align}
where the relative minus between the asymptotic two-sphere and infinitesimal ones is due to their orientation induced from $\UU$.  Supposing defect boundary conditions of the form
\begin{equation}\label{trialbc}
B^k = \frac{P_n}{2 \varepsilon_{n}^2} \hat{r}_{n}^k + O(\varepsilon_{n}^{-2 + \updelta'})~, \quad \Phi = -\sigma \frac{P_n}{2 \varepsilon_n} + O(\varepsilon_{n}^{-1 + \updelta'})~, \qquad \textrm{as $\varepsilon_n \to 0$}~,
\end{equation}
for some $\updelta' > 0$, such that the variations $\delta A^j, \delta \Phi = O(\varepsilon_{n}^{-1 + \updelta'})$, one finds that the boundary terms in $\delta S_{\rm bulk}$ from the infinitesimal two-spheres go as $\varepsilon_{n}^{-1 + \updelta'}$ and are divergent or finite even for $\updelta' = 1$.  We want to choose the boundary action such that its variation cancels these terms, and makes the defect boundary conditions consistent with $\delta S = 0$.  A simple and natural choice that does the job is
\begin{equation}\label{SVbndry}
S_{\rm bndry} = - \int \ed t V_{\rm bndry} =  \int \ed t \left( - \frac{2 \sigma}{g^2} \sum_n \int_{S_{\varepsilon_n}^2} \Tr \left\{ \Phi F \right\} \right)~.
\end{equation}
Noting that $F |_{S_{\varepsilon_n}^2} = \ed \Omega_n \hat{r}_{n}^i B_{i}$, this gives us
\begin{align}\label{Svar}
\delta S_{\rm YMH} =&~ \frac{2}{g_{0}^2} \int \ed t \bigg( \lim_{r \to \infty} \int_{S_{\infty}^2} \ed\Omega r^2 \hat{r}^i \Tr \left\{ \epsilon_{ijk} B^k \delta A^j + (D_i \Phi) \delta \Phi \right\} + \cr
& +\sum_n \lim_{\varepsilon_n \to 0} \int_{S_{\varepsilon_n}^2} \ed \Omega_n \varepsilon_{n}^2 \hat{r}_{n}^i  \Tr \left\{ (F_{ij} - \sigma \epsilon_{ijk} D^k \Phi )\delta A^j + (D_i \Phi - \sigma B_i) \delta \Phi \right\} \bigg)~. \qquad \quad
\end{align}
The terms in the second line vanish on a solution to the Bogomolny equation.  More generally, however, a consistent variational principle requires that $\delta S = 0$ on any solution to the (second order) equations of motion.  The leading order divergence of \eqref{trialbc} cancels out so that $D_i \Phi - \sigma B_i = O(\varepsilon^{-2 + \updelta'})$, whence the boundary variation $\delta S = O(\varepsilon_{n}^{-1 + 2\updelta'})$ as $\varepsilon_n \to 0$.  Naively, this means we should require $\updelta' > \half$ in \eqref{trialbc}.  However, we show in appendix \ref{app:admissible} that any solution to the equations of motion satisfying \eqref{trialbc} also satisfies $D_i \Phi - \sigma B_i = 0$ at the first subleading order.  Hence, $\updelta' = \half$ is also admissible.  The reason we stress this point is that later we will construct explicit zero-mode fluctuations $(\delta A_i, \delta \Phi)$ that have this behavior.  Furthermore such behavior can be observed in explicit solutions to the Bogomolny equation found in \cite{Cherkis:2007jm,Cherkis:2007qa}, representing one smooth monopole in the presence of a minimal \tHooft defect in $SO(3)$ gauge theory.  Thus we arrive at the boundary conditions \eqref{tHooftpole}.

In addition to providing a consistent variational principle, the boundary potential \eqref{SVbndry} also regulates the energy of a field configuration satisfying defect boundary conditions.  Plugging into \eqref{Vibp}, \eqref{bbound}, (and keeping in mind that we are now denoting the $\pm$ in that equation by $\sigma$), we find that the $S_{\varepsilon_n}^2$ boundary terms cancel, leaving only
\begin{align}\label{Estatic}
E =&~ \frac{1}{g^2} \int \ed^3 x \Tr | B_i - \sigma D_i \Phi|^2 + \frac{2\sigma}{g^2} \int_{S_{\infty}^2} \Tr\{ \Phi F \} \cr
\geq &~ \frac{2\sigma}{g^2} \int_{S_{\infty}^2} \Tr\{ \Phi F \}~,
\end{align}
for static field configurations.  This result has the same form whether or not defects are present, and therefore we impose the same asymptotic boundary conditions, \eqref{finen}.  In addition to ensuring finiteness of the energy, they imply that the $S_{\infty}^2$ boundary term in the variation \eqref{Svar} will vanish.  Furthermore, the bound \eqref{Estatic} is consistent with the (classical limit of the) BPS bound for framed BPS states found in \cite{Gaiotto:2010be}.

\subsection{Gauge transformations and the moduli space}

Let us discuss the role of gauge transformations.  Having fixed time dependence by working in $A_0 = 0$ gauge, the residual gauge symmetry consists of time-independent transformations, $\cg : \UU \to G$.  These act on the fields sending $(A,\Phi) \to (A',\Phi')$ with
\begin{equation}\label{finitegt}
A = \Ad_{\cg^{-1}}(A') + \cg^\ast \theta~, \qquad \Phi = \Ad_{\cg^{-1}}(\Phi')~,
\end{equation}
where $\theta$ is the Maurer--Cartan form on $G$; for matrix groups, $\cg^\ast \theta = \cg^{-1} \ed\cg$ and $\Ad_{\cg}(H) = \cg H \cg^{-1}$.  If $\cg = \exp(\epsilon)$ then these transformations correspond to the infinitesimal action $A \to A' =  - \eD \epsilon$, $\Phi \to \Phi' = \ad(\epsilon)(\Phi) = [\epsilon,\Phi]$.

When defining the moduli space as a set of gauge inequivalent field configurations we must distinguish between local gauge transformations such that $\lim_{r\to \infty} \cg = 1_{G}$ and global gauge transformations that can be asymptotically nontrivial.  Two field configurations related by the former are physically equivalent and we want to divide out by this equivalence relation.  In contrast we do not identify field configurations related by global gauge transformations.  Rather we can use global gauge transformations to infer properties of the moduli space.  For example, gauge covariance of the Bogomolny equation implies that, for a given set of 't Hooft defects, if two sets of asymptotic data $(M(\hat{r}),\tilde{\Phi}_\infty(\hat{r}))$, $(M'(\hat{r}),\tilde{\Phi}_{\infty}'(\hat{r}))$, are related by a global gauge transformation then the corresponding moduli spaces will be isometric.  Thus we want to use global gauge transformations to make the asymptotic data as simple as possible.

To each regular element of $\mathfrak{g}$ we can associate a unique Cartan subalgebra $\mathfrak{t}$.  Pick a point on $S_{\infty}^2$, say the north pole $\hat{p}_{\rm n}$, and let the value of the Higgs field there, $\Phi_\infty := \tilde{\Phi}(\hat{p}_{\rm n})$, define our Cartan subalgebra.  Since $\tilde{\Phi}(\hat{r})$ is covariantly constant on $S_{\infty}^2$ we can make a patch-wise gauge transformation that brings $\tilde{\Phi}(\hat{r})$ to $\Phi_\infty$ everywhere.  As $M(\hat{r})$ is also covariantly constant and commutes with $\tilde{\Phi}(\hat{r})$ these gauge transformations bring $M$ to a constant $\gm \in \mathfrak{t}$, so that
\begin{align}\label{asymptoticbcs}
\Phi_{\rm n,s}' =&~ \Ad_{\cg_{\rm n,s}}(\Phi) = \Phi_\infty - \sigma \frac{\gm}{2 r} + O(r^{-(1+\updelta)})~,  \cr
B_{\rm n,s}' =&~ \Ad_{\cg_{\rm n,s}}(B) = \frac{\gm}{2 r^2} \hat{r} + O(r^{-(2+\updelta)})~,
\end{align}
where ${\rm n,s}$ refer to patches covering the northern and southern hemisphere.  The magnetic field corresponds to an asymptotic two-form field strength $F_{\rm n,s}' \to \half \gm \sin{\theta} \ed \theta \ed \phi$.  This is the form of the asymptotic boundary conditions quoted in \eqref{largerbc}.  In terms of these data the energy \eqref{Estatic} of a static field configuration is
\begin{equation}\label{Ebound}
E =  \frac{1}{g^2} \int_\UU \ed^3 x \Tr \left| B_i - \sigma D_i \Phi \right|^2 + \sigma \frac{4\pi}{g^2} \Tr ( \Phi_\infty \gm ) \geq \sigma \frac{4\pi}{g^2} \Tr ( \Phi_\infty \gm )~.
\end{equation}

Unless $\gm$ is trivial, we will have a patch-dependent asymptotic gauge field
\begin{equation}
A_{\rm n}' \to \half \gm (1- \cos{\theta}) \ed \phi~, \qquad A_{\rm s}' \to \half \gm (-1- \cos{\theta}) \ed \phi~, \quad \textrm{as $r \to \infty$}~.
\end{equation}
These gauge fields are related by a gauge transformation with the transition function $\cg_{\rm sn} = \cg_{\rm s}^{-1} \cg_{\rm n} = \exp(\gm \phi)$ on the overlap of the patches.  Single-valuedness of the transition function requires $\exp{(2\pi \gm)} = 1_G$, and thus $\gm \in \Lambda_G \cong \Hom(U(1),T)$, where $T \subset G$ is the Cartan torus obtained by exponentiating $\mathfrak{t}$.  Further restrictions on $\gm$  arise from other considerations, both topological and dynamical, and their form depends strongly on whether \tHooft defects are present or not.

Further topological restrictions arise from demanding that the global gauge transformation in \eqref{asymptoticbcs} be extendable to all of $\UU$.  In the smooth case when $\UU = \mathbb{R}^3$, the principal $G$-bundle over $S_{\infty}^2$ defined by the transition function $\cg_{\rm sn}(\phi)$ must be trivial, since the radial coordinate provides a homotopy of $S_{\infty}^2$ to a point at $r = 0$.  It will be trivial if and only if the closed loop $\phi \mapsto \cg_{\rm sn}(\phi)$ is homotopically trivial in $G$, and this will be the case iff the loop lifts to a closed loop in $\tilde{G}$, the simply-connected cover.  Thus one concludes that $\gm$ sits in a coarser lattice: $\gm \in \Lambda_{\rm cr} \cong \Hom(U(1),\tilde{T})$, the co-root lattice.

Now suppose that a single \tHooft defect of charge $P$ is present at the origin, such that $\UU = \mathbb{R}^3 \setminus \{ 0 \}$.  Then there is a homotopy of the asymptotic two-sphere to the infinitesimal one surrounding the origin.  The $G$-bundle restricted to the infinitesimal two-sphere has a transition function around the equator given by $\cg_{\rm sn} = \exp(P \phi)$, with $P \in \Lambda_G$, while the $G$-bundle restricted to the asymptotic sphere is defined by the transition function $\cg_{\rm sn} = \exp(\gm \phi)$.   Hence we must have that $\gm = \gm' + P$ for some $\gm' \in \Lambda_{\rm cr}$.  Since $P$ need not be in the co-root lattice, $\gm$ need not be in the co-root lattice.  Rather, $\gm$ sits in a shifted copy of the co-root lattice which lacks a zero-element (if $P \notin \Lambda_{\rm cr}$).  Such a set is by definition a \emph{torsor} for the co-root lattice, and this is precisely the type of structure that is observed for the IR charge lattice in the low-energy Seiberg--Witten description of $\NN = 2$ theories probed by line defects \cite{Gaiotto:2010be}.  Note that two \tHooft charges related by a Weyl transformation differ by an element of the co-root lattice.  Therefore the torsor only depends on the Weyl orbit of the \tHooft charge, $\gm \in [P] + \Lambda_{\rm cr}$.  These arguments generalize to the case of multiple \tHooft defects such that $\gm \in \sum_n [P_n] + \Lambda_{\rm cr}$.

Let $\Phi_\infty$ be given. Not all magnetic charges $\gm$ allowed by the above topological classification are realized; \ie\ there do not exist solutions to the Bogomolny equation satisfying the asymptotic conditions for all pairs $(\gm;\Phi_\infty)$.  In the case without defects there is a straightforward restriction that follows from the energy bound.  Suppose that $(\gm;\Phi_\infty)$ are such that $\Tr(\Phi_\infty \gm) < 0$.  Then, by choosing $\sigma = -$ in \eqref{Ebound} we learn that $E \geq E_{\rm min} =  \frac{4\pi}{g^2} | \Tr(\Phi_\infty \gm) |$.  However a solution to $B_i = + D_i \Phi$ with these boundary conditions would have $E =  \frac{4 \pi}{g_{0}^2} \Tr(\Phi_\infty \gm) < E_{\rm min}$ --- a contradiction.  Thus one concludes that solutions to $B_i = D_i \Phi$ with $\Tr(\Phi_\infty \gm) < 0$ do not exist.

We stress that there is no such analogous argument in the case with defects, because one is not free to consider either sign of $\sigma$.  The choice of $\sigma$ is dictated by specifying the boundary conditions defining the defect.  An \tHooft line defect depends on three pieces of data:
\begin{equation}
L_{\rm tH} = L_{\rm tH}(\sigma,P; \vec{x}_0)~,
\end{equation}
its location, charge, and the choice of sign $\sigma$.  These data enter into the boundary conditions on the fields that define the defect as follows:
\begin{equation}\label{tHooftpole2}
\Phi = -\sigma \frac{P}{2 r} + O(r^{-1/2})~, \qquad F = \half P \sin{\theta} \ed \theta \ed \phi + O(r^{-3/2})~,
\end{equation}
with $\vec{x} - \vec{x}_0 = (r\sin{\theta}\cos{\phi},r \sin{\theta}\sin{\phi},r\cos{\theta})$.  If we have multiple \tHooft defects we require that the same choice of $\sigma$ be made for each.\footnote{This $\mathbb{Z}_2$ choice is promoted to the choice of a $U(1)$ phase, denoted $\zeta$ in \cite{Gaiotto:2010be}, in the embedding of the Yang--Mills--Higgs system into $\NN = 2$ theories where the Higgs field is complexified.}  With $\sigma$ given we must solve $B_i = \sigma D_i \Phi$; solutions to $B_i = -\sigma D_i \Phi$ would not satisfy \eqref{tHooftpole2}.  The energy bound will be $E \geq E_{\rm min} = \sigma \frac{4\pi}{g^2} \Tr(\Phi_\infty,\gm)$ and we cannot deduce restrictions on $\gm$ by comparing two different bounds as we did above.  In the language of $\NN = 2$ supersymmetry, the choice of $\sigma$ in the smooth case corresponds to the choice of whether we consider monopoles or anti-monopoles; they preserve different subsets of the supersymmetries and we are free to consider either.  In contrast, the \tHooft defect determines which subset of supersymmetries is to be preserved and there are no further choices to be made.

Despite this difference, we argue there is still a strong dynamical constraint on the charges $\gm$ for which there exist solutions to the Bogomolny equation.  Again, let us briefly recall the analogous result for the case without defects.  Let $\{ \alpha_I ~|~ I = 1,\ldots, \rnk{\mathfrak{g}} \}$ be a system of simple roots determined uniquely by the regular element $\Phi_{\infty} \in \mathfrak{t}$, and let $H_I$ be the corresponding simple co-roots.  Then solutions to $B_i = \sigma D_i \Phi$, subject to the boundary conditions $(\gm; \Phi_\infty)$ exist if and only if $\gm = \sigma \sum_I m^I H_I$ with all $m^I$ non-negative.  Note this is a much stronger statement than what one deduces from the simple argument involving the energy bound given above.  The physical motivation for it was discussed in the introduction following \eqref{dim2}, where we also discussed a conjectural analogous condition when \tHooft defects are present.

\begin{figure}
\begin{center}
\includegraphics{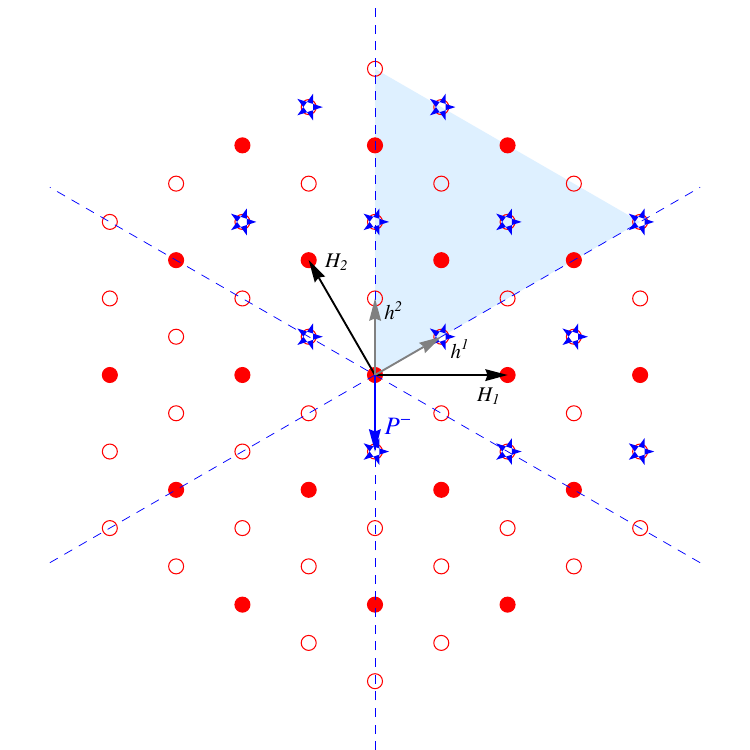}
\caption{The (co-)root diagram for $A_2$.  Filled red dots are elements of the co-root lattice and open red dots are elements of the magnetic weight lattice, which is the co-character lattice of $PSU(3)$.  The shaded region is the fundamental Weyl chamber.  $H_{1,2}$ are the simple co-roots and $h^{1,2}$ are the fundamental magnetic weights.  We have chosen an \tHooft defect with charge $P$ such that $P^- = -h^{2}$.  The stars represent the asymptotic magnetic charges $\gm$ for which we expect the moduli space $\fMM$ to be non-empty.}
\label{fig0}
\end{center}
\end{figure}

In the case with defects we conjecture the following.  Let $P_{n}^- (P_{n}^+)$ denote the representative of $[P_n]$ in the closure of the anti-fundamental (fundamental) Weyl chamber.  If $\sigma = +$ we choose $P^-$ and vice versa; we denote this as $P^{-\sigma}$.  Then define the \emph{relative magnetic charge} $\tilde{\gamma}_{\rm m} := \gm - \sum_n P_{n}^{-\sigma}$.  This is the generalization of Kronheimer's ``non-Abelian'' $SU(2)$ charge \cite{Kronheimer} to arbitrary compact simple $G$.  It is a measure of the charge due to the smooth monopoles in the system.  $\tilde{\gamma}_{\rm m}$ is an element of the co-root lattice, and we claim that solutions to the Bogomolny equation $B_i = \sigma D_i \Phi$ exist if and only if $\tilde{\gamma}_{\rm m} = \sigma \sum_I \tilde{m}^I H_I$ with all $\tilde{m}^I \geq 0$.  We show in \cite{MRVdimP2} that this claim is strongly motivated by brane configurations in string theory that realize singular monopoles.  Note that by construction this condition only depends on $\sigma$ and the Weyl orbit, $[P]$, of $P$.  In Figure \ref{fig0} we give an example of the set of allowed asymptotic magnetic charges for $G = PSU(3)$, $\sigma = +$, and $P^- = -h^{2}$, where $h^{1,2}$ are the fundamental magnetic weights of $\mathfrak{su}(3)$.

Having discussed when we expect the moduli space to be non-empty, it is high time that we define it.  In order to define the moduli space, we first define the group of local gauge transformations.  Consider the action of gauge transformations in the vicinity of an \tHooft defect.  Although two charges $P,P' \in \Lambda_G$ related by a Weyl transformation are physically equivalent, it will be convenient to define the moduli space for a given set of $P_n \in \Lambda_G$, rather than for a given set of Weyl orbits of \tHooft charges.  Thus we require elements in the group of local gauge transformations to leave the $P_n$ invariant.  If $\cg$ is a gauge transformation, let $\cg_n := \cg |_{S_{\varepsilon_n}^2}$ be the restriction to the infinitesimal two-sphere surrounding $\vec{x}_n$.  We define
\begin{equation}\label{localgts}
\GG_{\{P_n\}} := \left\{ \cg : \UU \to G ~|~ \Ad_{\cg_{n}}(P_n) = P_n~, \forall n~, \textrm{ and } \lim_{r \to \infty} \cg = 1_G \right\}~.
\end{equation}
Since the principal $G$-bundle over $\UU$ may be nontrivial,\footnote{It will be nontrivial iff any of the $P_n \in \Lambda_G$ satisfy $P_n \notin \Lambda_{\rm cr}$ --- \ie\ if there is nontrivial \tHooft flux.} we should really speak of a collection of smooth patch-wise transformations $\cg_\alpha : \UU_\alpha \to G$ with $\{ \UU_\alpha \}$ an open cover for $\UU$ and the $\cg_\alpha$ patched together appropriately via the transition functions $\cg_{\alpha\beta}$ of the bundle.  Similar remarks of course apply to the Higgs field and gauge field.  We understand ``$\cg,\Phi,A$'' to denote such collections.  Also, in order to be more precise about \eqref{localgts}, if $\GG_{\{P_n \}} \ni \cg = \exp(\epsilon)$, then we require $\epsilon = O(r^{-1})$ as $r \to \infty$ and $\epsilon = \epsilon_n + O(|\vec{x} - \vec{x}_n|^{1/2})$ as $\vec{x} \to \vec{x}_n$, where $\epsilon_n : S_{\varepsilon_n}^2  \to \mathfrak{g}$ satisfies $\cg_n = \exp(\epsilon_n)$ and $[\epsilon_n ,P_n] = 0$.

The moduli spaces of interest in this paper are then
\begin{align}\label{Mdefsigma}
& \fMM_{\sigma}\left( (\vec{x}_{n}, P_{n} )_{n = 1}^{N_t} ; \gamma_{\rm m} ; \Phi_{\infty} \right) := \cr
&~  \left\{ (A,\Phi) ~ \bigg| ~ B_i = \sigma D_i \Phi~, \begin{array}{l} \Phi = - \frac{\sigma }{2 |\vec{x} - \vec{x}_n|} P_n + O(|\vec{x}-\vec{x}_n|^{-1/2})~,~ \vec{x} \to \vec{x}_n~, \\  \Phi = \Phi_\infty - \frac{\sigma}{2|\vec{x}|} \gm + O(|\vec{x}|^{-(1 + \updelta)})~, ~ |\vec{x}| \to \infty \end{array}  \right\} \bigg/ \GG_{\{ P_n \}}~. \qquad ~ ~
\end{align}
This defines the space as a set of gauge equivalence classes of solutions to the Bogomolny equation satisfying prescribed boundary conditions.  In the next subsection we will recall the additional structure that makes $\fMM$ a hyperk\"ahler manifold.  The boundary conditions discussed above for the gauge field follow from the Bogomolny equation and the boundary conditions on the Higgs field.  

We have defined the moduli space for either case of the sign $\sigma$.  However we see from the definition that it only depends on the product $\sigma \Phi_\infty$.  This motivates the definition of a new Higgs field that absorbs the sign $\sigma$:
\begin{equation}\label{Xdef}
X := \sigma \Phi~.
\end{equation}
Then an equivalent definition of the moduli space is
\begin{align}\label{Mdef}
& \fMM\left( (\vec{x}_{n}, P_{n} )_{n = 1}^{N_t} ; \gamma_{\rm m} ; X_{\infty} \right) := \cr
& ~  \left\{ (A,X) ~ \bigg| ~ B_i = D_i X~,\begin{array}{l} X = - \frac{1}{2 |\vec{x} - \vec{x}_n|} P_n + O(|\vec{x}-\vec{x}_n|^{-1/2})~,~ \vec{x} \to \vec{x}_n~, \\  X = X_\infty - \frac{1}{2|\vec{x}|} \gm + O(|\vec{x}|^{-(1 + \updelta)})~, ~ |\vec{x}| \to \infty \end{array}  \right\} \bigg/ \GG_{\{ P_n \}}~. \qquad ~ ~
\end{align}
We work mostly with the definitions \eqref{Xdef} and \eqref{Mdef} in the remainder of the paper.

\subsection{Deformations and the tangent space}\label{sec:tangentspace}

To compute the dimension of $\fMM$ we compute the dimension of the tangent space, $T_{[(A,X)]} \fMM$, at a point $[(A,X)] \in \fMM$.  It is convenient to introduce the notation $\hat{A} = (A,X)$, which we think of as a $U(1)$-invariant gauge field on $\UU \times S^1$,
\begin{equation}
\hat{A} = \hat{A}_a \ed x^a = A + X \ed x^4~,
\end{equation}
where $x^a = (x^i,x^4)$ are coordinates on $\UU \times S^1$ with orientation such that $ \ed^3 x \wedge \ed x^4$ is positive.  The Bogomolny equation for $(A,X)$ is equivalent to the self-duality of the field strength $\hat{F} = \ed \hat{A} + \hat{A} \wedge\hat{A} = \hat{\star} \hat{F}$.  We take the circle bundle over $\UU$ to be trivial and the metric on the total space to be flat, $\ed s^2 = \ed x_i \ed x^i +(\ed x^4)^2$.

In order to compute the dimension of $T_{[\hat{A}]} \fMM$, we use the one-to-one correspondence between tangent vector fields and flows, or one-parameter families of diffeomorphisms.  We have that $[\hat{A}] \to [\hat{A}'] = [\hat{A}] + [\delta \hat{A}]$ will be the infinitesimal flow corresponding to a (non-zero) tangent vector $\delta \in T_{[\hat{A}]} \MM$, if and only if $[\hat{A}']$ satisfies $\hat{F} = \hat{\star} \hat{F}$ to $O(\delta^2)$ and $[\delta \hat{A}] \neq 0$, that is, $\delta \hat{A}$ is not pure gauge.  The first condition says that $\delta \hat{A}$ should satisfy the linearized self-duality equation:
\begin{equation}\label{linearsd}
\hat{D}_{[a} \delta \hat{A}_{b]} = \half \epsilon_{ab}^{\phantom{ab}cd} \hat{D}_c \delta \hat{A}_d~,
\end{equation}
where $\hat{D}$ is the covariant derivative with respect to background solution $\hat{A}$.  To quantify the second condition it is useful to introduce a metric on the space of finite-energy field configurations and require $\delta \hat{A}$ to be orthogonal to gauge transformations.  In fact the kinetic energy part of \eqref{Ham} defines the appropriate metric\footnote{The factor of two is a normalization convention.  This is so that in a collective coordinate expansion the kinetic terms would have canonical normalization, $\int \ed t \half g_{mn} \dot{z}^m \dot{z}^n$.}:
\begin{equation}\label{metC}
g(\delta_1,\delta_2) = \frac{2}{g^2} \int_{\UU} \ed^3 x \Tr \left\{ \delta_1 \hat{A}_a \delta_2 \hat{A}^a \right\}~.
\end{equation}
Note it is natural that $\delta \hat{A}_a \in \LL^2[\UU,\mathbb{R}^4 \otimes \mathfrak{g}]$, the space of square-normalizable $\mathbb{R}^4 \otimes \mathfrak{g}$-valued functions on $\UU$: $\delta \hat{A}_a$ is the difference between two solutions to the Bogomolny equation satisfying the same asymptotic and \tHooft defect boundary conditions, so it follows from \eqref{tHooftpole} that $\delta \hat{A} = O(\varepsilon_{n}^{-1/2})$ as $\varepsilon_n = |\vec{x} - \vec{x}_n| \to 0$, and from \eqref{largerbc} that $\delta \hat{A} = O(r^{-(1+\updelta)})$ as $|\vec{x}| = r \to \infty$.  These conditions are sufficient to ensure square-normalizability.  Now, choosing $\delta_2 = \delta_\epsilon$ to be the tangent vector corresponding to a local gauge transformation generated by $\epsilon(\vec{x}) \in \mathfrak{g}$, $\delta_{\epsilon} \hat{A} = - \hat{\eD} \epsilon$, we find that $g(\delta,\delta_\epsilon) = 0$ if and only if
\begin{equation}\label{gaugeorth}
\hat{D}^a \delta \hat{A}_a = 0~.
\end{equation}
Here we have used that $\exp{(\epsilon)} \in \GG_{ \{P_n \} }$ implies that $\lim_{r \to \infty} \epsilon(\vec{x}) \to 0$ fast enough to kill the boundary term at infinity, and $\lim_{\vec{x} \to \vec{x}_n} \epsilon(\vec{x})$ is regular such that the boundary terms from $S_{\varepsilon_n}^2$ vanish as well.

The number of linearly independent, $\LL^2$-normalizable solutions $\delta \hat{A}_a$ to \eqref{linearsd} and \eqref{gaugeorth} determines the dimension of $T_{[\hat{A}]}\fMM$.  Together they total four independent equations which can be combined into a chiral Dirac equation \cite{PhysRevD.16.417,Weinberg:1979ma}.  Let $(\tau^a)_{\alpha\dot{\alpha}} = (\vec{\sigma}, -i\mathbbm{1})_{\alpha\dot{\alpha}}$, and $(\bar{\tau}^a)^{\dot{\alpha}\alpha} = (\vec{\sigma}, i\mathbbm{1})^{\dot{\alpha}\alpha}$, where $\vec{\sigma}$ are Pauli matrices.  Then one can show
\begin{equation}\label{boszm}
\hat{D}_{[a} \delta \hat{A}_{b]} = \half \epsilon_{ab}^{\phantom{ab}cd} \hat{D}_c \delta \hat{A}_d \quad \& \quad \hat{D}^a \delta \hat{A}_a = 0 \qquad  \Longleftrightarrow \qquad (\bar{\tau}^a)^{\dot{\alpha}\alpha} \hat{D}_a (\delta \hat{A})_{\alpha\dot{\beta}}  = 0~,
\end{equation}
where $(\delta \hat{A})_{\alpha\dot{\beta}} := (\tau^b)_{\alpha\dot{\beta}} \delta \hat{A}_b$.  This is done by using $\bar{\tau}^a \tau^b = \delta^{ab} \mathbbm{1} + \bar{\tau}^{ab}$, where $\bar{\tau}^{ab} := \bar{\tau}^{[a} \tau^{b]} = \half (\bar{\tau}^a \tau^b - \bar{\tau}^b \tau^a)$ is anti-self-dual.  We will denote $L := i \bar{\tau}^a \hat{D}_a$ and write this equation as $L \delta \hat{A} = 0$.

The right side of \eqref{boszm} is a good starting point for showing that the Riemannian manifold $(\fMM,g)$ admits a hyperk\"ahler structure.   We observe that right multiplication of $\delta \hat{A}$ by any $2\times 2$ constant matrix commutes with the action of $L$.  Thus if $\delta \hat{A}$ is a solution then so is $\delta \hat{A} \sigma^r$, $r = 1,2,3$.  Using the identity $\tau_b \sigma^r = i \tau^a \bar{\eta}^{r}_{ab}$, where $\bar{\eta}^{r}_{ab}$ are the anti-self-dual 't Hooft symbols, we conclude that if $\delta \hat{A}_a$ is a solution to \eqref{linearsd}, \eqref{gaugeorth}, then so is $\bar{\eta}^{r}_{ab} \delta \hat{A}^b$.  This defines a triplet of endomorphisms $J^r : T_{[\hat{A}]} \fMM \to T_{[\hat{A}]} \fMM$ through
\begin{equation}\label{quatstructure}
J^r(\delta \hat{A}_a) = -\bar{\eta}^{r}_{ab} \delta \hat{A}^b~,
\end{equation}
that satisfy the quaternionic algebra
\begin{equation}\label{quatalgebra}
J^r J^s = - \delta^{rs} \mathbbm{1} + \epsilon^{rst} J^t~,
\end{equation}
where $\mathbbm{1}$ is the identity map on $T_{[\hat{A}]} \fMM$.  This construction is completely analogous to the case of smooth monopoles and, by the same manipulations as there \cite{MR934202,Weinberg:2006rq}, one can show that this triplet of complex structures is compatible with the metric and integrable.  Thus, if $\fMM$ is non-empty and finite dimensional, then locally---\ie\ away from any singular loci---it is a hyperk\"ahler manifold.

Returning to the question of the dimension, we are after the number of linearly independent solutions, $\delta \hat{A}_{\alpha\dot{\beta}}$, to $L \delta \hat{A} = 0$.  Let us recall how this can be cast into an index for a Dirac operator.  Let
\begin{equation}
\Gamma^a = \left( \begin{array}{c c} 0 & \tau^a \\ \bar{\tau}^a & 0 \end{array} \right)~,
\end{equation}
and define
\begin{equation}\label{hatDirac}
i \hat{\slashed{D}} := i \Gamma^a \hat{D}_a = \left( \begin{array}{c c} 0 & i \tau^a \hat{D}_a \\ i \bar{\tau}^a \hat{D}_a & 0 \end{array} \right) =: \left( \begin{array}{c c} 0 & L^\dag \\ L & 0 \end{array} \right)~.
\end{equation}
$i\hat{\slashed{D}}$ is a self-adjoint operator on a dense domain of the Hilbert space $\LL^2[\UU,\mathbb{C}^4 \otimes \mathfrak{g}]$.  Since \tHooft line defects behave like singular Dirac monopoles, one might worry that $i \hat{\slashed{D}}$ is merely symmetric and that one needs to make a choice of self-adjoint extension as in \cite{Kazama:1976fm,Goldhaber:1977xw,Callias:1977cc}.  However, the difference between those references and the situation considered here is that here the Higgs field also has a $1/|\vec{x} - \vec{x}_n|$ singularity.  Our analysis in appendix \ref{app:Dirac} demonstrates that this singularity actually removes the subtleties that were present with the lowest angular momentum mode in those references.  The operators $L^\dag := i \tau^a \hat{D}_a$ and $L = i \bar{\tau}^a \hat{D}_a$ are closed, densely defined operators acting on $\LL^2[\UU, \mathbb{C}^2 \otimes \mathfrak{g}]$ and are the adjoints of each other, as indicated by the notation.  If the $G$-bundle over $\UU$ is nontrivial then we should really speak of $\LL^2$-sections of the adjoint bundle (tensor $\mathbb{C}^4$ or $\mathbb{C}^2$).  In either case we take $(f,g) = \int_{\UU} \ed^3 x \Tr \{ \bar{f} g \}$ as the innerproduct on these Hilbert spaces, where the overbar denotes the standard transpose-conjugate on $\mathbb{C}^4$ or $\mathbb{C}^2$.\footnote{In general we use the overbar to denote transpose-conjugation for finite-dimensional vector spaces while $\dag$ is reserved for the adjoint on infinite-dimensional Hilbert spaces.}

Note that $\ker{L} = \ker{L^\dag L}$ and $\ker{L^\dag} = \ker{L L^\dag}$.  Using the self-duality of the background $\hat{F}_{ab}$, the (anti-) self-duality properties of $(\bar{\tau}^{ab}) \tau^{ab}$, and the Bogomolny equation we find
\begin{align}\label{positiveL}
& L L^\dag = -\hat{D}^a \hat{D}_a \equiv -\hat{D}^2~, \cr
& L^\dag L = -\hat{D}^2 - \half \tau^{ab} \ad(\hat{F}_{ab}) = - \hat{D}^2 -2 i \vec{\sigma} \cdot \ad(\vec{B})~.
\end{align}
Both of these are positive operators since they are of the form $Q Q^\dag$ for some operator $Q$.  However in the first case it is easy to argue that $-\hat{D}^2 \otimes \mathbbm{1}_2$ is a positive-definite operator acting on $\LL^2[\UU,\mathbb{C}^2 \otimes \mathfrak{g}]$; thus $\ker{L^\dag} = \ker{L L^\dag} = \{ 0 \}$.  To see this suppose $\psi \in \ker{ \hat{D}^2}$.  Then
\begin{align}
& 0 = \int_{\UU} \ed^3 x \Tr \left\{ \overline{\psi} \hat{D}^2 \psi \right\} = - \int_{\UU} \ed^3 x \Tr \left\{ \overline{\hat{D}^a \psi} \hat{D}_a \psi \right\} \quad\Rightarrow \quad \hat{D}_a \psi = 0~.
\end{align}
Thus $\psi(\vec{x}) = \Ad_{g(\gamma)}(\psi(\vec{x}_0))$ where $\gamma$ is a path in $\UU$ connecting $\vec{x}_0$ to $\vec{x}$ and $g(\gamma) \in G$ is the path-ordered exponential.  In order that $\psi \in L^2[\UU,\mathfrak{g}]$ we require $\lim_{|\vec{x}| \to  \infty} \psi(\vec{x}) = 0$, but this implies $\psi(\vec{x}_0) = 0$, $\forall \vec{x}_0 \in \UU$ and thus $\psi = 0$.

To connect the dimension of the tangent space with the index of $L$, note that there is a two-to-one mapping between bosonic zero-modes $\delta \hat{A}_a$ and the kernel.  If $\psi_\alpha \in \ker{L}$ then we can get two linearly independent solutions for $\delta \hat{A}$ by taking $\psi_{\alpha} = \delta \hat{A}_{\alpha \dot{1}}$ or $\psi_\alpha = \delta \hat{A}_{\alpha \dot{2}}$.  We conclude that
\begin{align}\label{Dbarindex}
\dim T_{[\hat{A}]} \fMM =&~ 2 \dim{\ker{L}} = 2 \left( \dim{\ker{L}} - \dim{\ker{L^\dag}} \right)~.
\end{align}
Thus it would appear that the dimension is twice the index of the operator $L$.  In the case without line defects the right-hand side of \eqref{Dbarindex} was computed long ago by Weinberg \cite{Weinberg:1979ma,Weinberg:1979zt}, employing the methods of Callias \cite{Callias:1977kg}.  We will recall Weinberg's calculation and generalize it to the case with \tHooft defect insertions in the next section.

There is, however, one issue we would like to address before concluding this section.  It is sometimes remarked that Weinberg's calculation is not a mathematically rigorous one.  This complaint stems from the fact that, although $L$ appears to be an operator of the type considered by Callias, it fails a technical condition stated in \cite{Callias:1977kg} that is required for an operator to be Fredholm.  This condition says that the matrix representation of the asymptotic Higgs field must not have a null space.  It fails for the adjoint representation, (as pointed out in \cite{Weinberg:1979ma}), since $\ad(X_\infty)(H) = [X_\infty,H] = 0$ for any $H \in \mathfrak{t}$.  This leads to some cause for concern on both a conceptual and technical level, but the concern can be alleviated in both cases.

On a technical level, the fact the operator fails to be Fredholm\footnote{when defined on the the particular domain used in \cite{Callias:1977kg}.} means that it can---and, as it turns out, does---have a continuous spectrum extending down to $0$.  (If $\lambda = 0$ is in the spectrum of a Fredholm operator, it is necessarily isolated from any continuous part of the spectrum \cite{MR0284837}.)  Physically the continuous spectrum is due to the massless fluctuations of the Higgs and gauge field along the Cartan directions.  One might worry that the continuous part of the spectrum could contribute to the trace over the kernel.  By studying the asymptotics of the linearized Bogomolny equation, Weinberg determined the leading behavior of the spectral density function $d(\lambda)$ and showed that it is not singular enough to contribute, provided one is in the case of maximal symmetry breaking.  With the aid of this supplementary result, the techniques of Callias can be used to compute the right-hand side of \eqref{Dbarindex}.

On a conceptual level one worries that if $L$ is not Fredholm, then \eqref{Dbarindex} need not be invariant under small perturbations of the operator; the dimension of the tangent space might jump discontinuously.  One may wonder how a closed, densely defined operator $L : \HH \to \HH$ on Hilbert space can fail to be Fredholm if $\ker{L}$ and $\ker{L^\dag}$ are both finite-dimensional.  The point is that an operator is Fredholm when $\ker{L}$ and $\coker{L} \cong \HH/\Rng(L)$ are finite-dimensional, where $\Rng(L)$ is the range (image) of the operator $L$.  The cokernel of $L$ and the kernel of $L^\dag$ are the same for operators on finite-dimensional Hilbert space, but this need not be true for operators on infinite-dimensional spaces.  The closed range theorem states that it will be true if and only if $\Rng(L)$ is a closed subspace of $\HH$.  Indeed, if $\Rng(L)$ is not closed then $L$ cannot be Fredholm.  (See, for example, chapter IV of \cite{MR2446016}.)  This is precisely what goes wrong for $L = i \bar{\tau}^a \hat{D}_a$.  In general, closedness of the range is related to a certain ``boundedness away from zero'' property of the operator, which can be formulated in terms of the reduced minimum modulus of the operator.  In the context of the Dirac-type operators considered in \cite{Callias:1977kg}, this condition implies the above-mentioned condition on the asymptotic Higgs field.

Now that we've understood the problem, let us describe the resolution.  A short answer is that we can compute the right side of \eqref{Dbarindex} and see that the result does not depend on the details of the background field configuration so, a posteriori, it is stable against perturbations.  However this result begs for a better explanation.  An explanation was provided by Taubes, \cite{Taubes:1983bm}, in the case of smooth monopoles.  He showed that $L$ and its adjoint can be made Fredholm by choosing an appropriate domain of definition, $\tilde{\HH} \subset \LL^2[\mathbb{R}^3,\mathbb{C}^2 \otimes \mathfrak{g}]$.  The domain $\tilde{\HH}$ introduced by Taubes is the Hilbert space completion in the metric \eqref{metC} of the space of compactly supported sections of the $\mathbb{C}^2 \otimes \mathfrak{g}$ bundle.  He went on to compute the index of the Fredholm operator $L |_{\tilde{\HH}}$ and recovered Weinberg's result via a different method.  One expects the results to agree since it is also shown in \cite{Taubes:1983bm} that the $\LL^2$-kernels and $\tilde{\HH}$-kernels of $L,L^\dag$ agree.  We expect similar arguments can be made in the case of singular monopoles, and we will write the right-hand side of \eqref{Dbarindex} as an index, so that
\begin{align}\label{Dbarindex2}
\dim T_{[\hat{A}]} \fMM =&~ 2 \ind{L}~.
\end{align}
In the next section we will follow the approach of Callias--Weinberg to compute this quantity since this approach readily generalizes to the case with line defect insertions.

\section{The index computation}\label{sec:index}

\subsection{Reduction to boundary terms}

The operators $L L^\dag$ and $L^\dag L$ are self-adjoint and positive on $\LL^2[\UU,\mathbb{C}^2 \otimes \mathfrak{g}]$.  Following Callias\footnote{See also \cite{Niemi:1985ht}.} we consider
\begin{equation}\label{Bz}
B_z := \tr_{\mathfrak{g} \otimes \mathbb{C}^2} \left( \frac{z}{L^\dag L +z} - \frac{z}{ L L^\dag + z} \right)~,
\end{equation}
which is a bounded linear operator on $\LL^2[\UU,\mathbb{C}] \equiv \LL^2[\UU]$ for $z \in  \mathbb{C}$ away from the negative real axis.  Let $\{ \phi_m \}_{m=1}^\infty$ be an orthonormal basis for $\LL^2[\UU]$.  If $B_z$ is traceclass on a domain $C \subset \mathbb{C}$ which has $z = 0$ as a limit point, then we can compute $I(z) = \Tr_{\LL^2[\UU]} B_z \equiv \sum_m (\phi_m, B_z \phi_m)$ and take the limit $\lim_{z \to 0} I(z)$.  In this limit we see that $\phi_m \in \ker{L}$ contributes $+1$ to $I(0)$, $\phi_m \in \ker{L}^\dag$ contributes $-1$, while the contribution from any other $\phi_m$ vanishes.  Therefore, under this assumption about $B_z$, $I(0)$ computes the index we are after.  Here is where it is important to augment the original arguments of Callias with Weinberg's analysis of the large $|\vec{x}|$ asymptotics of the linearized Bogomolny equation \cite{Weinberg:1979ma}, since $L^\dag L$ and $L L^\dag$ have continuum spectra on the positive real axis extending down to $\lambda = 0$.  These arguments go through identically in the case with defect insertions since they are concerned with the large distance behavior of the background fields, which is the same.

The main thrust of \cite{Callias:1977kg} is to show that $B_z$ is traceclass for all $z$ in a common domain $C$ whose boundary contains $z = 0$.  The strategy involves writing the kernel (in the sense of the Green's function) of the integral operator representation of $B_z$ in a sufficiently explicit way such that this property can be demonstrated and as a byproduct a practical formula is obtained for the index $I(0)$.  Here we review some of the key formulae; this material can also be found in \cite{Weinberg:1979ma}, or the review \cite{Weinberg:2006rq}.

We consider a slight generalization of \eqref{Bz}, as in \cite{Callias:1977kg}, where we replace $\mathfrak{g}$ with the representation space $V_\rho$ of an arbitrary finite-dimensional representation $\rho: \mathfrak{g} \to \mathfrak{gl}(V_\rho)$ that lifts to a representation of the gauge group $G$; \eqref{Bz} corresponds to $\rho = \ad$ with $V_{\ad} \cong \mathfrak{g}$.  The analysis is no harder and the result is useful when considering generalizations of the Yang--Mills--Higgs system adding flavor degrees of freedom.  The condition that $\rho$ lifts to a representation of $G$ is important.  It means that the weights $\mu$ of the representation must sit in the character lattice $\Lambda_{G}^\vee \subset \mathfrak{t}^\ast$.  This is the integral dual of the co-character lattice where the \tHooft charges reside, and it is necessary that $\mu \in \Lambda_{G}^\vee$ in order that the transition functions $\exp(\rho(P_n) \phi_n)$ on the infinitesimal two-spheres be single-valued.  For example, if our gauge group is $G = SO(3)$ and we have an \tHooft defect with charge equal to the fundamental magnetic weight, then it is not consistent to couple the Dirac operator to the fundamental representation of $\mathfrak{su}(2)$, or any representation with half-integer spin.

When coupling to the representation $\rho$, the operators $L,L^\dag$, and $i\hat{\slashed{D}}$ are modified as follows:
\begin{align}\label{LdagLexp}
(i \hat{\slashed{D}})_\rho = \left( \begin{array}{c c} 0 & L_{\rho}^\dag \\ L_\rho & 0 \end{array} \right)~, \qquad \begingroup \renewcommand{\arraystretch}{1.2}  \begin{array}{l} L_\rho = i \sigma^i \otimes \left( \pd_i + \rho(A_i) \right) - \mathbbm{1}_2 \otimes \rho(X)~, \\
L_{\rho}^\dag =  i \sigma^i \otimes \left( \pd_i + \rho(A_i) \right) + \mathbbm{1}_2 \otimes \rho(X)~. \end {array} \endgroup
\end{align}
Recall that $\vec{\sigma}$ are the Pauli matrices, we represent Lie algebra elements with anti-Hermitian matrices, and of course $(\pd_i)^\dag = - \pd_i$.  In this paper we always assume maximal symmetry breaking: $\langle \mu, X_\infty \rangle \neq 0$, $\forall \mu \in \Delta_{\rho}$, $\mu \neq 0$, where $\Delta_\rho \subset \mathfrak{t}^\ast$ is  the set of weights of the representation $\rho$.  With these definitions we then consider
\begin{equation}\label{Bzrho}
B_{z,\rho} := \tr_{\mathbb{C}^2 \otimes V_\rho} \left( \frac{z}{L_{\rho}^\dag L_\rho +z} - \frac{z}{ L_\rho L_{\rho}^\dag + z} \right)~,
\end{equation}
so that $B_z = B_{z, {\rm ad}}$ and $L = L_{\rm ad}$.

Consider the resolvent operator $G_\lambda$ for $(i \hat{\slashed{D}})_\rho$ acting on $\LL^2[\UU,\mathbb{C}^2 \otimes  V_\rho]$,
\begin{equation}
G_\lambda := \left( (i \hat{\slashed{D}})_\rho + \lambda \right)^{-1}~.
\end{equation}
On the one hand this must be a right inverse for $(i \hat{\slashed{D}})_\rho + \lambda$, which can be expressed in the form
\begin{align}\label{GreenDhat}
G_{i\lambda} =&~ \left( (i \hat{\slashed{D}})_\rho - i \lambda \right) \left( (i \hat{\slashed{D}})_{\rho}^2 + \lambda^2 \right)^{-1}  \cr
=&~ \left( \begin{array}{c c} -i\lambda & L_{\rho}^\dag \\ L_{\rho} & -i\lambda \end{array}\right) \left( \begin{array}{c c} (L_{\rho}^\dag L_\rho + \lambda^2)^{-1}  & 0 \\ 0 & (L_\rho L_{\rho}^\dag + \lambda^2)^{-1} \end{array} \right)~.
\end{align}
Now let $\bar{\Gamma} := \Gamma^1 \Gamma^2 \Gamma^3 \Gamma^4 = \diag( - \mathbbm{1}_2, \mathbbm{1}_2 )$.  Left multiplying $G_{i\lambda}$ by $\bar{\Gamma} \otimes \mathbbm{1}_{V_\rho}$ and manually taking the trace over the explicit $\mathbb{C}^2$ block structure, one finds a result of the same form as \eqref{Bzrho}.  The precise relation is
\begin{equation}\label{BGrel}
B_{z,\rho} = -i \sqrt{z} \tr_{\mathbb{C}^4 \otimes V_\rho} \left( \bar{\Gamma} G_{i \sqrt{z}} \right)~.
\end{equation}

On the other hand we can obtain a useful expression for $G_{\lambda}$ by considering the Green's function, $G_\lambda(\vec{x},\vec{y})$ associated with its integral operator representation.  For $\vec{x} \neq \vec{y}$ the Green's function must satisfy the equations
\begin{align}\label{Greens}
0 =&~ \left[ i \Gamma^i \otimes \left( \frac{\pd}{\pd x^i} + \rho(A_i)(\vec{x}) \right) + i \Gamma^4 \otimes \rho(X)(\vec{x}) + \lambda \right] G_{\lambda}(\vec{x},\vec{y})~, \cr
0 =&~ -i \left( \frac{\pd}{\pd y^i} G_{\lambda}(\vec{x},\vec{y}) \right)  \Gamma^i + G_\lambda(\vec{x},\vec{y}) \bigg[ i \Gamma^i \otimes \rho(A_i)(\vec{y}) + i \Gamma^4 \otimes \rho(X)(\vec{y}) + \lambda \bigg]~. \qquad
\end{align}
The second equation can be obtained by writing an equation for the Green's function associated with the Hilbert space adjoint $G_{\lambda}^\dag$, which is a right inverse of $(i \hat{\slashed{D}})_\rho + \lambda^\ast$.  Next we use the fact that the Green's function for $G_{\lambda}^\dag$ is related to the Green's function for $G_\lambda$ by $G_{\lambda}^\dag(\vec{x},\vec{y}) = \overline{G_\lambda(\vec{y},\vec{x})}$, where the bar means transpose conjugate on $\mathbb{C}^4 \otimes V_\rho$.  Finally, we take the the transpose conjugate of this equation with respect to the $\mathbb{C}^4 \otimes V_\rho$ structure to arrive at the second of \eqref{Greens}.  Now we left-multiply both of these equations by $\bar{\Gamma}$, add the result, and take the trace over $\mathbb{C}^4 \otimes V_\rho$.  Using cyclicity of the trace and $\{ \Gamma^a, \bar{\Gamma} \} = 0$, we find
\begin{align}
2\lambda \tr_{\mathbb{C}^4 \otimes V_\rho}\left\{ \bar{\Gamma} G_{\lambda}(\vec{x},\vec{y}) \right\} =&~ -i \left( \frac{\pd}{\pd x^i} + \frac{\pd}{\pd y^i} \right) \tr_{\mathbb{C}^4 \otimes V_\rho} \left\{ \bar{\Gamma} \Gamma^i G_{\lambda}(\vec{x},\vec{y}) \right\} + \cr
&~ - i \tr_{\mathbb{C}^4 \otimes V_\rho} \left\{ \bar{\Gamma} \Gamma^a \left( \rho(\hat{A}_a)(\vec{x}) - \rho(\hat{A}_a)(\vec{y}) \right) G_{\lambda}(\vec{x},\vec{y}) \right\}~.
\end{align}
Comparing with \eqref{BGrel}, we see that $B_{z,\rho}$ is an integral operator with associated Green's function
\begin{equation}\label{Bkernel}
2 B_{z,\rho}(\vec{x},\vec{y}) = \left( \frac{\pd}{\pd x^i} + \frac{\pd}{\pd y^i} \right) J_{z,\rho}^i(\vec{x},\vec{y}) + C_{z,\rho}(\vec{x},\vec{y})~,
\end{equation}
where
\begin{align}
J_{z,\rho}^i(\vec{x},\vec{y}) :=&~ i \tr_{\mathbb{C}^4 \otimes V_\rho} \left\{ \bar{\Gamma} \Gamma^i G_{i\sqrt{z}}(\vec{x},\vec{y}) \right\}~, \cr
C_{z,\rho}(\vec{x},\vec{y}) :=&~ i \tr_{\mathbb{C}^4 \otimes V_\rho} \left\{ \bar{\Gamma} \Gamma^a \left( \rho(\hat{A}_a)(\vec{x}) - \rho(\hat{A}_a)(\vec{y}) \right) G_{i\sqrt{z}}(\vec{x},\vec{y}) \right\}~.
\end{align}

Equation \eqref{Bkernel} and the analogous operator relation are the starting point for showing that $B_{z,\rho}$ is traceclass.  The idea is to show that $B_{z,\rho}(\vec{x},\vec{y})$ is continuous as $\vec{y} \to \vec{x}$, and that the trace of $B_{z,\rho}$ exists and is computed by $\int \ed^3 x B_{z,\rho}(\vec{x},\vec{x})$.  The reason one might expect $B_{z}(\vec{x},\vec{y})$ to be well defined as $\vec{y} \to \vec{x}$, even though the Green's function $G_{\lambda}(\vec{x},\vec{y})$ is clearly not, is that the singularity in $G_{\lambda}(\vec{x},\vec{y})$ is compensated for in $J_{z,\rho}^i,C_{z,\rho}$ by zeros coming from tracing over $\gamma$-matrix structure and/or from $\rho(\hat{A}_a)(\vec{y}) \to \rho(\hat{A}_a)(\vec{x})$.   It is proven in \cite{Callias:1977kg} that these expectations are borne out; the same arguments can be applied here since they are concerned with analyzing the structure of the singularity in $G_{\lambda}(\vec{x},\vec{y})$ as $\vec{y} \to \vec{x}$, and this structure does not depend on the boundary conditions defining $G_\lambda$.  In particular $\lim_{\vec{y} \to \vec{x}} C_{z,\rho}(\vec{x},\vec{y}) = 0$, and
\begin{equation}\label{indexdensity}
I_{\rho}(z) := \Tr_{\LL^{2}[\UU]} B_{z,\rho} = \half \int_{\UU} \ed^3 x \pd_i J_{z,\rho}^i(\vec{x},\vec{x})~.
\end{equation}
Thus we find that the index reduces to a sum of boundary contributions,
\begin{align}\label{index1}
\ind{L_\rho} =&~ \lim_{z \to 0^+} I_\rho(z)~, \qquad \textrm{where} \cr
I_\rho(z) =&~ \half \left( \ \lim_{r = |\vec{x}|\to \infty} - \sum_{n=1}^{N_t} \ \lim_{r = |\vec{x} - \vec{x}_n| \to 0} \right) \int_{S^2} \vol_{S^2} r^2 \hat{r} \cdot \vec{J}_{\rho,z}(\vec{x},\vec{x})~,
\end{align}
where the minus sign takes into account the relative orientation of the boundary components of $\UU$, $\vol_{S^2}$ is the volume form on the unit two-sphere, and we are using a spherical coordinate system centered on $\vec{x} = 0$ for the asymptotic sphere and $\vec{x} = \vec{x}_n$ for the infinitesimal ones.\footnote{The form of this result, as a sum of boundary contributions, suggests that it should be applicable on a generic Riemannian three-manifold, $(M_3,g)$, with boundary.  Indeed it is straightforward to generalize the analysis of this subsection.  We let $L_\rho$ act on $\LL^2$-normalizable sections, $f \in \Gamma(\SS(M_3) \otimes \EE_\rho)$, of the Dirac spinor bundle $\SS(M_3)$ of $M_3$ tensored with the associated $G$-bundle $\EE_\rho$ corresponding to representation $\rho$, with fiber $\mathbb{C}^2 \otimes V_\rho$.  Furthermore we impose the boundary conditions $\int_{\pd M_3} {\rm vol}_{\pd} \  n_i \tr_{\mathbb{C}^2 \otimes V_\rho} \{ \bar{f} \sigma^i f \} = 0$, where ${\rm vol}_\pd$ is the induced volume form and $n_i$ the unit normal of $\pd M_3$.  Then in local coordinates, $L_\rho = i \sigma^i( \pd_i + \frac{1}{4} \omega_{i,\uj\uk} \sigma^{\uj\uk}) \otimes \mathbbm{1}_{V_\rho} + i \sigma^i \otimes \rho(A_i) - \mathbbm{1}_2 \otimes \rho(X)$, where the $\{\sigma^i \}$ are related to flat-space Pauli matrices $\{ \sigma^{\underline{i}} \}$ through an orthonormal frame, $\sigma^{\underline{i}} = {e^{\underline{i}}}_i \sigma^i$, and $\omega_{i,\uj\uk} = {e_{\uj}}^j \nabla_i e_{\uk j}$ are the components of the frame connection.  Here $\nabla_i$ is the covariant derivative with respect to the Levi--Civit\'a connection.  $L_{\rho}^\dag$ has the same form but with a sign flip on the $\rho(X)$ term.  Then with these $L_\rho, L_{\rho}^\dag$, we define the Dirac operator $(i \hat{\slashed{D}})_\rho$, its resolvent $G_{\lambda}$, and the current $J_{z,\rho}^i$ as above.  After analogous manipulations one finds $I_\rho(z) = \half \int_{M_3} \ed^3 x \sqrt{g} \nabla_i J_{z,\rho}^i(\vec{x},\vec{x}) = \half \int_{M_3} \ed^3 x \pd_i \left( \sqrt{g} J_{z,\rho}^i(\vec{x},\vec{x}) \right)$.}

\subsection{The contribution from the two-sphere at infinity}

In this subsection we recall Weinberg's computation \cite{Weinberg:1979ma,Weinberg:1979zt} giving the contribution to \eqref{index1} from the two-sphere at infinity.  In general we are after the Green's function for the operator
\begin{equation}
\hat{r} \cdot \vec{J}_{z,\rho} = i \tr_{\mathbb{C}^4 \otimes V_\rho} \left\{ \bar{\Gamma} (\hat{r} \cdot \vec{\Gamma}) G_{i\sqrt{z}} \right\}~.
\end{equation}
Using \eqref{GreenDhat} and recalling \eqref{positiveL}~,
\begin{align}\label{rhatdotJ1}
\hat{r} \cdot \vec{J}_{z,\rho} =&~ i \tr_{\mathbb{C}^4 \otimes V_\rho} \left\{  \left( \begin{array}{c c} 0 & - \hat{r} \cdot \vec{\sigma} \\ \hat{r} \cdot \vec{\sigma} & 0 \end{array}\right) \left( \begin{array}{c c} - i\sqrt{z} & L_{\rho}^\dag \\ L_{\rho} & -i\sqrt{z} \end{array} \right) \left( \begin{array}{c c} \left( R_{z,\rho} - 2i \vec{\sigma} \cdot \rho(\vec{B}) \right)^{-1} & 0 \\ 0 & R_{z,\rho}^{-1}  \end{array} \right) \right\} \cr
=&~ i \tr_{\mathbb{C}^2 \otimes V_\rho} \left\{ (\hat{r} \cdot \vec{\sigma}) \left[ L_{\rho}^\dag R_{z,\rho}^{-1}  - L_\rho \left( R_{z,\rho} - 2i \vec{\sigma} \cdot \rho(\vec{B}) \right)^{-1} \right] \right\}~, \raisetag{24pt}
\end{align}
where
\begin{equation}
R_{\rho,z} := -\hat{D}_{\rho}^2 + z = - (\vec{\pd} + \rho(\vec{A}))^2 - \rho(X)^2 + z~,
\end{equation}
and is proportional to the identity on the $\mathbb{C}^2$ factor.  Since $R_{z,\rho}^{-1}$ is bounded and $\vec{B} \propto \frac{\hat{r}}{r^2}$ the series
\begin{equation}\label{Ropseries}
R_{z,\rho}^{-1} + R_{z,\rho}^{-1} \left(2i \vec{\sigma} \cdot \rho(\vec{B})\right) R_{z,\rho}^{-1} -  R_{z,\rho}^{-1} \left(2i \vec{\sigma} \cdot \rho(\vec{B})\right) R_{z,\rho}^{-1} \left(2i \vec{\sigma} \cdot \rho(\vec{B})\right) R_{z,\rho}^{-1} + - \cdots~
\end{equation}
is absolutely convergent for large enough $r$, and by acting with $R_{z,\rho} - 2i \vec{\sigma} \cdot \rho(\vec{B})$ on the left we see that it converges to the inverse, $( R_{z,\rho} - 2i \vec{\sigma} \cdot \rho(\vec{B}) )^{-1}$.  Then by plugging this series into \eqref{rhatdotJ1} and noting from \eqref{LdagLexp} that $L_{\rho}^\dag - L_\rho = 2 \mathbbm{1}_2 \otimes \rho(X)$, we see that the trace over the $\mathbb{C}^2$ tensor factor leads to a cancelation of the leading order terms:
\begin{equation}
\tr_{\mathbb{C}^2 \otimes V_\rho} \left\{ (\hat{r} \cdot \vec{\sigma}) \left[ (L_{\rho}^\dag - L_\rho) R_{z,\rho}^{-1} \right] \right\} = 0~.
\end{equation}
Thus,
\begin{equation}\label{rhatdotJ2}
\hat{r} \cdot \vec{J}_{z,\rho} = 2 \tr_{\mathbb{C}^2 \otimes V_\rho} \left\{ (\hat{r} \cdot \vec{\sigma}) L_\rho R_{z,\rho}^{-1} \left( \vec{\sigma} \cdot \rho(\vec{B}) \right) R_{z,\rho}^{-1} \right\} + \cdots~,
\end{equation}
where the ellipses correspond to the third and higher terms of the series \eqref{Ropseries}.  These terms will give contributions to the diagonal, $\hat{r} \cdot \vec{J}_{z,\rho}(\vec{x},\vec{x})$, that are subleading to the contribution from the displayed term at large $r =|\vec{x}|$.

Now consider the leading large $r$ behavior of \eqref{rhatdotJ2}.  Using \eqref{asymptoticbcs} we have
\begin{align}
\vec{\sigma} \cdot \rho(\vec{B}) =&~ \frac{\rho(\gm)}{2 r^2} (\hat{r} \cdot \vec{\sigma})  + O(r^{-(2+\updelta)})~, \cr
R_{z,\rho}^{-1} =&~ \left( - \vec{\pd}^2 - \rho(X_\infty)^2 + z + O(r^{-(1+\updelta)}) \right)^{-1} \cr
=&~ \left(  - \vec{\pd}^2 - \rho(X_\infty)^2 + z \right)^{-1} \left(1 + O(r^{-(1+\updelta)}) \right)~.
\end{align}
We can commute $\vec{\sigma} \cdot \rho(\vec{B})$ past $R_{z,\rho}^{-1}$ and $L_\rho$ at leading order.  Tracing over the remaining $\mathbb{C}^2$ picks out the $\rho(X)$ term in $L_\rho$, so that
\begin{equation}
\hat{r} \cdot \vec{J}_{z,\rho} = -\frac{2}{r^2} \tr_{V_\rho} \left\{ \frac{ \rho(X_\infty) \rho(\gm) }{ ( - \vec{\pd}^2 - \rho(X_\infty)^2 + z )^2} \right\} + O(r^{-(2+\updelta)})~.
\end{equation}
The representation matrices $\rho(X_\infty), \rho(\gm)$ are pure-imaginary diagonal matrices.  We carry out the trace over $V_\rho$ by employing an orthonormal basis associated with the decomposition into weight spaces, $V_\rho = \oplus_\mu V_\rho[\mu]$, where $\mu \in \Delta_{\rho} \subset \Lambda_{G}^\vee \subset \mathfrak{t}^\ast$ are the weights of the representation.  We will denote the dimension of each weight space $n_\rho(\mu) := \dim V_{\rho}[\mu]$.  For any $v \in V_{\rho}[\mu]$ we have $i \rho(X_\infty) v = \langle \mu, X_\infty \rangle v$, where $\langle~,~\rangle$ denotes the canonical pairing $\mathfrak{t}^\ast \otimes \mathfrak{t} \to \mathbb{R}$.  Thus we have
\begin{equation}
\hat{r} \cdot \vec{J}_{z,\rho} =  \frac{2}{r^2} \sum_{\mu \in \Delta_{\rho}}  \frac{ n_{\rho}(\mu) \langle \mu, X_\infty \rangle \langle \mu, \gm \rangle }{ ( - \vec{\pd}^2 + \langle \mu, X_\infty \rangle^2 + z )^2}  + O(r^{-(2+\updelta)})~.
\end{equation}

The diagonal of the integral kernel is evaluated by Fourier transform,
\begin{align}
\hat{r} \cdot \vec{J}_{z,\rho}(\vec{x},\vec{x}) =&~  \frac{2}{r^2} \sum_{\mu \in \Delta_{\rho}} n_{\rho}(\mu) \int \frac{\ed^3 k}{(2\pi)^3} \frac{ \langle \mu, X_\infty \rangle \langle \mu, \gm \rangle }{ ( k^2 + \langle \mu, X_\infty \rangle^2 + z )^2} +  O(r^{-(2+\updelta)}) \cr
=&~  \frac{1}{4\pi r^2}  \sum_{\mu \in \Delta_{\rho}} n_{\rho}(\mu) \frac{ \langle \mu, X_\infty \rangle \langle \mu, \gm \rangle }{ \sqrt{ \langle \mu, X_\infty \rangle^2 + z} } +  O(r^{-(2+\updelta)})~.
\end{align}
Plugging this expression into \eqref{index1} we get the following contribution to $I_{\rho}(z)$:
\begin{equation}\label{SinftyI}
\half \lim_{r \to \infty} \int_{S^2} \vol_{S^2} r^2 \hat{r} \cdot \vec{J}_{z,\rho}(\vec{x},\vec{x}) = \half  \sum_{\mu \in \Delta_{\rho}} n_\rho(\mu) \frac{ \langle \mu, X_\infty \rangle \langle \mu, \gm \rangle }{ \sqrt{ \langle \mu, X_\infty \rangle^2 + z} }~.
\end{equation}
In the case of the adjoint representation the weights are the roots, $\mu \to \alpha \in \Delta_{\rm ad} \equiv \Delta$, and $n_{\rm ad}(\alpha) = 1$, $\forall \alpha$.  Recalling the factor of $2$ in the relation between $T_{[\hat{A}]}\fMM$ and $I(0)$, we see that \eqref{SinftyI} is consistent with the corresponding term in the dimension formula, \eqref{dim1}.

\subsection{The contribution from an infinitesimal two-sphere}\label{sec:infS2index}

To compute the contribution from one of the $S_{\varepsilon_n}^2$ we work in spherical coordinates centered on $\vec{x}_n = 0$ and we set $P_n = P$ to simplify notation.  Again we are after the diagonal of the integral kernel, $\hat{r} \cdot \vec{J}_{z,\rho}(\vec{x},\vec{x})$, where
\begin{equation}\label{rhatdotJ3}
\hat{r} \cdot \vec{J}_{z,\rho} =  i \tr_{\mathbb{C}^4 \otimes V_\rho} \left\{ \bar{\Gamma} (\hat{r} \cdot \vec{\Gamma}) G_{i\sqrt{z}} \right\}  = i \tr_{\mathbb{C}^4 \otimes V_\rho} \left\{ \bar{\Gamma} (\hat{r} \cdot \vec{\Gamma}) \left( (i\hat{\slashed{D}})_\rho +  i\sqrt{z} \right)^{-1} \right\}~.
\end{equation}
Consider the small $r$ expansion of the Dirac operator.  With
\begin{align}
X =&~ - \frac{P}{2r} + O(r^{-1/2}) \equiv X^{(0)} + \delta X~, \cr
A =&~ \frac{P}{2} (\pm 1 - \cos{\theta})\ed\phi + O(r^{-1/2}) \equiv A^{(0)} + \delta A~,
\end{align}
where the $\pm$ refer to the northern or southern patch of the two-sphere, we have
\begin{align}
(\hat{\slashed{D}})_{\rho} = (\hat{\slashed{D}})_{\rho}^{(0)} + \slashed{\delta}~,
\end{align}
where
\begin{align}\label{DiracDirac}
(\hat{\slashed{D}})_{\rho}^{(0)} :=&~ \vec{\Gamma} \cdot \vec{\pd} + \Gamma^{\phi} \otimes \frac{\rho(P)}{2} (\pm 1 - \cos{\theta}) - \Gamma^4 \otimes \frac{\rho(P)}{2 r}~, \cr
\slashed{\delta} :=&~ \Gamma^a \rho(\delta \hat{A}_a)~.
\end{align}
$\slashed{\delta}$ is an anti-Hermitian multiplication operator with leading behavior $O(r^{-1/2})$ as $r \to 0$.  

The operator $(\hat{\slashed{D}})_{\rho}^{(0)}$ is essentially the Dirac operator on $\mathbb{R}^3 \setminus \{0 \}$ in a Dirac monopole background.  The spectral problem associated with this operator is a classic problem first studied independently by Banderet \cite{Banderet} and Harish-Chandra \cite{HarishChandra:1948zz} in the 1940's, and then with renewed interest following the discovery of the 't Hooft--Polyakov solution in a series of papers \cite{Kazama:1976fm,Goldhaber:1977xw,Callias:1977cc}.  The Dirac operator \eqref{DiracDirac} appears to be slightly different than the operators considered in these references, in that the background Higgs field also has a pole at $r = 0$.  As we remarked previously this point turns out to be crucial for avoiding the issue of self-adjoint extensions dealt with in \cite{Kazama:1976fm,Goldhaber:1977xw,Callias:1977cc}.  Nevertheless the same techniques can be used to find a completely explicit solution.  (See also \cite{Cheng:2013mla}.)  In particular the spectrum of the self-adjoint operator $(i \hat{\slashed{D}})_{\rho}^{(0)}$ is purely continuous and consists of the entire real line.  Since $i \slashed{\delta}$ is Hermitian, 
\begin{equation}
\hat{R}_{z,\rho} := (i\hat{\slashed{D}})_{\rho}^{(0)} + i\sqrt{z} + i\slashed{\delta} ~,
\end{equation}
is invertible for $z$ away from the negative real axis, and 
\begin{equation}\label{rhatdotJ4}
\lim_{r \to 0} r^2 \hat{r} \cdot \vec{J}_{z,\rho}(\vec{x},\vec{x}) = \lim_{r \to 0} i r^2  \tr_{\mathbb{C}^4 \otimes V_\rho} \left\{ \bar{\Gamma} (\hat{r} \cdot \vec{\Gamma}) \hat{R}_{z,\rho}^{-1}(\vec{x},\vec{x}) \right\}~.
\end{equation}
Our strategy is to evaluate this expression by employing the explicit spectral representation of $(i \hat{\slashed{D}})_{\rho}^{(0)}$.  We will see that the result is independent of $z$ and $\slashed{\delta}$ in the $r \to 0$ limit.

We review the solution of the spectral problem for $(i \hat{\slashed{D}})_{\rho}^{(0)}$ in appendix \ref{app:Dirac} and summarize the results here.  (See especially \ref{app:Dirac1}.)  Let $(i \hat{\slashed{D}})_{\rho}^{(0)}$ act on $\Psi_{\pm}(\vec{x})$; these are $\mathbb{C}^4 \otimes V_\rho$-valued functions on $\mathbb{R}_+ \times S_{\pm}$, where $S_{\pm}$ are northern and southern patches covering $S^2$.  They will be patched together on the overlap by the transition function $\exp{(\rho(P) \phi)}$.  Let $\{ {\bf e}_{i_\mu} \}$ denote an orthonormal basis associated with the weight decomposition $V_\rho = \oplus_{\mu} V_{\rho}[\mu]$.  Here $\mu$ runs over the set of weights $\Delta_{\rho}$ and for each $\mu$ the label $i_\mu = 1,\ldots, n_\rho(\mu) = \dim{V_{\rho}[\mu]}$ takes into account the degeneracy.   We expand $\Psi_{\pm}$ in this basis, writing $\Psi_{\pm} = \sum_{\mu} \sum_{i_\mu} \Psi_{\pm}^{(i_\mu)} {\bf e}_{i_\mu}$, so that each $\Psi_{\pm}^{(i_\mu)}$ is a $\mathbb{C}^4$-valued function.  We will also write $\Psi_{\pm}^{(\mu)}$ to denote the collection of $\Psi_{\pm}^{(i_\mu)}$ for a given $\mu$; these will be $4 n_\mu(\rho)$ component objects.  We have that $\rho(P) {\bf e}_\mu = - i \langle \mu, P \rangle {\bf e}_\mu$; the $\langle \mu, P \rangle$ are integers because we require $\mu \in \Lambda_{G}^\vee$ as explained above \eqref{LdagLexp}.  The Dirac equation $(i \hat{\slashed{D}})_{\rho}^{(0)} \Psi = E \Psi$ splits into $\dim V_{\rho}$ Dirac equations for the $\Psi^{(i_\mu)}$:
\begin{equation}\label{Diracmu}
\left[ \vec{\Gamma} \cdot \vec{\pd} -i \Gamma^\phi  \frac{p_\mu}{2} (\epsilon 1 - \cos{\theta}) -i \Gamma^4  \left( x_\mu - \frac{p_\mu}{2 r} \right) \right] \Psi_{\epsilon}^{(i_\mu)} = -i E \Psi_{\epsilon}^{(i_\mu)}~,
\end{equation}
where we have introduced the shorthand $\langle \mu, P \rangle \equiv p_\mu \in \mathbb{Z}$, $x_\mu \equiv \langle \mu, X_\infty \rangle \in \mathbb{R}$, and $\epsilon = \pm$ keeps track of the patch we are working in.\footnote{We emphasize that the $x_\mu$ here is not a coordinate on Euclidean space.}

Equation \eqref{Diracmu} does not possess any $\LL^2$-normalizable solutions (bound states), but it does possess a continuum of plane-wave normalizable solutions for any real $E$ (scattering states).  The scattering states can be used to construct the spectral measure associated with $(i \hat{\slashed{D}})_{\rho}^{(0)}$, which leads to an explicit representation of the integral kernel for $\hat{R}_{z,\rho}$.  The scattering states are as follows.  Let
\begin{equation}\label{bfUdef}
{\bf U}(\theta,\phi) = \mathbbm{1}_2 \otimes U(\theta,\phi) = \mathbbm{1}_2 \otimes e^{-i \phi \sigma^3/2} e^{-i \theta \sigma^2/2}~.
\end{equation}
Then we have two families of solutions
\begin{equation}\label{Psifam}
(\Psi_{\epsilon}^{(i_\mu)})_{j,m,1} = e^{i\epsilon p_\mu \phi/2} {\bf U}(\theta,\phi) \tilde{\Psi}^{(i_\mu)}_{j,m,1}~, \qquad (\Psi_{\epsilon}^{(i_\mu)})_{j,m,2} = e^{i\epsilon p_\mu \phi/2} {\bf U}(\theta,\phi) \tilde{\Psi}^{(i_\mu)}_{j,m,2}~, ~
\end{equation}
where
\begin{align}\label{genjPsisol}
\tilde{\Psi}_{j,m,1}^{(i_\mu)}(E;\vec{x}) =&~ \frac{\sqrt{|E|} e^{-i m \phi}}{2 \sqrt{2\pi r}} \left( \begingroup \renewcommand{\arraystretch}{1.2} \begin{array}{c} a_- J_{j+1}(|E| r)  d^{j}_{m,\half(p_\mu - 1)}(\theta) \\ -a_+ J_{j+1}(|E| r)  d^{j}_{m,\half(p_\mu + 1)}(\theta) \\ i \sgn(E) a_- J_{j}(|E| r)  d^{j}_{m,\half(p_\mu - 1)}(\theta) \\ i \sgn(E) a_+ J_{j}(|E| r)  d^{j}_{m,\half(p_\mu +1)}(\theta) \end{array} \endgroup \right)~, \displaybreak[3]  \nonumber \\
\tilde{\Psi}_{j,m,2}^{(i_\mu)}(E;\vec{x}) =&~ \frac{\sqrt{|E|} e^{-i m \phi}}{2 \sqrt{2\pi r}} \left( \begingroup \renewcommand{\arraystretch}{1.2} \begin{array}{c} a_+ J_{j}(|E| r)  d^{j}_{m,\half(p_\mu - 1)}(\theta) \\ a_- J_{j}(|E| r)  d^{j}_{m,\half(p_\mu + 1)}(\theta) \\ -i \sgn(E) a_+ J_{j+1}(|E| r)  d^{j}_{m,\half(p_\mu - 1)}(\theta) \\ i \sgn(E) a_- J_{j+1}(|E| r)  d^{j}_{m,\half(p_\mu +1)}(\theta) \end{array} \endgroup \right)~,
\end{align}
with $a_{\pm} \equiv \sqrt{j + \half \pm \frac{p_\mu}{2}}$.  For these solutions the allowed values of $m$ run from $-j$ to $j$ in integer steps and the allowed values of $j$ start at $j = \half (|p_\mu| + 1) = j_\mu +1$ and increase in integer steps.\footnote{${\bf U}(\theta,\phi)$ is not single-valued under $\phi \to \phi + 2\pi$, but it follows from the relations among $j,m,p_\mu$ that $p_\mu + 2m$ is odd and hence the functions \eqref{Psifam} are well-defined.}  The notation $j_\mu := \half(|p_\mu| - 1)$ will be useful below.  The $J_\nu$ are Bessel functions and the $d^{j}_{m,m'}$ are Wigner (small) $d$ functions.\footnote{We follow the conventions of \cite{Sakurai} for Wigner $d$ functions and $SU(2)$ representation matrices.  The combination $e^{-i m \phi} d^{j}_{m,m'}(\theta)$ can also be expressed in terms of spin-weighted spherical harmonics, ${}_{m'}Y_{jm}$.}

Additionally, when $p_\mu \neq 0$, there is one more family of solutions with fixed $j = j_\mu$.  Their form depends on the sign of $p_\mu$ and we denote the two possibilities with a $\pm$:
\begin{equation}\label{Psifam2}
(\Psi_{\epsilon}^{(i_\mu)})_{m,+} = e^{i\epsilon p_\mu \phi/2} {\bf U}(\theta,\phi) \tilde{\Psi}^{(i_\mu)}_{m,+}~, \quad (\Psi_{\epsilon}^{(i_\mu)})_{m,-} = e^{i\epsilon p_\mu \phi/2} {\bf U}(\theta,\phi) \tilde{\Psi}^{(i_\mu)}_{m,-}~,
\end{equation}
with
\begin{align}\label{specialjPsisol}
\tilde{\Psi}_{m,+}^{(i_\mu)}(E;\vec{x}) =&~ \frac{\sqrt{|p_\mu| |E|} e^{-i m \phi}}{2 \sqrt{2\pi r}} \left( \begin{array}{c} J_{j_\mu}(|E| r)  d^{j_\mu}_{m,j_\mu}(\theta) \\ 0 \\ -i \sgn(E) J_{j_\mu+1}(|E| r)  d^{j_\mu}_{m,j_\mu}(\theta) \\   0 \end{array}\right)~, \quad ( p_\mu > 0 )~, \displaybreak[3] \nonumber \\
\tilde{\Psi}_{m,-}^{(i_\mu)}(E;\vec{x}) =&~ \frac{\sqrt{|p_\mu| |E|} e^{-i m \phi}}{2 \sqrt{2\pi r}} \left( \begin{array}{c} 0 \\ J_{j_\mu}(|E| r)  d^{j_\mu}_{m,-j_\mu}(\theta) \\ 0 \\ i \sgn(E) J_{j_\mu+1}(|E| r)  d^{j_\mu}_{m,-j_\mu}(\theta) \end{array}\right)~, \quad ( p_\mu < 0 )~.
\end{align}
Here $m$ runs from $-j_\mu$ to $j_\mu$.  If $p_\mu = 0$ then these solutions do not exist.

Together, these wavefunctions form an orthonormal set in the sense that
\begin{equation}
\int_{\UU} \ed^3 x \overline{ \Psi_{j_1,m_1,s_1}^{(i_\mu)}(E_1; \vec{x})}  \Psi_{j_2,m_2,s_2}^{(i_\mu)}(E_2; \vec{x}) = \delta(E_1 - E_2) \delta^{j_1 j_2} \delta_{m_1m_2} \delta_{s_1 s_2}~.
\end{equation}
Here $s$ takes values in $\{1,2,\sgn(p_\mu)\}$ with the understanding that $j$ is fixed to $j_\mu$ when $s = \sgn(p_\mu)$.  If $p_\mu = 0$ then $s$ only runs over $\{1,2\}$.  The $\{ \Psi_{j,m,s} \}$ are also complete; in appendix \ref{app:Dirac1} we show that they furnish a resolution of the identity operator on $\LL^2[\UU,V_\rho \otimes \mathbb{C}^4]$.  More generally we can construct integral kernels for functions of $(i \hat{\slashed{D}})_{\rho}^{(0)}$; for example,
\begin{align}\label{Rhatkernel}
\hat{R}_{z,\rho}^{-1}(\vec{x},\vec{y}) =&~  \sum_{j,m,s} \int_{-\infty}^{\infty} \frac{ \ed E}{E + i \sqrt{z} + i \slashed{\delta}} \Psi_{j,m,s}(E;\vec{x}) \overline{ \Psi_{j,m,s}(E;\vec{y}) }~,
\end{align}
where $\Psi \bar{\Psi}$ is considered as an operator on $\mathbb{C}^4 \otimes V_\rho$.  We plug this expression into \eqref{rhatdotJ4}.  From \eqref{Psifam}, \eqref{Psifam2} we have that $\Psi, \tilde{\Psi}$ are related by a unitary transformation on $\mathbb{C}^4 \otimes V_\rho$:  
\begin{equation}
(\Psi_\epsilon)_{j,m,s} = ({\bf U}(\theta,\phi) \otimes \exp(\epsilon \rho(P) \phi/2) ) \tilde{\Psi} \equiv \hat{{\bf U}}(\theta,\phi) \tilde{\Psi}_{j,m,s}~,
\end{equation}
where the unitary matrix ${\bf U}$ is given in \eqref{bfUdef}.  Then
\begin{align}\label{rhatdotJ5}
r^2 \hat{r} \cdot \vec{J}(\vec{x},\vec{x}) =&~ i r^2 \sum_{j,m,s}   \tr_{\mathbb{C}^4 \otimes V_\rho} \left\{ \bar{\Gamma} (\hat{r} \cdot \vec{\Gamma})  \int_{-\infty}^{\infty}  \frac{\ed E}{E + i \sqrt{z} + i \slashed{\delta}} \hat{{\bf U}} \tilde{\Psi}_{j,m,s} \overline{\tilde{\Psi}_{j,m,s}} \hat{{\bf U}}^{-1} \right\} \cr
=&~ i r^2 \sum_{j,m,s}  \ \tr_{\mathbb{C}^4 \otimes V_\rho} \left\{ \bar{\Gamma} \Gamma^3  \int_{-\infty}^{\infty} \frac{\ed E}{E + i \sqrt{z} + i (\hat{{\bf U}} \slashed{\delta} \hat{{\bf U}}^{-1}) } \tilde{\Psi}_{j,m,s} \overline{\tilde{\Psi}_{j,m,s}} \right\}~.
\end{align}
Here we used that ${\bf U}$ commutes with $\bar{\Gamma}$ and that the adjoint action of $U(\theta,\phi)^{-1}$ corresponds precisely to the $SO(3)$ rotation sending the $\hat{r}$-axis to the $\hat{z}$-axis; see \eqref{framerotation}.

Now we evaluate the integral over energy.  First we extract the $r$ and $E$ dependence from $\tilde{\Psi}$ by defining the $\mathbb{C}^2 \otimes V_\rho$-valued spinors $\tilde{\psi}(\theta,\phi),\tilde{\chi}(\theta,\phi)$ such that
\begin{align}\label{Eextract}
&  \tilde{\Psi}_{j,m,1} = \frac{\sqrt{|E|}}{\sqrt{r}} \left( \begin{array}{c} J_{j+1}(|E| r) \tilde{\psi}_{j,m,1} \\ \sgn(E) J_{j}(|E| r) \tilde{\chi}_{j,m,1} \end{array} \right)~, ~    \tilde{\Psi}_{j,m,2} = \frac{\sqrt{|E|}}{\sqrt{r}} \left( \begin{array}{c} J_{j}(|E| r) \tilde{\psi}_{j,m,1} \\ \sgn(E) J_{j+1}(|E| r) \tilde{\chi}_{j,m,1} \end{array} \right)~, \cr
&  \tilde{\Psi}_{m,\pm} = \frac{\sqrt{|E|}}{\sqrt{r}} \left( \begin{array}{c} J_{j_\mu}(|E| r) \tilde{\psi}_{m,\pm} \\ \sgn(E) J_{j_\mu + 1}(|E| r) \tilde{\chi}_{m,\pm} \end{array} \right)~, \raisetag{16pt}
\end{align}
Next we let $\SS$ be the unitary similarity transformation that diagonalizes the Hermitian matrix $i \hat{{\bf U}} \slashed{\delta} \hat{{\bf U}}^{-1}$, such that $i \hat{{\bf U}} \slashed{\delta} \hat{{\bf U}}^{-1} = \SS D \SS^{-1}$, where $D$ is the diagonal matrix of eigenvalues.  Let $d(\vec{x})$ denote a generic (real) eigenvalue.  Since $\slashed{\delta} = O(r^{-1/2})$ we have $d(\vec{x}) = O(r^{-1/2})$.  Then, given the form of $\tilde{\Psi}$, the two types of integrals we encounter are
\begin{align}
\II^{(1)}_{\nu}(a) :=&~  \int_{-\infty}^{\infty}\ed E \frac{ r |E|}{E + i\sqrt{z} + d(\vec{x}) } J_{\nu}(|E| r)^2 = -2 i a \int_{0}^{\infty} \ed \xi \frac{\xi}{\xi^2 + a^2} J_\nu(\xi)^2~, \\
\II_{\nu}^{(2)}(a) :=&~ \int_{-\infty}^{\infty} \ed E \frac{r E}{E +i\sqrt{z} + d(\vec{x})} J_\nu(|E| r) J_{\nu+1}(|E| r)  \cr
=&~ 2 \int_{0}^{\infty} \ed \xi \frac{\xi^2}{\xi^2 + a^2} J_{\nu}(\xi) J_{\nu+1}(\xi)~,
\end{align}
Here we have incorporated the factor of $r^2$ out in front of the summations in \eqref{rhatdotJ5} into the definition of $\II^{(1,2)}$, so that all $r$-dependence of $r^2 \hat{r} \cdot \vec{J}(\vec{x},\vec{x})$ is accounted for.  In the second step we changed variables to $\xi = |E| r$.  Although we are interested in the $r \to 0$ limit of this expression, $|E|$ can be arbitrarily large, so we must consider the full range of $\xi$.  We have $\nu = j$ or $j+1$ while 
\begin{equation}
a := r \sqrt{z} - i r d(\vec{x}) = O(r^{1/2})~, 
\end{equation}
Notice the crucial factor of $r$ that appears in $a$, such that $a \to 0$ as $r \to 0$.  

The integrals $\II^{(1,2)}$ are finite for any $\nu \geq -1$ and $\Re(a) \neq 0$.  This will be the case for us as long as $z$ is off the negative real axis.  Their leading behavior at small $a$ is
\begin{equation}\label{integrals}
\II_{\nu}^{(1)}(a) = -2i a \left\{ \begin{array}{l l} \frac{1}{2\nu} + O(a^{\min(2,2\nu)})~, & \nu > 0 \\ - \ln(a) + O(1)~ & \nu = 0~, \end{array} \right. ~, \qquad \II_{\nu}^{(2)}(a) = 1 + O(a^2)~.
\end{equation}
Thus only $\II^{(2)}$ is nonzero in the $r \to 0$ limit.  This picks out the cross terms $\tilde{\psi} \overline{\tilde{\chi}}$ and $\tilde{\chi} \overline{\tilde{\psi}}$ from $\tilde{\Psi} \overline{\tilde{\Psi}}$.  Plugging in \eqref{Eextract}, using \eqref{integrals}, and carrying out the trace over $V_\rho$ and the $\mathbb{C}^2$ block structure, we find
\begin{align}\label{rhatdotJ6}
r^2 \hat{r} \cdot \vec{J}(\vec{x},\vec{x}) =&~ i \sum_{\mu \in \Delta_\rho} n_\rho(\mu) \sum_{j,m,s} \tr_{\mathbb{C}^2} \left\{ \sigma^3 \left( \tilde{\psi}_{j,m,s} \overline{\tilde{\chi}}_{j,m,s} - \tilde{\chi}_{j,m,s} \overline{\tilde{\psi}}_{j,m,s} \right) \right\} + O(r^{1/2}) \cr
=&~ \frac{1}{4\pi} \sum_{\mu \in \Delta_\rho} n_{\rho}(\mu) \bigg\{ p_\mu \sum_{j,m} \left( d^{j}_{m,\half(p_\mu+1)}(\theta)^2 - d^{j}_{m,\half(p_\mu-1)}(\theta)^2 \right) + \cr
& \qquad \qquad \qquad \qquad \qquad \quad - | p_\mu | \sum_{m} d^{j_\mu}_{m,\sgn(p_\mu) j_\mu}(\theta)^2 + O(r^{1/2}) \bigg\}~.
\end{align}
In the second step we have explicitly carried out the final trace and the sum over $s$, determining $\tilde{\psi},\tilde{\chi}$ by comparing \eqref{Eextract} with \eqref{genjPsisol}, \eqref{specialjPsisol}.  The line with the $p_\mu$ prefactor originates from the $s = 1,2$ terms where we used $a_{+}^2 - a_{-}^2 = p_\mu$, while the line with the $|p_\mu|$ prefactor originates from the $s = \sgn(p_\mu)$ term.

For any physical $j,m'$, we have that $\sum_{m=-j}^j d_{m,m'}^j(\theta)^2 = 1$.  This follows from thinking of $d^{j}_{m,m'}(\theta)$ as a special case of the components of a Wigner $D$ matrix: $d^{j}_{m,m'}(\theta) = D^{j}_{m,m'}(0,\theta,0) \equiv ({\bf D}^j(R_y(\theta)))_{m,m'}$, which are the $m$-$m'$ matrix elements of the spin-$j$ representation of a rotation $R_{y}(\theta)$ by angle $\theta$ about the $y$-axis.  Then
\begin{align}
\sum_{m = -j}^j d_{m,m'}^j(\theta)^2 =&~ \sum_{m=-j}^j ( \overline{ {\bf D}^j(R_y(\theta)) } )_{m' m} ({\bf D}^j(R_y(\theta)))_{m m'} = ({\bf D}^j( R_{y}(\theta)^{-1} R_{y}(\theta)))_{m' m'} \cr
=&~ ({\bf D}^j(1))_{m' m'} = 1~. \raisetag{26pt}
\end{align}
It follows that the sum in the first line of \eqref{rhatdotJ6} vanishes and we are left with
\begin{equation}
r^2 \hat{r} \cdot \vec{J}_{z,\rho}(\vec{x},\vec{x}) = - \sum_{\mu \in \Delta_{\rho}} n_{\rho}(\mu) \frac{|p_\mu|}{4\pi} + O(r^{1/2})~,
\end{equation}
or
\begin{equation}\label{defectcontribution}
\lim_{r \to 0} \int_{S^2} \vol_{S^2} r^2 \hat{r} \cdot \vec{J}_{z,\rho}(\vec{x},\vec{x}) = - \sum_{\mu \in \Delta \rho} n_\rho(\mu) | \langle \mu, P \rangle |~.
\end{equation}
%

\subsection{The index}\label{sec:indexresult}

The results \eqref{SinftyI} and \eqref{defectcontribution} can be combined to give the index, \eqref{index1}:
\begin{align}\label{indLrho}
I_{\rho}(z) =&~ \half \sum_{\mu \in \Delta_{\rho}} n_\rho(\mu) \left\{ \frac{ \langle \mu, X_\infty \rangle \langle \mu, \gm \rangle }{ \sqrt{ \langle \mu, X_\infty \rangle^2 + z} } + \sum_{n=1}^{N_t} | \langle \mu, P_n \rangle | \right\} \cr
\Rightarrow \ind{L_\rho} =&~ \lim_{z \to 0^+} I_\rho(z) = \half  \sum_{\mu \in \Delta_{\rho}} n_\rho(\mu) \left\{ \frac{ \langle \mu, X_\infty \rangle \langle \mu, \gm \rangle }{ | \langle \mu, X_\infty \rangle | } + \sum_{n=1}^{N_t} | \langle \mu, P_n \rangle | \right\}~.
\end{align}
Recall we assume maximal symmetry breaking so that $\langle \mu, X_\infty \rangle \neq 0$ holds for all weights $ \mu \in \Delta_{\rho}$, such that $\mu \neq 0$.  The $\mu = 0$ terms are well-defined and vanishing in $I_\rho(z)$; thus we understand the contribution of the zero weight to the index to be zero.

As we have discussed, \tHooft charges sit in the co-character lattice $\Lambda_{G} \subset \mathfrak{t}$, the asymptotic magnetic charge is shifted from $\sum_n P_n$ by an amount in the co-root lattice and thus also sits in the co-character lattice, and finally the weights $\mu$ are required to sit in the integral-dual character lattice.  Thus it is clear that each term in the summand is an integer.  However there is a one-half out in front of the sum and it is not immediately clear that the sum is an even integer.  We expect that this is the case since $\ind{L_\rho}$ should be an integer.  In fact, we know that $\ind{L_\rho}$ must be a non-negative integer when the data $(P_n;\gm;X_\infty;)$ are such that $\fMM$ is non-empty.  The reason is that we have the vanishing theorem $\ker{L_{\rho}^\dag} = \{0 \}$, and therefore $\ind{L_\rho}$ gives the dimension of the kernel of $L_\rho$ (assuming the background monopole configuration $(A,X)$ used to construct $L_\rho$ exists).  Recall that our conjecture for when $\fMM$ is non-empty is the following: the relative magnetic charge defined by $\tilde{\gamma}_{\rm m} := \gm - \sum_n P_{n}^-$ should be a non-negative integral linear combination of simple co-roots, where $P_{n}^-$ is the representative of $P_n$ in the anti-fundamental Weyl chamber and the basis of simple roots is determined from $X_\infty$.  A purely Lie algebra-based proof that \eqref{indLrho} is an integer, when this condition holds, goes as follows.

First note that $\sum_{\mu \in \Delta_\rho} n_\rho(\mu) | \langle \mu, P_n \rangle |$ is the trace of the diagonal matrix $| \rho(P_n) |$, and is thus Weyl invariant.  Therefore we can replace $P_n$ with $P_{n}^-$ in this term.\footnote{Thus we continue to expect that line defects with \tHooft charge $P$ only depend on the Weyl orbit of $P$.}  Then for the other term write $\gm = \tilde{\gamma}_{\rm m} + \sum_n P_{n}^-$, so that
\begin{equation}\label{nonnegP1}
\ind{L_\rho} = \half \sum_{\mu \in \Delta_\rho} n_{\rho}(\mu) \sgn(\langle \mu, X_\infty \rangle) \langle \mu, \tilde{\gamma}_{\rm m} \rangle + \sum_{n=1}^{N_t} \half \sum_{\mu \in \Delta_\rho} n_{\rho}(\mu) \left( \langle \mu, P_{n}^- \rangle + |\langle \mu, P_{n}^- \rangle | \right)~.
\end{equation}
Now $\langle \mu, P_{n}^- \rangle + |\langle \mu, P_{n}^- \rangle |$ is either zero or $2 |\langle \mu, P_{n}^- \rangle |$.  It follows that the contribution of the \tHooft charges to \eqref{nonnegP1} is a non-negative integer, and we can focus on the $\tilde{\gamma}_{\rm m}$ term.  Define
\begin{equation}\label{posnegweights}
A := \sum_{ \mathclap{\substack{ \mu \in \Delta_\rho \\ \langle \mu, X_\infty \rangle > 0 }} } n_{\rho}(\mu) \langle \mu, \tilde{\gamma}_{\rm m} \rangle~, \qquad  B := \sum_{ \mathclap{\substack{ \mu \in \Delta_\rho \\ \langle \mu, X_\infty \rangle < 0 }} } n_{\rho}(\mu) \langle \mu, \tilde{\gamma}_{\rm m} \rangle~.
\end{equation}
Then, on the one hand, the first term of \eqref{nonnegP1} is $\half (A - B)$.  On the other hand $A + B = \tr_{V_\rho} \left( i \rho(\tilde{\gamma}_{\rm m}) \right)$, but $\tilde{\gamma}_{\rm m}$ is a linear combination of co-roots and the trace of a co-root in any representation vanishes.  Therefore $A + B = 0$ and we have
\begin{equation}\label{nonnegP2}
\half \sum_{\mu \in \Delta_\rho} n_{\rho}(\mu) \sgn(\langle \mu, X_\infty \rangle) \langle \mu, \tilde{\gamma}_{\rm m} \rangle = \sum_{ \mathclap{\substack{ \mu \in \Delta_\rho \\ \langle \mu, X_\infty \rangle > 0 }} } n_{\rho}(\mu) \langle \mu, \tilde{\gamma}_{\rm m} \rangle~.
\end{equation}
This establishes that \eqref{nonnegP1} is an integer.  

In fact, our index result together with the vanishing of $\ker{L_{\rho}^\dag}$ implies more: the right-hand side of \eqref{nonnegP2} must be a non-negative integer.  This is equivalent to showing that
\begin{equation}
 \sum_{ \mathclap{\substack{ \mu \in \Delta_\rho \\ \langle \mu, X_\infty \rangle > 0 }} } n_{\rho}(\mu) \langle \mu, H_I \rangle
 \end{equation}
is non-negative for all simple co-roots $H_I$, which is in turn equivalent to showing that
\begin{equation}
 \sum_{ \mathclap{\substack{ \mu \in \Delta_\rho \\ \langle \mu, X_\infty \rangle > 0 }} } n_{\rho}(\mu) \mu
 \end{equation}
 lies in the closure of the fundamental Weyl chamber.  This must hold for any representation $\rho$.  It would be interesting to give a purely Lie algebra-based proof of this statement.

Going back to \eqref{indLrho}, we note that if $\rho = \ad$ so that $\{\mu\} \to \{\alpha\}$, then $n_{\ad}(\alpha) = 1$ $\forall \alpha \in \Delta_{\rm ad} \equiv \Delta$.  We recover the expected result, \eqref{dim1}, for $\dim_{\mathbb{R}}{T_{[\hat{A}]}} \fMM = 2 \ind{L}$.

Finally, we note a very curious aspect of our formula for $\ind{L_\rho}$, namely
that it exhibits wall-crossing behavior as a function of $X_\infty$. Indeed
the formula is discontinuous as a function of $X_\infty$ across walls
where   $\langle \mu, X_\infty\rangle =0 $ for some
weight in the representation $\rho$. Consider such a wall and let $\mu_a$
be the (parallel) weights which all define the same wall.
As $X_\infty$ crosses this wall some quantities
$\sgn(\langle \mu_a, X_\infty \rangle)$ change from $-1$ to $+1$
and some change from $+1$ to $-1$. Let $\chi_a=+1$ in the former case
and $\chi_a=-1$ in the latter case.  Then the difference in the index
after the wall minus before the wall is
\be\label{eq:IndexJump}
\Delta \ind L_\rho = \sum_{a} \chi_a n_\rho(\mu_a) \langle \mu_a, \gm \rangle.
\ee
This has some interesting physical implications and interpretations. One immediate implication is
the following: If the Yang-Mills-Higgs system is coupled to fermions in the representation $\rho$ of the
gauge group then the low energy effective quantum mechanics on the moduli space of monopoles is
modified to include a coupling to an ``index bundle'' whose rank is $\ind{L_\rho}$ \cite{Manton:1993aa,Sethi:1995zm,Cederwall:1995bc,Gauntlett:1995fu,Henningson:1995hj}. If we consider families of theories with fixed $\gm, P_n$ but variable $X_\infty$ then the rank of this bundle
will jump.  In the case of the adjoint representation, the index bundle is simply the tangent bundle.  Jumping of its rank corresponds to a change in the moduli space itself.  In the case of gauge algebra $\mathfrak{g} = \mathfrak{su}(N)$, there is a nice way of understanding this jump
using the brane pictures of \cite{MRVdimP2}.  See section section 8 of that work.

\section{Zero-modes about the Cartan-valued solutions}\label{sec:cartanzms}

In this section we will check the formula \eqref{dim2} for the case of a single
defect with $\gm = P$ using an explicit construction of the tangent space
around a distinguished point in the moduli space. Note that $P$ is not
necessarily in the anti-fundamental chamber and hence $\tilde{\gamma}_{\rm m} = \gm - P^- = P- P^-$
might well be nontrivial. Indeed in this case equation \eqref{dim2} becomes
\begin{equation}\label{Cartandim}
\dim  \fMM  = 2 \sum_{\alpha \in \Delta^+} \left( \langle \alpha, P \rangle + | \langle \alpha, P \rangle | \right) = \sum_{\mathclap{\substack{ \alpha \in \Delta^+ \\ \langle \alpha, P \rangle > 0 }} } 4 \langle \alpha, P \rangle~.
\end{equation}
In particular, if $\langle \alpha, P \rangle \leq 0$, $\forall \alpha \in \Delta^+$, then the dimension is zero, meaning that the moduli space is a point.  This condition means that $P$ is in the closure of the anti-fundamental Weyl chamber, $P = P^-$.

Our goal is to verify \eqref{Cartandim} by explicit construction of the zero-modes around a distinguished point in the moduli space with $\gm = P$.
This point is the Abelian solution given by placing the defect at $\vec{x}_0 = 0$ and
taking
\begin{equation}\label{Cartanvalued}
X = X_\infty - \frac{P}{2r}~, \qquad A = \frac{P}{2}( \pm 1 - \cos{\theta}) \ed \phi~,
\end{equation}
where as before the $\pm$ refers to the solution in the northern and southern patches, $\mathbb{R}_+ \times S_{\pm}$ of $\UU \cong \mathbb{R}_+ \times S^2$.  This background is sufficiently simple to allow for a complete solution to the deformation problem, so we will be able to construct an explicit basis of the tangent space $T_{[\hat{A}]}\fMM$ and check the dimension against the prediction from \eqref{dim1} or \eqref{dim2}.

Before embarking on this computation we make three  remarks. First, the result that the Cartan-valued solution with $P= P^-$ is isolated is what motivated our definition of the relative magnetic charge $\tilde{\gamma}_{\rm m} = \gm - P^-$.  The relative charge is zero for this solution, suggesting it represents a ``pure'' \tHooft defect without any smooth monopoles.  In contrast a Cartan-valued solution with $P \neq P^-$ will have a non-zero $\tilde{\gamma}_{\rm m}$, so we would interpret that solution as describing an \tHooft defect with some number of smooth monopoles sitting on top of it.  The deformations about the solution correspond to moving these monopoles off the defect or exciting their phases.

Secondly we remark that one can write more general examples of Abelian (i.e. Cartan-valued) solutions.   For an Abelian field configuration we have $F = \star \ed X$; the Bianchi identity, $\ed F = 0$, implies $\ed \star \ed X = 0$.  Thus $X$ is harmonic.  On $\mathbb{R}^3$ the only harmonic function satisfying the asymptotic boundary condition is the constant function, $X = X_\infty$.  On $\UU$, however, we can allow for simple poles at the points $\vec{x}_n$, leading to a natural generalization of equation
\eqref{Cartanvalued}. We cannot easily run our check in this more general case because we can
no longer employ spherical symmetry.\footnote{See however \cite{Lamy-Poirier:2015fmo} where the requisite technology has recently been developed.}

Our third remark is that  two \tHooft charges $P,P'$ related by a Weyl transformation are physically equivalent; however, two solutions of the form \eqref{Cartanvalued} which differ only by the interchange $P \leftrightarrow P'$ are \emph{physically distinct}.  One way to see this is that two such solutions can not be related by a local gauge transformation.  A local gauge transformation which implements a Weyl transformation on the infinitesimal two-sphere around the origin and which goes to the identity at infinity can not be valued purely in the Cartan torus.  If it acts on a Cartan-valued solution the result will be a new, physically equivalent solution that is not Cartan-valued.\footnote{An example of such a gauge transformation in $SU(2)$ theory, written in the defining representation, is $\cg = \mathbbm{1}_2 \sin{(\vartheta(r))} -i \sigma^2 \cos{(\vartheta(r))}$, where $\tan{\vartheta} = r$.  As $r \to 0$, $\cg \to -i \sigma^2$ which implements the Weyl transformation $\Ad(\cg)(H) = - H$ for $H \in \mathfrak{su}(2)$, and as $r \to \infty$, $\cg \to \mathbbm{1}_2$.  This gauge transformation will map a Cartan-valued solution to one which has non-zero components along the roots.}  A simpler way to see this is that, while two such solutions have physically equivalent \tHooft charges, they have inequivalent asymptotic magnetic charges, $\gm = P$ and $\gm' = P'$.

Turning to the computation, recall from the discussion of subsection \ref{sec:tangentspace} that bosonic zero-modes can be constructed from $\psi \in \ker{L}$ via the relation $\delta \hat{A}_a = (\bar{\tau}_a)^{\dot{\alpha}\alpha} \psi_\alpha$, where $L = i\bar{\tau}^a \hat{D}_a$.  For each linearly independent $\psi_\alpha$ we get two linearly independent zero-modes by taking $\dot{\alpha} = \dot{1}$ or $\dot{2}$.  Therefore we are interested in finding the complete set of $\LL^2[\UU,\mathbb{C}^2 \otimes \mathfrak{g}]$-normalizable solutions to
\begin{equation}
-i L \psi = \left\{ \sigma^i \otimes \left( \pd_i + \ad(A_i) \right) + i \mathbbm{1}_2 \otimes \ad(X) \right\} \psi = 0~.
\end{equation}
Much of the analysis parallels the construction of the scattering states, \eqref{genjPsisol} and \eqref{specialjPsisol} in the previous subsection, and details can be found in appendix \ref{app:Dirac}.  (See especially \ref{app:zeromodes}.)

We make a root decomposition of the Lie algebra,
\begin{equation}\label{rootdecomp}
\mathfrak{g}_{\mathbb{C}} = \mathfrak{t}_{\mathbb{C}} \oplus \bigoplus_{\alpha \in \Delta} (-iE_\alpha) \cdot \mathbb{C}~,
\end{equation}
where the $E_\alpha$ are raising/lowering operators.  (See appendix \ref{app:Lie} for our Lie algebra conventions.)  We take $\{H_I \}_{I = 1}^{\rnk{\mathfrak{g}}}$ to be a basis for $\mathfrak{t}$ consisting of the simple co-roots; if $H \in \mathfrak{t}$ we have $ \ad(i H)(E_\alpha) := [i H, E_\alpha] = \langle \alpha, H \rangle E_{\alpha}$, and $\ad(H)(H_I) = 0$.  The $H_I$ together with the $-iE_\alpha$ form a basis for $\mathfrak{g}_{\mathbb{C}}$.  We expand $\psi$ in this basis, writing $\psi = \sum_{\alpha} \psi^{(\alpha)}(-i E_\alpha) + \sum_I \psi^{(I)} H_I$.  We let $\epsilon = \pm$ keep track of the patch we are in so that we have $\mathbb{C}^2$-valued functions $\psi^{(\alpha,I)}_\epsilon$ satisfying
\begin{align}\label{zmcompeqn}
& \left\{ \vec{\sigma} \cdot \vec{\pd}  - i \tau^\phi \frac{p_\alpha}{2} (\epsilon 1 - \cos{\theta}) + \mathbbm{1}_2 \left( x_\alpha - \frac{p_\alpha}{2 r} \right) \right\} \psi_{\epsilon}^{(\alpha)} = 0~, \qquad \vec{\sigma} \cdot \vec{\pd} \psi_{\epsilon}^{(I)} = 0~,
\end{align}
where $p_\alpha \equiv \langle \alpha, P \rangle \in \mathbb{Z}$ and $x_\alpha = \langle \alpha, X_\infty \rangle \in \mathbb{R}$.  Maximal symmetry breaking implies $x_\alpha \neq 0$ for all roots $\alpha$.  We can view the equation for $\psi^{(I)}$ as a special case of that for $\psi^{(\alpha)}$ with $p_\alpha = x_\alpha = 0$.

Separation of variables leads one to
\begin{equation}
\psi_{\epsilon,j,m}^{(\alpha)}(\vec{x}) = e^{i \epsilon p_\alpha \phi/2} U(\theta,\phi) \tilde{\psi}_{j,m}^{(\alpha)}(\vec{x})~,
\end{equation}
where $U(\theta,\phi) = e^{-i \phi \sigma^3/2} e^{-i \theta \sigma^2/2}$ and the form of $\tilde{\psi}$ depends on whether $j > j_\alpha$ or $j = j_\alpha$, where $j_\alpha := \half(|p_\alpha| - 1)$.  In the first case,
\begin{equation}\label{jmfam}
\tilde{\psi}_{j,m}^{(\alpha)}(\vec{x}) = \left( \begingroup \renewcommand{\arraystretch}{1.3} \begin{array}{c} \hat{\psi}_{1}^{(\alpha)}(r) d^{j}_{m,\half(p_\alpha +1)}(\theta) \\ \hat{\psi}_{2}^{(\alpha)}(r) d^{j}_{m,\half(p_\alpha -1)}(\theta) \end{array} \endgroup \right) e^{-i m \phi}~.
\end{equation}
In the second case, which exists only when $p_\alpha \neq 0$, the form depends on the sign of $p_\alpha$:
\begin{align}
\tilde{\psi}_{m}^{(\alpha)}(\vec{x}) =&~ \left\{ \begingroup \renewcommand{\arraystretch}{1.2}  \begin{array}{l l}  \left( \begin{array}{c} \hat{\psi}_{1}^{(\alpha)}(r) d^{j_\alpha}_{m,j_\alpha}(\theta) \\ ~\, 0 \end{array} \right) e^{-i m \phi}~,~ &  (p_\alpha > 0)~,  \\
\left( \begin{array}{c} 0 \\ \hat{\psi}_{2}^{(\alpha)}(r) d^{j_\alpha}_{m,-j_\alpha}(\theta) \end{array} \right) e^{-i m \phi}~,~ &  (p_\alpha < 0)~. \end{array} \endgroup \right.
\end{align}
Plugging these back into \eqref{zmcompeqn} yields the radial equations
\begin{equation}\label{radialzmeqn}
\left[ r \pd_r + x_\alpha r - \frac{p_\alpha}{2} \right] f_1 = k f_2~, \qquad \left[ r \pd_r - x_\alpha r + \frac{p_\alpha}{2} \right] f_2 = k f_1~.
\end{equation}
where $\hat{\psi}_{1,2}^{(\alpha)} = \frac{1}{r} f_{1,2}$ and $k = \sqrt{ (j+\half)^2 - \frac{p_{\alpha}^2}{4} }$.  Note that $k = 0$ when $j = j_\alpha$.

We analyze the radial equations in appendix \ref{app:zeromodes} and find the following:
\begin{itemize}
\item There are no $\LL^2$ solutions of \eqref{radialzmeqn} when $k > 0$; \ie\ there are no $\LL^2$ solutions of the form \eqref{jmfam}.
\item There are $\LL^2$ solutions of \eqref{radialzmeqn} when $k = 0$ if and only if $p_\alpha$ and $x_\alpha$ have the same sign.  In this case the solutions are
\begin{equation}
f_1 \propto r^{p_\alpha/2} e^{-x_\alpha r}~, \quad (x_\alpha,p_\alpha > 0)~, \qquad f_2 \propto r^{-p_\alpha/2} e^{x_\alpha r}~, \quad (x_\alpha, p_\alpha < 0)~.
\end{equation}
\end{itemize}
In the appendix we also analyze the equation for $\LL^2$ zero-modes of the adjoint operator, $L^\dag \chi = 0$, and show that there are none, in agreement with the general vanishing theorem discussed around \eqref{positiveL}.  Note also that it is crucially important that $|x_\alpha| > 0$ for the existence of bound states.  This is consistent with our analysis in the previous subsection where we did not find any zero energy $\LL^2$ eigenfunctions of $(i \hat{\slashed{D}})^{(0)}$.  In summary, we have $\LL^2$ solutions to \eqref{zmcompeqn} if and only if $x_\alpha, p_\alpha \neq 0$ and $\sgn(p_\alpha) \cdot \sgn(x_\alpha) = 1$.  In this case there are $2 j_\alpha + 1 = |p_\alpha|$ solutions labeled by $-j_\alpha \leq m \leq j_\alpha$ and given by
\begin{align}\label{kerLsols}
\psi_{\epsilon,m}^{(\alpha)}(\vec{x}) =& \left\{ \begingroup \renewcommand{\arraystretch}{1.2}  \begin{array}{l l} \left(  \begin{array}{c} e^{-i \phi/2} \cos{\frac{\theta}{2}} \\ e^{i\phi/2} \sin{\frac{\theta}{2}} \end{array} \right) b_{m}^{(\alpha)} r^{-1+ p_\alpha/2} e^{-x_\alpha r}d^{j_\alpha}_{m,j_\alpha}(\theta) e^{i (\epsilon p_\alpha -2m) \phi/2}~,~ &  (x_\alpha,p_\alpha > 0)~,  \\
\left( \begin{array}{c} -e^{-i\phi/2} \sin{\frac{\theta}{2}} \\ e^{i\phi/2} \cos{\frac{\theta}{2}} \end{array} \right) c_{m}^{(\alpha)} r^{-1 - p_\alpha/2} e^{x_\alpha r} d^{j_\alpha}_{m,-j_\alpha}(\theta) e^{i (\epsilon p_\alpha -2m) \phi/2}~,~ &  (x_\alpha,p_\alpha < 0)~, \end{array} \endgroup \right.
\end{align}
where $b_{m}^{(\alpha)}, c_{m}^{(\alpha)} \in \mathbb{C}$ are arbitrary.

Before applying this result to the construction of bosonic zero-modes we would like to comment on the relation to previous work.  This result is consistent with previous investigations in the literature considering bound states of spin $1/2$ particles interacting with a classical Dirac monopole, \cite{Kazama:1976fm,Goldhaber:1977xw,Callias:1977cc}, in so far as we also find bound states.  However the details of the wavefunctions are slightly different because we have a background Higgs field that is also singular.  As we mentioned before this actually removes the need to choose a self-adjoint extension of the Dirac operator by specifying a boundary condition at $r = 0$.  The Dirac operator we work with is already self-adjoint and does not require the specification of a boundary condition.  The mechanism at work for this can be seen by considering the solutions presented above.  The most singular behavior we find as $r \to 0$ is $\psi \sim r^{-1/2}$, which occurs for $|p_\alpha| = 1$.  Thus $\ed^3 x || \psi||^2 \sim r^2 || \psi ||^2 = O(r)$ and no boundary terms arise from integrating by parts when checking the self-adjointness of $i \hat{\slashed{D}}$.  If the $1/r$ term in the background Higgs field were not present, the only change in the differential equation would be the absence of the $p_\alpha$ terms in \eqref{radialzmeqn}.  With these gone, the normalizable solution when $k = 0$ would be $\hat{\psi} \propto r^{-1} e^{- |x_\alpha| r}$, in which case $\ed^3 x || \psi||^2 \sim r^2 || \psi ||^2 = O(1)$ and we would pick up boundary terms at $r = 0$ from integration by parts.  This same type of reasoning extends to the entire spectrum of $(i \hat{\slashed{D}})$.  The scattering states of $(i \hat{\slashed{D}})^{(0)}$ constructed in the previous subsection have the same property that the most singular behavior as $r \to 0$ is $r^{-1/2}$.  Note that the $r^{-1/2}$ behavior for a normalizable deformation, $\delta \hat{A}_a$, is consistent with the subleading behavior in the defect boundary condition \eqref{tHooftpole}.

Returning to the enumeration of bosonic zero-modes, we use $X_\infty$ to define a polarization of the root system, $\Delta = \Delta^+ \cup \Delta^-$ with $\alpha \in \Delta^+ \iff x_\alpha > 0$.  For each $\alpha \in \Delta^+$, there are $2 p_\alpha$ solutions $\psi \in \ker{L}$ if $p_\alpha > 0$ and none if $p_\alpha < 0$.  This is because when $\alpha \in \Delta^+$ and $p_\alpha > 0$, we get $p_\alpha$ solutions of the first type in \eqref{kerLsols}, but we also get $p_\alpha$ solutions of the second type since $x_\alpha,p_\alpha > 0$ implies $x_{-\alpha},p_{-\alpha} < 0$.  This shows that
\begin{equation}
\dim{\ker{L}} = \sum_{ \mathclap{\substack{ \alpha \in \Delta^+ \\ \langle \alpha, P \rangle > 0 }} } 2 \langle \alpha, P \rangle~.
\end{equation}
Then since $\dim{T_{[\hat{A}]} \fMM} = 2 \dim{\ker{L}}$, we recover \eqref{Cartandim}.

Let us be a little more explicit about how the bosonic zero-modes are constructed from the $\psi^{(\alpha)}$.  Suppose that $\alpha \in \Delta^+$ such that $p_\alpha > 0$.  Then we have the solutions $\psi =  \psi^{(\alpha)} (-iE_\alpha)$ and $\psi^{(-\alpha)}(-i E_{-\alpha}) \in \ker{L}$.  We can construct bosonic zero modes by setting either $\delta \hat{A}_a = (\bar{\tau}_a)^{\dot{1}\alpha}\psi_\alpha$ or $\delta \hat{A}_a = (\bar{\tau}_a)^{\dot{2}\alpha} \psi_\alpha$.  Consider the the first type.  We also use the fact that the gauge field is valued in the real (compact) form of the Lie algebra, implying $\delta \hat{A}^{(-\alpha)} = \delta \hat{A}^{(\alpha) \ast}$.  Then we have
\begin{align}\label{type1bzm}
& \delta A_{1}^{(\alpha)} = \half \left( \psi^{(\alpha)}_2 + \psi_{2}^{(-\alpha)\ast} \right)~, \quad \delta A_{2}^{(\alpha)} =  \frac{1}{2i} \left( \psi^{(\alpha)}_2 - \psi_{2}^{(-\alpha)\ast} \right)~, \cr
& \delta A_{3}^{(\alpha)} = \half \left( \psi^{(\alpha)}_1 + \psi_{1}^{(-\alpha)\ast} \right)~, \quad \delta X^{(\alpha)} = \frac{i}{2} \left( \psi^{(\alpha)}_1 - \psi_{1}^{(-\alpha)\ast} \right)~.
\end{align}
There is a $2 p_\alpha$-dimensional space of bosonic zero-modes $\delta \hat{A}_a = \delta \hat{A}_{a}^{(\alpha)} (-i E_\alpha)$, with $\delta \hat{A}_{a}^{(\alpha)}$ of the form \eqref{type1bzm}: the label $m$ runs over $p_\alpha$ values and for each $m$ there is a two-dimensional solution space corresponding to the freely specifiable constants $b_{m}^{(\alpha)}$ and $c_{m}^{(-\alpha)\ast}$.

Analogously, for the second type we find
\begin{align}\label{type2bzm}
& \delta A_{1}^{(\alpha)} = \half \left( \psi^{(\alpha)}_1 + \psi_{1}^{(-\alpha)\ast} \right)~, \quad \delta A_{2}^{(\alpha)} =  \frac{i}{2} \left( \psi^{(\alpha)}_1 - \psi_{1}^{(-\alpha)\ast} \right)~, \cr
& \delta A_{3}^{(\alpha)} = -\half \left( \psi^{(\alpha)}_2 + \psi_{2}^{(-\alpha)\ast} \right)~, \quad \delta X^{(\alpha)} = \frac{i}{2} \left( \psi^{(\alpha)}_2 - \psi_{2}^{(-\alpha)\ast} \right)~.
\end{align}
These give another $2p_\alpha$ linearly independent zero-modes $\delta \hat{A}_a = \delta \hat{A}_{a}^{(\alpha)} (-i E_\alpha)$.  Thus we have a total of $4 p_\alpha$ bosonic zero-modes associated with each positive root $\alpha \in \Delta^+$ such that $p_\alpha > 0$.  This again confirms \eqref{Cartandim}.  One can also show that for fixed $\alpha$ and $m$ the four-dimensional space of solutions given by the two of type \eqref{type1bzm} and the two of type \eqref{type2bzm} form an invariant subspace under the action of the quaternionic structure \eqref{quatstructure}.

\section{A two-parameter family of spherically symmetric singular monopoles}\label{sec:cfamily}

In this section we discuss a simple generalization of the renowned Prasad-Sommerfield solution to the case of singular monopole solutions. The physical interpretation of this solution (which for some time puzzled the authors) is greatly facilitated by the dimension formula, and indeed this example was part of the motivation for deriving that formula. We are especially indebted to Sergey Cherkis for a useful discussion on the very rich relations of this solution to previous literature on singular monopoles.

We begin by recalling the derivation of the Prasad--Sommerfield solution, \cite{Prasad:1975kr}, for the smooth $\mathfrak{su}(2)$ monopole. Let $\{H, E_{\pm} \}$ denote the co-root and raising and lowering operators of $\mathfrak{sl}(2)$, (see Appendix \ref{app:Lie} for conventions).  If we input the ansatz
\begin{align}\label{tHooftansatz}
X =&~ \half h(r) H~, \cr
A =&~ \half (\pm 1 - \cos{\theta}) \ed \phi H + \cr
& + \half f(r) \left[ e^{\pm i \phi} \left( -\ed\theta - i \sin{\theta} \ed \phi \right) E_+ + e^{\mp i \phi} \left( \ed \theta - i \sin{\theta} \ed \phi \right) E_- \right]~,
\end{align}
we find that the Bogomolny equation, $F = \star \eD X$, is equivalent to
\begin{equation}\label{fopair}
f'(r)+f(r) h(r) = 0~, \qquad r^2 h'(r) + f(r)^2 - 1 = 0~.
\end{equation}
Equation \eqref{tHooftansatz} is just the 't Hooft--Polyakov spherically symmetric ansatz, gauge transformed from hedgehog to string gauge. In solving \eqref{fopair} there are two integration constants. The first one is fixed to the asymptotic Higgs vev, $X_\infty = \half m_W H$, where $m_{W}$ is the mass of the elementary $W$-boson. The second one is usually set to zero so that the solution is regular at $r = 0$. However if we leave this integration constant, denoted $c$, in the solution then we find
\begin{equation}\label{cfam}
h(r) = m_W \coth{\left(m_W r + c\right)} - \frac{1}{r}~, \qquad f(r) = \frac{m_W r}{\sinh{\left( m_W r + c \right)} }~.
\end{equation}
When $c =0$ we recover the Prasad--Sommerfield solution.

\begin{figure}
\begin{center}
\includegraphics{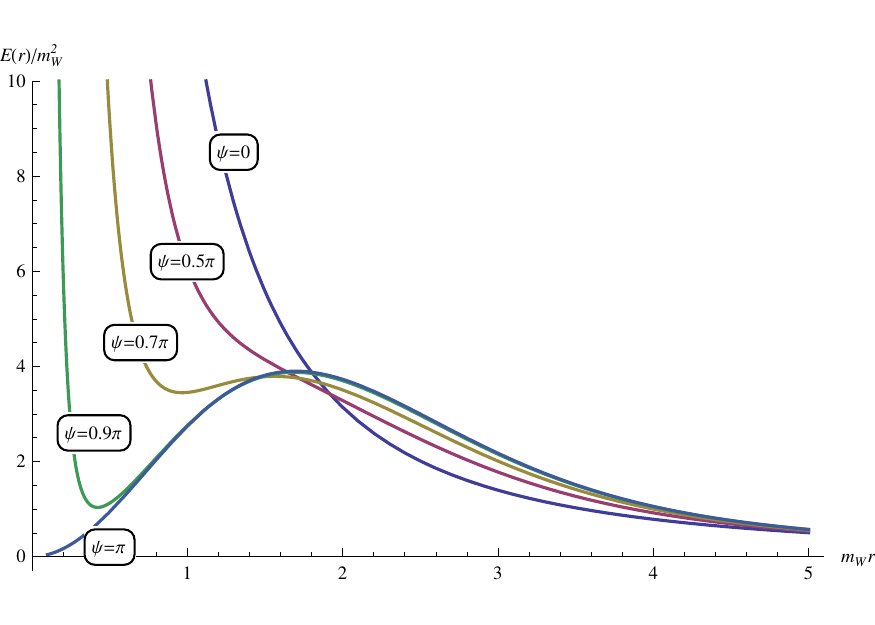}
\caption{The energy density \eqref{endens} for the $c$-family of solutions, \eqref{tHooftansatz}, \eqref{cfam}, with $c = -\ln(\sin(\psi/2))$ for the values of $\psi$ shown. Allowing $\psi$ to vary linearly in time, the energy density of the spherically symmetric field configuration oscillates back and forth through the profiles shown.}
\label{fig2}
\end{center}
\end{figure}

However for any $c \in (0,\infty)$ we obtain a field configuration that has a singularity at $r = 0$ consistent with the \tHooft defect boundary conditions! Field configurations with different values of $c$ are clearly gauge-inequivalent --- for example, the gauge-invariant energy density depends on $c$. (See Figure \ref{fig2} below for a plot.)  Hence $c$ parameterizes a one-parameter family of gauge-inequivalent, spherically symmetric singular monopole configurations. Note well that the limit $c \to \infty$ makes sense and simply yields a Cartan-valued solution.

This one-parameter family can be extended to a two-parameter family by acting with an asymptotically non-trivial gauge transformation that preserves the asymptotic Higgs field. The infinitesimal action $\delta_\epsilon \hat{A} = -\hat{\eD} \epsilon$ gives a deformation that solves the linearized Bogomolny equation \eqref{linearsd} around the background \eqref{tHooftansatz}, \eqref{cfam}. Demanding that the deformation also satisfy the orthogonality condition, \eqref{gaugeorth}, implies $\hat{D}^2 \epsilon = 0$. This equation is analogous to the one that occurs in the study of the Julia--Zee dyon \cite{Julia:1975ff}. After imposing the boundary condition $\delta_\epsilon \hat{A} = O(r^{-1/2})$ at $r = 0$, as stipulated by \eqref{tHooftpole}, we find the solution
\begin{equation}\label{gaugegen}
\epsilon(r) = \frac{1}{2} \left[ \coth(m_W r + c) + \frac{1}{m_W r} \left( \frac{ \coth(m_W r + c)}{\coth(c)} - 1 \right) \right] H~.
\end{equation}
The corresponding gauge transformation $\mathpzc{g} = \exp(\chi \epsilon)$ asymptotes to $\exp(\chi H/2)$ and thus we may take $\chi \sim \chi + 2\pi$. After acting on \eqref{tHooftansatz} with $\mathpzc{g}$ through \eqref{finitegt} we obtain a two-parameter family of spherically symmetric singular monopoles, parameterized by $(c,\chi)$.

How should we interpret this family? The dimension formula provides some clarification. The asymptotic magnetic charge of the $(c,\chi)$ family of solutions is $\gm = H$ while the \tHooft charge of the singularity is $P = H$. Hence the relative charge is $\tilde{\gamma}_{\rm m} = 2H$. We conclude that the $(c,\chi)$ family of solutions is a two-dimensional locus of spherically symmetric solutions within the eight-dimensional moduli space of two smooth $\mathfrak{su}(2)$ monopoles in the presence of an $SU(2)$ \tHooft defect of charge $P = H$. In the notation we have introduced, the eight-dimensional manifold is $\fMM\left( (\vec{x}_0,H);H; X_\infty \right)$. In the following we will simply refer to this manifold as $\fMM_8$. We will refer to the two-dimensional surface parameterized by $(c,\chi)$ as $\Sigma \hookrightarrow \fMM_8$.

We can now interpret the parameters $(c,\chi)$. Our generalized Prasad-Sommerfield solutions  represent configurations where the two smooth monopoles are spread out and surrounding the defect in a spherical shell.   If we place the defect at $\vec{x}_0$ then $r$ measures the distance from it, $r = |\vec{x} - \vec{x}_0|$.  The parameters $(c,\chi)$ represent, respectively, the relative and overall phase of the constituent smooth monopoles.  Allowing either of these parameters to become time-dependent results in a configuration of dyons. Six  other parameters can be taken to be the displacement of   the two smooth monopoles in $\mathbb{R}^3$. We cannot obtain these configurations starting from within the ansatz \eqref{tHooftansatz} because they do not possess spherical symmetry about $\vec{x}_0$.

The restriction to $\Sigma$ of the metric on $\fMM_8$ can be obtained from the explicit solutions above. First we note that $\pd_c (A,X) = \pd_c \hat{A}$ satisfies the gauge orthogonality condition and therefore the associated zero-mode is $\delta_c \hat{A} = \pd_c \hat{A}$. This allows us to compute the component $g_{cc} = g(\delta_c, \delta_c)$ of the metric directly from the definition, \eqref{metC}. Furthermore we find $g_{c\chi} = g(\delta_c, \delta_\epsilon) = 0$. Finally $g_{\chi\chi}$ can be reduced to a boundary term,
\begin{equation}
g(\delta_\epsilon, \delta_\epsilon) = \frac{2}{g^2} \int_{\UU} \ed^3 x \Tr \left\{ \hat{D}_a \epsilon \hat{D}^a \epsilon \right\} = \frac{2}{g^2} \lim_{r \to \infty} \int_{S_{\infty}^2} r^2 \hat{r} \cdot \Tr \left\{ \epsilon \vec{\pd} \epsilon \right\}~,
\end{equation}
which can also be explicitly evaluated. This leads to the metric
\begin{equation}
\ed s_{\Sigma}^2 = \frac{4\pi}{g^2 m_W} \left( \frac{2}{e^{2c} - 1} \ed c^2 + (1-\tanh{c}) \ed \chi^2 \right)~.
\end{equation}
After changing variables according to $e^{-c} = \sin(\psi/2)$ the metric takes the form
\begin{equation}\label{ds2Sigma}
\ed s_{\Sigma}^2 = \frac{4\pi}{g^2 m_W} \left( \ed \psi^2 + \tan^2(\psi/2) \ed \chi^2 \right)~.
\end{equation}
Decreasing $c$ corresponds to increasing $\psi$, and as $c$ ranges down from $\infty$ to $0$, $\psi$ ranges up from $0$ to $\pi$. Hence we may take $\psi \in [0,\pi)$, while again, $\chi \sim \chi + 2\pi$.

Observe that the $\chi$ circle shrinks to zero size as $\psi \to 0$, corresponding to $c \to \infty$, where we approach the Cartan-valued solution. There is a nice physical explanation of this. Recall that $\chi$ parameterizes asymptotically nontrivial gauge transformations that act effectively on the field configuration \eqref{tHooftansatz}. In the case of a Cartan-valued background, however, the gauge transformation $\mathpzc{g} = \exp(\chi \epsilon)$, which takes values in the Cartan torus, is not effective. No new solutions are generated and thus the corresponding Killing vector $\frac{\pd}{\pd\chi}$ should have vanishing norm at this point. The point $\psi = 0$ is an orbifold singularity of the two-dimensional metric \eqref{ds2Sigma}, however in order to determine whether or not it is a singular point of the full eight-dimensional space we need to know how the surface $\Sigma$ is embedded in $\fMM_8$.

In fact $\fMM_8$ has been previously studied in the context of singular monopoles, but in order to compare with the literature we first describe a slight generalization. As we noted above, the \tHooft charge of the defect is $P = H$. In the case of $G = SO(3)$ gauge group this is twice the minimal charge, and we can view $\fMM_8$ as a special case of a more general eight-manifold $\fMM_{8}'(\vec{x}_1,\vec{x}_2)$, which is the moduli space of two smooth $SO(3)$ monopoles in the presence of two minimal defects of charges $P_{1,2} = \half H$, located at positions $\vec{x}_{1,2}$. $\fMM_8$ corresponds to the case where the minimal defects are coincident: $\fMM_8 = \fMM_{8}'(\vec{x}_0,\vec{x}_0)$.

The manifold $\fMM_{8}'(\vec{x}_1,\vec{x}_2)$ is analyzed in section IV of Houghton's work \cite{Houghton:1997ei}, where it is obtained from a higher-dimensional moduli space of smooth monopoles in the limit where some monopole masses become infinite. It is also described in terms of a (finite-dimensional) hyperk\"ahler quotient in the work of Cherkis and Kapustin \cite{Cherkis:1997aa}; it corresponds to the case $n=2$ and $k=2$ in the notation of that paper.

$\fMM_{8}'$ possesses a tri-holomorphic $U(1)$ isometry, \ie\ a $U(1)$ isometry that preserves the hyperk\"ahler structure, \eqref{quatstructure}. This isometry is none other than the one generated by the asymptotically nontrivial gauge transformation $\delta_\epsilon$ discussed above. (Equation \eqref{gaugegen} gives the gauge generator $\epsilon(\vec{x};c)$ only on a special two-dimensional locus $\Sigma \subset \fMM_8$ parameterized by $c$ and $\chi$.) In \cite{Houghton:1997ei} it is shown that this isometry has no fixed points provided $\vec{x}_1 \neq \vec{x}_2$. In this case one can take a hyperk\"ahler quotient with respect to this $U(1)$ and obtain a smooth four-dimensional hyperk\"ahler manifold, which is the \emph{centered} moduli space of the two monopoles in the presence of the defects \cite{Houghton:1997ei,Cherkis:1997aa}. The geodesics of this moduli space capture the motion of the two smooth monopoles relative to their center of mass, $\vec{x}_{\rm cm}$. The relative motion of the two smooth monopoles is influenced by the presence of the defects.\footnote{Much as the motion of the earth and moon relative to the center of mass of the earth-moon system is influenced by the sun, which plays the role of a defect in this analogy.} The coordinates of the displacements of the center of mass from the defects, $\vec{d}_{1,2} = \vec{x}_{\rm cm} - \vec{x}_{1,2}$, will appear as parameters in the centered moduli space. This is the moduli space denoted $N(\vec{d}_1, \vec{d}_2)$ in \cite{Houghton:1997ei}. When $\vec{d}_1 = \vec{d}_2$ however, it was pointed out in \cite{Houghton:1997ei} that there are fixed points of the $U(1)$ action, in which case $N(\vec{d},\vec{d})$ will be singular. This is the relevant case for us, where $\vec{d} = \vec{x}_{\rm cm} - \vec{x}_0$. Indeed the general argument we gave above shows that the Cartan-valued solution is a fixed-point of the $U(1)$ action generated by $\delta_\epsilon$.

$N(\vec{d}_1,\vec{d}_2)$ is the third member in a sequence of (families of) four-dimensional hyperk\"ahler manifolds. The first member of this sequence is the Atiyah--Hitchin manifold \cite{MR934202}, the geodesics of which capture the relative motion of two smooth $\mathfrak{su}(2)$ monopoles in the absence of defects. The second member is Dancer's four-dimensional hyperk\"ahler manifold \cite{Dancer:1992kn,Dancer:1992km}, the geodesics of which describe relative motion of two smooth $\mathfrak{su}(2)$ monopoles in the presence of a single, minimal $SO(3)$ defect \cite{Houghton:1997ei,Cherkis:1997aa,Houghton:1999qu}. Dancer's manifold is a one-parameter generalization of the Atiyah--Hitchin manifold where, in this context, the parameter is interpreted as the distance of the center-of-mass of the two smooth monopole system relative to the defect.

These manifolds are examples of $D_k$ ALF spaces. The first three members of the sequence discussed above correspond to $k=0$ (the Atiyah--Hitchin manifold), $k=1$ (Dancer's manifold), and $k=2$ (Houghton's manifold, $N(\vec{d}_1,\vec{d}_2)$). See \cite{Cherkis:1998xca,Cherkis:1998hi} for a construction of $D_k$ ALF spaces from the point of view of singular monopoles and Nahm data, and \cite{Cherkis:2003wk} for an explicit construction of their metrics. These manifolds also appear in other, though as it turns out, related contexts. For example, they are the transverse metric in the $M$-theory description of $k$ D6-branes in the presence of an $\textrm{O6}^-$-plane \cite{Sen:1997kz}. In this context the displacement parameters $\vec{d}_{1,2}$ of $N(\vec{d}_1,\vec{d}_2)$ have a different interpretation, as the positions of the two D6-branes relative to the orientifold plane. Locally, in the vicinity of the (M-theory lifted) D6-brane, the transverse metric looks like Taub-NUT, with the location of the D6-brane corresponding to the nut. This makes it easy to understand the singularity structure of $N(\vec{d},\vec{d})$, which corresponds to taking the two D6-branes coincident. Locally this gives a two-centered Taub-NUT space with degenerate centers; in other words a $\mathbb{Z}_2$ orbifold singularity. $N(\vec{d}_1,\vec{d}_2)$ are also the metrics on the Coulomb branch of three-dimensional $SU(2)$ gauge theory with $\NN = 4$ supersymmetry and $k$ matter hyper-multiplets in the fundamental representation \cite{Seiberg:1996nz}. Here the $\vec{d}_i$ are the bare masses of the hyper-multiplets (which are real three-vectors).

Now let us return to the case of the solutions \eqref{tHooftansatz} with \eqref{cfam}. The relevant centered moduli space is $N(0,0)$, as these solutions have the center of mass of the two smooth monopoles coincident with the defect. However it can be seen from several points of view that $N(0,0)$ is simply the flat orbifold space $(\mathbb{R}_{t}^3 \times S_{\psi}^1)/\mathbb{Z}_2$, where the $\mathbb{Z}_2$ acts by flipping the sign of all coordinates, $(\vec{t},\psi) \to (-\vec{t},-\psi)$. In the three-dimensional field theory context of \cite{Seiberg:1996nz} this corresponds to the case with 2 hypermultiplets of vanishing mass. In this situation there are no perturbative or non-perturbative corrections to the potential and the metric on the Coulomb branch is the classical metric, $(\mathbb{R}^3 \times S^1)/\mathbb{Z}_2$, where the $\mathbb{Z}_2$ quotient arises from the action of the Weyl group of $SU(2)$. In the M-theory context of \cite{Sen:1997kz}, $N(0,0)$ corresponds to having both D6-branes coincident with the $\textrm{O6}^-$-plane. When this is done, the sources that generate a nontrivial transverse metric cancel out. All that remains is the $\mathbb{Z}_2$ identification due to the orientifold plane.\footnote{A more general degeneration for the $D_k$ ALF space when two of the displacement parameters are set to zero has been demonstrated in \cite{Chalmers:1998pu}, using the Legendre transform construction of \cite{Lindstrom:1987ks}.}

In the context of singular monopoles, the $\mathbb{R}_{t}^3$ factor of $N(0,0)$ corresponds to displacing the two smooth monopoles from the defect by equal and opposite amounts, so that the their center of mass remains coincident with the defect. Since the solutions \eqref{tHooftansatz} are spherically symmetric, they all have $\vec{t} = 0$. The $S^1$ factor of $N(0,0)$, however, corresponds to the relative $U(1)$ phases of the constituents, and hence we see that the interval $\{ \vec{0} \} \times S^1/\mathbb{Z}_2 \hookrightarrow N(0,0)$ is precisely the locus being parameterized by $\psi \in [0,\pi)$.

What is the interpretation of the limit $\psi \to \pi$? We have excluded the point $\psi = \pi$, corresponding to $c =0$, because at this point the solution \eqref{tHooftansatz} is smooth. There is no longer an \tHooft defect and so this configuration does not correspond to any point in $\fMM_8$. The $\psi \to \pi$ limit is an example of the monopole bubbling phenomenon \cite{Kapustin:2006pk}, in which an \tHooft defect emits or absorbs a smooth monopole, changing its charge in the process. In this case an \tHooft defect of charge $P = H$ absorbs a single smooth monopole and is reduced to the trivial defect, $P = 0$. One smooth monopole remains, the field configuration of which is the original Prasad--Sommerfield solution. Note that $(\vec{t},\psi) = (\vec{0},\pi)$ is also a fixed point of the $\mathbb{Z}_2$ orbifold action. One can consider a completion $\fMM_{8}^\star$, where $\fMM_8$ corresponds to the space obtained from $\fMM_{8}^\star$ after removal of the four-dimensional locus of points corresponding to $(\vec{t},\psi) = (\vec{0},\pi)$. One anticipates that this four-dimensional locus should correspond to the moduli space of the one remaining smooth monopole. Similar phenomena were observed in \cite{Kapustin:2006pk} where a smaller moduli space provided a compactification of a larger one related to it by monopole bubbling.\footnote{Here, in contrast, $\fMM_{8}^\star$ is not a compactification of $\fMM_8$; but this is simply because we are studying moduli spaces of monopoles on a \emph{non-compact} three-space.} Also, the complete screening of an $SU(2)$ defect by a smooth monopole has been nicely demonstrated by the exact solutions of \cite{Cherkis:2007jm} describing one smooth monopole in the presence of a defect.

Finally, we cannot restrain ourselves from making a few brief comments on monopole scattering. As we discussed above $N(\vec{d},\vec{d})$ is a natural generalization of the Atiyah--Hitchin manifold; its geodesics describe the scattering of two smooth monopoles in the presence of the $SU(2)$ \tHooft defect, $P = H$, where $\vec{d}$ gives the displacement of the center of mass of the two monopole system from the defect. One expects that as $|\vec{d}| \to \infty$ $N(\vec{d},\vec{d})$ should approach the Atiyah--Hitchin manifold, describing the scattering of the monopoles in the absence of the defect. We'll focus on the opposite extreme, $\vec{d} = 0$, since it is easy to analyze and is the case of relevance for our solutions \eqref{tHooftansatz}. Then we are interested in the geodesics on flat $(\mathbb{R}^3 \times S^1)/\mathbb{Z}_2$, which are simply the images of the geodesics of $\mathbb{R}^3 \times S^1$ under the projection.

We comment on two types of scattering processes. First consider the head-on collision of the two monopoles, such that they meet at a point coincident with defect. At the instant they meet, the field configuration will be described by one of the solutions \eqref{tHooftansatz}, \eqref{cfam} for a fixed $c$ (or $\psi$). Since the geodesics are straight lines, the monopoles pass right through each other and continue on their way, or equivalently they scatter back-to-back at $180^\circ$. These are equivalent under the $\mathbb{Z}_2$ projection, which is accounting for the fact that the two monopoles are indistinguishable. Note that this is a dramatically different scattering process than when two $\mathfrak{su}(2)$ monopoles collide in the absence of a defect. In that case, the monopoles famously scatter at $90^\circ$ \cite{MR934202}, a phenomenon that can be attributed to the exchange of massive $W$-bosons when the two monopoles approach each other, leading to an absorption of angular momentum. Apparently the defect completely inhibits this exchange process, for head-on scattering directly atop the defect. One might anticipate that the geodesics of $N(\vec{d},\vec{d})$ corresponding to head-on collisions interpolate between these two behaviors---the $180^\circ$ scattering and the $90^\circ$ scattering---as $|\vec{d}|$ goes from $0$ to $\infty$. However, a detailed analysis of $N(\vec{d},\vec{d})$ should be carried out to confirm this.

The second process is of a rather different character. We consider the linear time evolution of the modulus $\psi$ along the geodesic $\{ \vec{0} \} \times S^1/\mathbb{Z}_2 \hookrightarrow N(0,0)$. This corresponds to evolution of the relative phase of the two constituent dyons, as they sit coincident with the defect. In order to describe this motion fully we must work in the completion $\fMM_{8}^\star$ which includes the point $\psi = \pi$ where the defect is perfectly screened. Then linear motion on $S_{\psi}^1$ projects to a bouncing motion along the interval $S_{\psi}^1/\mathbb{Z}_2$ in which the relative phase progresses back and forth between the two extremes corresponding to the Cartan-valued solution and the complete-screening or monopole-bubbling solution. Complete screening occurs when the phase of one of the two constituents is perfectly opposite to the phase associated with the defect. The field configurations for this process are completely captured by the $c$-family of solutions; recall that the relationship between $c$ and the phase $\psi$ is $e^{-c} = \sin(\psi/2)$.

It is instructive to plot the radial, gauge-invariant energy density,\footnote{Recall that the total energy of singular monopole solutions is defined by \eqref{Estatic}, which is related to the $r$-integral of \eqref{endens} through the subtraction of a constant term that regularizes the divergence.  This constant term originates from the boundary terms of the action, \eqref{SVbndry}, which we were required to add in order to have a well-defined variational principle.  Every member of the $c$-family of solutions, including the smooth case, $c = 0$, has the same asymptotic magnetic charge and thus the same total energy.}
\begin{equation}\label{endens}
E(r) := 4\pi r^2 \Tr \left\{ B_i B^i + D_i X D^i X \right\}~,
\end{equation}
for various values of $\psi$. See Figure \ref{fig2}. The profile oscillates back and forth between the purely Cartan-valued solution, where we have a simple $1/r^2$ behavior due to the infinite energy of the defect, and the completely screened solution, where the energy profile is that of a single smooth monopole. As $\psi$ approaches $\pi$ and the screening effect becomes stronger, a dip begins to form separating the $1/r^2$ density of the defect from the localized shell corresponding to a smooth monopole. The $1/r^2$ behavior becomes narrower and narrower until it disappears at $\psi = \pi$.

\section{Further Directions}\label{sec:further}

In this paper we have defined the moduli space $\fMM$ of BPS monopoles on $\mathbb{R}^3$ in the presence of \tHooft defects in Yang--Mills--Higgs theory for arbitrary compact simple gauge groups $G$.  This moduli space depends on the locations and charges of the defects, $\{ (\vec{x}_n, P_n) \}$, as well as the asymptotic Higgs vev and magnetic charge of the system, $(\gm;\Phi_\infty)$.  The defects also come with a sign choice, $\sigma$, which is a $\mathbb{Z}_2$ remnant of the phase $\zeta$ labeling line defects in $\NN = 2$ supersymmetric extensions of this model \cite{Gaiotto:2010be}.  We have computed the formal dimension of $\fMM_{\sigma}\left( (\vec{x}_{n}, P_{n} )_{n = 1}^{N_t} ; \gamma_{\rm m} ; \Phi_{\infty} \right)$ by generalizing the original computation of Weinberg using the Callias index theorem.  We continued by studying some examples: first, a simple class of field configurations---the Cartan-valued solutions to the Bogomolny equation---where a basis for the tangent space was explicitly constructed, and second, a two parameter family of spherically symmetric configurations that we argued sits inside an eight-dimensional moduli space corresponding to two smooth $SU(2)$ monopoles in the presence of a defect with $P = H$.

There are several directions to pursue.

\begin{itemize}
\item Global properties of these moduli spaces should be investigated.  In several places we assumed on physical grounds that the $\fMM$ are connected spaces --- for example, when stating that the zero-dimensional moduli space for configurations with $\gm = P = P^-$ is a point, rather than a collection of points.  Connectedness of the moduli space holds in the case of smooth monopoles via the correspondence between monopoles and rational maps \cite{Donaldson:1985id}.  It would be interesting to develop an analogous correspondence here.  

\item In this paper we have only considered the case of maximal symmetry breaking where the asymptotic value of the Higgs field is generic, breaking the global gauge group to the Cartan torus.  A great deal is known in the non-maximal case for smooth monopoles\footnote{See \eg\ Chapter 6 of \cite{Weinberg:2006rq} for a review with references.} and the extension to singular monopoles should be considered.

\item The $\fMM$ are expected to be hyperk\"ahler manifolds, possibly with singularities of at least orbifold type, and it would be nice to get one's hands on some explicit metrics.  One approach would be to determine their asymptotic form which would be valid when the smooth monopoles are widely separated both from each other and from the fixed defects.  One could carry out an analysis along the lines of \cite{Gibbons:1995yw,Lee:1996kz}, approximating the monopoles as point dyons and additionally including the fixed background fields of the defects.  Alternatively one could obtain a class of singular monopole moduli spaces by taking limits of smooth monopole moduli spaces in which some of the mass parameters associated with the Higgs vev become infinite.  This is not a new idea and it has been discussed in the context of some specific examples in \cite{Houghton:1997ei}.  Brane realizations of the Cheshire bow construction may provide a third approach as in \cite{Cherkis:2011ee}, where the asymptotic form of the moduli space metric for instantons on Taub-NUT space was determined via mirror symmetry and a one-loop computation in a four-dimensional supersymmetric gauge theory with defects.

\item It would be interesting to understand the connection between the dimension formula derived here and the one given in \cite{Cherkis:2010bn} for moduli spaces of instantons on Taub-NUT space.  The restriction of that formula to the case of Cheshire bows should be equivalent to ours.

\end{itemize}

\section*{Acknowledgements}

We thank Sergey Cherkis, Kimyeong Lee, and Edward Witten for helpful discussions, and Sergey Cherkis for comments on the draft as well.  GM and ABR acknowledge the Aspen Center for Physics  and the NSF grant \#1066293 for hospitality during the completion of this paper.  DVdB is partially supported by TUBITAK grant 113F164 and BAP grant 13B037, GM is supported by the U.S. Department of Energy under grant DE-FG02-96ER40959, and ABR is supported by the Mitchell Family Foundation.  ABR also thanks the NHETC at Rutgers University, and GM also thanks the KITP for hospitality during the final phases of the preparation of this paper (National Science Foundation under Grant No. NSF PHY11-25915).

\bigskip

\appendix

\section{Lie algebra and Lie group conventions}\label{app:Lie}

Our conventions for $\mathfrak{su}(2)$ are as follows.  A basis of anti-Hermitian generators in the fundamental representation is $T^i = - \frac{i}{2} \sigma^i$, where $\sigma^i$ are the Pauli matrices.  The simple co-root is $H = 2 T^3$ and the raising and lowering operators are $E_{\pm} = i (T^1 \pm i T^2)$.  An element $A \in \mathfrak{su}(2)$ can be expanded as $A = A^i T^i$ with the $A^i$ real, or equivalently $A = \AA H + A^+ (-iE_+) + (A^+)^\ast (-i E_-)$, where $\AA = \half A^3$ and $A^+ = \frac{1}{2} (A^1 - i A^2)$.  $\{ iH, E_{\pm} \}$ is a basis for the complexified algebra $\mathfrak{sl}(2,\mathbb{C}) = \mathfrak{su}(2)_{\mathbb{C}} \equiv \mathfrak{su}(2) \otimes \mathbb{C}$, satisfying $[ iH, E_{\pm}] = \pm 2 E_{\pm}$ and $[E_+, E_-] = iH$.  These relations are equivalent to the relations $[T^a,T^b] = \epsilon^{abc} T^c$ for $\mathfrak{su}(2)$.  We will typically instead use the basis $\{ H, -i E_{\pm} \}$ for $\mathfrak{su}(2)_{\mathbb{C}}$, as we did when expanding $A$ above.

In general we use $\mathfrak{g}_{\mathbb{C}}$ to denote a complexified simple Lie algebra and $\mathfrak{g}$ to denote its compact real form.  $\mathfrak{g}_{\mathbb{C}}$ has a root decomposition into a Cartan subalgebra $\mathfrak{t}_{\mathbb{C}}$ and one-dimensional root spaces spanned by elements $E_\alpha$:
\begin{equation}
\mathfrak{g}_{\mathbb{C}} = \mathfrak{t}_{\mathbb{C}} \oplus \bigoplus_{\alpha \in \Delta} (-iE_\alpha) \cdot \mathbb{C}~,
\end{equation}
$\alpha$ denotes a root, and the set of roots, $\Delta$, sits inside the dual space, $\mathfrak{t}^\ast$, of $\mathfrak{t}$.  $\mathfrak{g}_{\mathbb{C}}$ can be viewed as the representation space for the adjoint representation, which acts as $\ad(T^a)(T^b) = [T^a,T^b]$, and the roots are the weights of this representation: if $H \in \mathfrak{t}$ then $\ad(H)(E_\alpha) = -i \langle \alpha, H \rangle E_\alpha$, where $\langle~,~\rangle : \mathfrak{t}^\ast \otimes \mathfrak{t} \to \mathbb{R}$ is the canonical pairing between a vector space and its dual, and the factor of $i$ is present because we take $\ad(H)$ anti-Hermitian.

If $\alpha \in \Delta$ then $-\alpha \in \Delta$ is the only other linear multiple of $\alpha$ in $\Delta$.  For each root $\alpha$ there is a co-root $H_\alpha \in \mathfrak{t}$ such that $\{ iH_\alpha,  E_{\pm \alpha} \}$ form an $\mathfrak{sl}(2,\mathbb{C})$ subalgebra: $[i H_\alpha, E_{\pm \alpha}] = \pm 2 E_{\pm \alpha}$, $[E_\alpha, E_{-\alpha}] = H_\alpha$.  A choice of Killing form (positive definite, bi-invariant form) on $\mathfrak{g}$, denoted by $(~,~)$, determines one on $\mathfrak{t}$ by restriction.  Given an $H \in \mathfrak{t}$, we can use this form to define the dual element $H^\ast \in \mathfrak{t}^\ast$, such that $\langle H^\ast, H' \rangle = (H, H')$, $\forall H' \in \mathfrak{t}$.  The Killing form on $\mathfrak{t}$ then induces one on $\mathfrak{t}^\ast$, which we also denote by $(~,~)$.  Using the dual root, $\alpha^\ast \in \mathfrak{t}$, the co-root can be expressed as $H_\alpha = \frac{2 \alpha^\ast}{(\alpha,\alpha)}$.  It follows that $\langle \alpha, H_\beta \rangle = \frac{2 (\alpha,\beta)}{(\beta,\beta)}$.  This result is independent of the Killing form since the Killing form is unique up to rescaling (for simple $\mathfrak{g})$.

Given a polarization of $\mathfrak{t}^\ast$---a splitting into positive and negative half-spaces---we can define a basis $\{ \alpha_I ~|~ I = 1,\ldots, r \equiv \rnk{\mathfrak{g}} \}$ of simple roots, such that all remaining positive roots are positive linear combinations of these and no simple root can be written as a linear combination of other positive roots.  The corresponding simple co-roots, $H_I \equiv H_{\alpha_I}$, form a basis for $\mathfrak{t}$. Given such a basis we can define the Cartan matrix $C_{IJ}$ of the Lie algebra:
\begin{equation}\label{Cartanmatrix}
C_{IJ} := \langle \alpha_I, H_J \rangle = \frac{2 (\alpha_I, \alpha_J)}{(\alpha_J,\alpha_J)}~.
\end{equation}
It is a fundamental fact that all elements of $C_{IJ}$ are integral, taking values in the set $\{2,0,-1,-2,-3\}$.

The root lattice $\Lambda_{\rm rt} \subset \mathfrak{t}^\ast$ is the set of all integer linear combinations of the roots.  With a basis of simple roots given, we can write it as $\Lambda_{\rm rt} = \oplus_I \alpha_I \cdot \mathbb{Z}$.  Similarly, the co-root lattice, $\Lambda_{\rm cr} \subset \mathfrak{t}$, is defined as the set of all integer linear combinations of co-roots and, given a basis of simple co-roots, we have $\Lambda_{\rm cr} = \oplus_I H_I \cdot \mathbb{Z}$.  Two other important lattices associated with the Lie algebra are the integral dual lattices to these.  We define the weight lattice, $\Lambda_{\rm wt} \subset \mathfrak{t}^\ast$ as the integral dual of the co-root lattice, $\Lambda_{\rm wt} = \Lambda_{\rm cr}^\vee$, and the magnetic weight $\Lambda_{\rm mw} \subset \mathfrak{t}$ as the integral dual of the root lattice, $\Lambda_{\rm mw} = \Lambda_{\rm rt}^\vee$.  More explicitly,
\begin{align}
\Lambda_{\rm wt} :=&~ \left\{ \lambda \in \mathfrak{t}^\ast ~ \bigg| ~ \langle \lambda, H_\alpha \rangle \in \mathbb{Z}~,~ \forall H_\alpha \in \Lambda_{\rm cr} \right\}~, \cr
\Lambda_{\rm mw} :=&~ \left\{ h \in \mathfrak{t} ~ \bigg| ~ \langle \alpha, h \rangle \in \mathbb{Z}~,~ \forall \alpha \in \Lambda_{\rm rt} \right\}~.
\end{align}
Here we used that it is sufficient to require that the pairing with all co-roots or roots is integral.  Given a basis of simple co-roots and simple roots, we define the fundamental weights $\{ \lambda^I \}$ and the fundamental magnetic weights $\{ h^I \}$ by the conditions
\begin{equation}\label{fundweights}
\langle \lambda^I, H_J \rangle = {\delta^I}_J~, \qquad \langle \alpha_I, h^J \rangle = {\delta_I}^J~.
\end{equation}
These form bases for the weight and magnetic weight lattices, $\Lambda_{\rm wt} = \oplus_I \lambda^I \cdot \mathbb{Z}$, and $\Lambda_{\rm mw} = \oplus_I h^I \cdot \mathbb{Z}$.  From \eqref{Cartanmatrix} and \eqref{fundweights} we have
\begin{equation}
H_I = h^J C_{JI}~, \qquad \alpha_I = C_{IJ} \lambda^J~, \qquad \langle \lambda^I, h^J \rangle = C^{IJ}~,
\end{equation}
where $C^{IJ}$ are the components of the inverse Cartan matrix (which are in general fractional).  It follows from the first two relations, and the fact that the $C_{IJ}$ are integers, that $\Lambda_{\rm cr} \subset \Lambda_{\rm mw}$ and $\Lambda_{\rm rt} \subset \Lambda_{\rm wt}$.

A second characterization of the weight lattice $\Lambda_{\rm wt}$ is that it is the union of all sets of weights for all representations of $\mathfrak{g}$.  If $\rho : \mathfrak{g} \to \mathfrak{gl}(V_\rho)$ is a representation of the Lie algebra with representation space $V_\rho$, then associated to $\rho$ we have a set of weights $\Delta_\rho \subset \Lambda_{\rm wt}$.  $V_\rho$ can be decomposed into a direct sum of eigenspaces, $V_\rho = \oplus_{\lambda \in \Delta_\rho} V_\rho[\lambda]$, where if ${\bf e}_\lambda \in V_{\rho}[\lambda]$ and $H \in \mathfrak{t}$, then $\rho(H) {\bf e}_\lambda = -i \langle \lambda, H \rangle {\bf e}_\lambda$.  (Compare with the action of the adjoint representation.)

Thus far we have only discussed the algebra.  Lie's theorem guarantees that to each compact, real, simple Lie algebra $\mathfrak{g}$ there is a unique compact, connected, simply-connected Lie group $\tilde{G}$.  All compact simple Lie groups are of the form $G \cong \tilde{G}/\Gamma$ where $\Gamma \subset \mathcal{Z}(\tilde{G})$ is a subgroup of the center of the simply-connected cover.  We refer to the Lie group obtained by quotienting the simply-connected cover by its full center as the adjoint form of the group, $G_{\rm ad} \cong \tilde{G}/\mathcal{Z}(\tilde{G})$, because this is the group for which the adjoint representation is faithful.  In general $\tilde{G}/\ZZ \cong G$, where $\ZZ \subset \ZZ(\tilde{G})$ is a subgroup of the center of $\tilde{G}$, and $\ZZ \cong \pi_1(G)$.

Associated to the Lie group $G$ are two further lattices, the co-character lattice $\Lambda_{G} \subset \mathfrak{t}$, and its integral dual, the character lattice $\Lambda_{G}^\vee \subset \mathfrak{t}^\ast$.  We have
\begin{equation}
\Lambda_{G} := \left\{ H \in \mathfrak{t} ~ \bigg| ~ \exp(2\pi H) = 1_{G} \right\} \cong \Hom\left(U(1),T \right)~,
\end{equation}
where $1_G$ denotes the identity element in $G$ and $T \subset G$ is the Cartan torus.  This lattice precisely encodes allowed \tHooft charges since the transition function $\cg = \exp(\phi H)$ will be single-valued around the equator of the infinitesimal two-sphere when $H \in \Lambda_G$.  We have the inclusions $\Lambda_{\rm cr} \subset \Lambda_{G} \subset \Lambda_{\rm mw}$, with $\Lambda_{\rm mw}/ \Lambda_{G} \cong \ZZ(G)$ and $\Lambda_{G}/\Lambda_{\rm cr} \cong \pi_1(G)$.  We have that $\Lambda_G = \Lambda_{\rm cr}$ when $G = \tilde{G}$ and $\Lambda_G = \Lambda_{\rm mw}$ when $G = G_{\rm ad}$.

Dually,
\begin{equation}
\Lambda_{G}^\vee := \left\{ \mu \in \mathfrak{t} ~ \bigg| ~ \langle \mu, H \rangle \in \mathbb{Z}~,~ \forall H \in \Lambda_{G} \right\} \cong \Hom \left(T,U(1) \right)~,
\end{equation}
with the inclusions $\Lambda_{\rm rt} \subset \Lambda_{G}^\vee \subset \Lambda_{\rm wt}$.  We have $\Lambda_{G}^\vee = \Lambda_{\rm wt}$ when $G = \tilde{G}$ and $\Lambda_{G}^\vee =\Lambda_{\rm rt}$ when $G = G_{\rm ad}$.  $\Lambda_{G}^\vee$ can also be defined as the union of all sets of weights of representations $\rho$ of $G$.  These are representations $\rho$ of $\mathfrak{g}$ that lift to true (\ie\ not projective) representations of the group $G$.  The simplest example is $G = SO(3)$ with $\mathfrak{g} = \mathfrak{so}(3) \cong \mathfrak{su}(2)$.  The half-integer spin representations are representations of $\mathfrak{g}$ but not of $G$.  In this case $\Lambda_{\rm wt}/\Lambda_{G}^\vee = \mathbb{Z}/2\mathbb{Z} = \pi_1(SO(3))$.  We also have that $G = G_{\rm ad}$ in this case so, in particular, $\Lambda_G = \Lambda_{\rm mw}$.

\section{Defect boundary conditions with singular subleading behavior}\label{app:admissible}

In this appendix we consider the subleading behavior of a solution to the equations of motion following from the action \eqref{action}, satisfying the boundary conditions \eqref{trialbc} for small $r = |\vec{x} - \vec{x}_n|$.  We write, setting $X = \sigma \Phi$,
\begin{equation}\label{leading1}
A_i = A_{i}^{(0)} + \delta A_i~, \qquad X = X^{(0)} + \delta X~,
\end{equation}
with
\begin{equation} \label{leading2}
A^{(0)} = \frac{P}{2} (\pm 1 - \cos{\theta}) \ed\phi~, \qquad X^{(0)} = -\frac{P}{2r}~,
\end{equation}
and $\delta \hat{A}_a = (\delta A_i, \delta X) = O(r^{-1+\updelta'})$.  We assume that $A_0 = \delta A_0 = O(r^{-1+\updelta'})$ as well, and that $(A_\mu, X)$ together solve the second-order equations
\begin{equation}
D^\mu F_{\mu\nu} + [X, D_\nu X] = 0~, \qquad D^\mu D_\mu X = 0~.
\end{equation}
Additionally we work in a background gauge where $\pd^0 \delta A_0 + \hat{D}^{(0)a} \delta \hat{A}_a = 0$.

It follows that the leading order behavior of $\delta \hat{A}_a$ as $r \to 0$ is controlled by the linearized equations of motion which can be written in the form
\begin{equation}\label{lin2order}
\left( \delta_{ab} \hat{D}^{(0)c} \hat{D}_{c}^{(0)} + 2 \ad(\hat{F}_{ab}^{(0)}) \right) \delta A^b = 0~,
\end{equation}
together with the constraint $\hat{D}^{(0)a} \delta \hat{A}_a = 0$.  Now define the bi-spinor $\delta \hat{A} = \tau^a \delta \hat{A}_a$ and the operators $L,L^\dag$ as in \eqref{hatDirac}.  Then, starting with \eqref{positiveL}, we observe that
\begin{align}
& 0 = L^{(0) \dag} L^{(0)} \delta \hat{A} = - \left( \hat{D}^{(0) c} \hat{D}_{c}^{(0)} + \half \tau^{bc} \ad(\hat{F}_{bc}^{(0)}) \right) \tau^d \delta \hat{A}_d  \cr
\iff \quad & 0 = \tr \left\{ \bar{\tau}^a \left( \hat{D}^{(0) c} \hat{D}_{c}^{(0)} + \half \tau^{bc} \ad(\hat{F}_{bc}^{(0)}) \right) \tau^d \delta \hat{A}_d \right\}~, \quad \forall a~, \cr
\iff \quad& 0 = \left( 2 \delta^{ad} \hat{D}^{(0) c} \hat{D}_{c}^{(0)} + 4 \ad (\hat{F}^{(0)ad}) \right) \delta \hat{A}_d~,
\end{align}
where $\tr$ denotes a trace over the $\mathbb{C}^2$ spinor space, and we used the identity
\begin{equation}
\tr \{ \tau^{bc} \tau^{da} \} = -2 (\delta^{bd} \delta^{ca} - \delta^{ba} \delta^{cd} + \epsilon^{bcda} )~,
\end{equation}
together with the self-duality of the background: $\epsilon^{bcda} \hat{F}_{bc}^{(0)} = 2 \hat{F}^{(0)da}$.  Thus \eqref{lin2order} is equivalent to $L^{(0)\dag} L^{(0)} \delta \hat{A} = 0$, and $\ker \left(L^{(0)\dag} L^{(0)} \right) = \ker L^{(0)}$.  Hence \eqref{lin2order} holds if and only if $L^{(0)} \delta \hat{A} = 0$, and this is equivalent to the linearized Bogomolny equation, or self-duality equation in four-dimensional language, and the gauge constraint; see \eqref{boszm}.  

This is what we wanted to show.  If $\hat{A}_a$ is a general solution to the equations of motion with leading behavior given by \eqref{leading1}, \eqref{leading2}, then the first correction to this behavior is controlled by \eqref{lin2order} which implies the linearized Bogomolny equations.  Hence there exists a $\updelta'' >0$ such that
\begin{equation}
B_i - D_i X = O(r^{-2 + \updelta' + \updelta''})~,
\end{equation}
and therefore we in fact have $\delta S = O(r^{-1 + 2\updelta' + \updelta''})$ in \eqref{Svar}, implying $\updelta' = \half$ is consistent with the variational principle.

\section{Diagonalizing the Dirac operator in a Cartan-valued background}\label{app:Dirac}

In this appendix we consider the Dirac operator $(\hat{\slashed{D}})_\rho$, \eqref{hatDirac}, in a representation $\rho : \mathfrak{g} \to \mathfrak{gl}(V_\rho)$ that lifts to a true representation of $G$, coupled to the background
\begin{equation}
X = X_{\infty} - \frac{P}{2r}~, \qquad A = \frac{P}{2} ( \pm 1 - \cos{\theta}) \ed \phi~,
\end{equation}
where $X_\infty, P \in \mathfrak{t}$.  This is a purely Cartan-valued solution to the Bogomolny equation with a single \tHooft defect of charge $P$ inserted at $\vec{x}_0 = 0$.  Thus
\begin{equation}\label{CartanDirac}
(\hat{\slashed{D}})_\rho = \vec{\Gamma} \cdot \vec{\pd} + \Gamma^\phi \otimes \frac{\rho(P)}{2} (\pm 1 - \cos{\theta}) + \Gamma^4 \otimes \left( \rho(X_\infty) - \frac{\rho(P)}{2 r} \right)~,
\end{equation}
and we are interested in studying the eigenvalue problem
\begin{equation}\label{DiracCartan}
(\hat{\slashed{D}})_\rho \Psi = - i E \Psi~.
\end{equation}

We introduce a basis $\{ {\bf e}_{i_\mu} \}$ of $V_{\rho}$ associated with the weight decomposition $V_\rho = \oplus_{\mu} V_{\rho}[\mu]$, and we expand $\Psi = \sum_{\mu} \sum_{i_\mu} \Psi^{(i_\mu)} {\bf e}_{i_\mu}$.  Here $\mu$ runs over the weights of the representation, $\mu \in \Delta_{\rho}$, and for each weight, $i_\mu = 1,\ldots, n_\rho(\mu) = \dim{V_\rho[\mu]}$, takes into account the degeneracy of that weight.  The basis is such that $\rho(P) {\bf e}_{i_\mu} = -i \langle \mu,P\rangle {\bf e}_{i_\mu}$ and similarly for $\rho(X_\infty)$, and we have that the $\langle \mu,P \rangle$ are integers.  We also decompose $\Psi$ into local sections so that $\Psi_{\pm}^{(i_\mu)}$ is a $\mathbb{C}^4$-valued function on $\mathbb{R}_+ \times S_{\pm}$, where $S_{\pm}$ are the northern and southern patches covering $S^2$.  The $\Psi_{\pm}^{(i_\mu)}$ are related by the transition functions $e^{-i \langle \mu, P \rangle \phi}$ on the overlaps.  We use the shorthand $\langle \mu,P \rangle \equiv p_\mu \in \mathbb{Z}$, $\langle \mu, X_\infty \rangle = x_\mu \in \mathbb{R}$.  The Dirac equation \eqref{DiracCartan} is then equivalent to the following $\dim{V_\rho}$ Dirac equations:
\begin{equation}
\left[ \vec{\Gamma} \cdot \vec{\pd} -i \Gamma^\phi  \frac{p_\mu}{2} (\epsilon 1 - \cos{\theta}) -i \Gamma^4  \left( x_\mu - \frac{p_\mu}{2 r} \right) \right] \Psi_{\epsilon}^{(i_\mu)} = -i E \Psi_{\epsilon}^{(i_\mu)}~.
\end{equation}
Here we have introduced $\epsilon = \pm $ to keep track of which patch we are in.  Writing $\Psi^{(i_\mu)} = (\psi^{(i_\mu)},\chi^{(i_\mu)})^T$ where $\psi^{(i_\mu)},\chi^{(i_\mu)}$ are $\mathbb{C}^2$-valued, this equation is equivalent to the coupled equations
\begin{align}\label{Weyl1}
\left[ \vec{\sigma} \cdot \vec{\pd} -i \sigma^\phi \ \frac{p_\mu}{2} (\epsilon 1 - \cos{\theta}) - \mathbbm{1}_2 \left( x_\mu - \frac{p_\mu}{2 r} \right) \right] \chi_{\epsilon}^{(i_\mu)} = -i E \psi_{\epsilon}^{(i_\mu)}~, \cr
\left[ \vec{\sigma} \cdot \vec{\pd} -i \sigma^\phi  \frac{p_\mu}{2} (\epsilon 1 - \cos{\theta}) + \mathbbm{1}_2  \left( x_\mu - \frac{p_\mu}{2 r} \right) \right] \psi_{\epsilon}^{(i_\mu)} = -i E \chi_{\epsilon}^{(i_\mu)}~.
\end{align}

Following \cite{HarishChandra:1948zz}, we express the derivative operator in spherical coordinates using $\vec{\sigma} \cdot \vec{\pd} = \sigma^r \pd_r + \sigma^\theta \pd_\theta + \sigma^\phi \pd_\phi$, where
\begin{equation}
\sigma^r = \sigma^{\hat{r}}~, \qquad \sigma^{\theta} = \frac{1}{r} \sigma^{\hat{\theta}}~, \qquad \sigma^\phi = \frac{1}{r\sin{\theta}} \sigma^{\hat{\phi}}~,
\end{equation}
and the $\sigma^{\hat{r},\hat{\theta},\hat{\phi}}$ are related to the Pauli matrices by the $SU(2)$ element sending the $\{ \hat{z}, \hat{x}, \hat{y} \}$ frame to the $\{ \hat{r}, \hat{\theta}, \hat{\phi} \}$ frame:
\begin{align}\label{framerotation}
\sigma^{\hat{r}} =&~ \sigma^3 \cos{\theta}  + ( \sigma^1 \cos{\phi} + \sigma^2 \sin{\phi} )  \sin{\theta} =  U(\theta,\phi) \sigma^3 U(\theta,\phi)^{-1} ~, \cr
\sigma^{\hat{\theta}} =&~ - \sigma^3 \sin{\theta} + ( \sigma^1 \cos{\phi} + \sigma^2 \sin{\phi} )  \cos{\theta} =  U(\theta,\phi) \sigma^1 U(\theta,\phi)^{-1} ~, \cr
\sigma^{\hat{\phi}} =&~ - \sigma^1 \sin{\phi} + \sigma^2 \cos{\phi} = U(\theta,\phi) \sigma^2 U(\theta,\phi)^{-1}~,
\end{align}
with $U(\theta,\phi)^{-1} = e^{i \theta \sigma^2/2} e^{ i \phi \sigma^3/2}$.  With the aid of
\begin{align}
U(\theta,\phi)^{-1} \pd_\theta =&~ \left( \pd_\theta - \frac{i}{2} \sigma^2 \right) U(\theta,\phi)^{-1}~, \cr
U(\theta,\phi)^{-1} \pd_\phi =&~ \left( \pd_\phi - \frac{i}{2} (\sigma^3 \cos{\theta} - \sigma^1 \sin{\theta}) \right) U(\theta,\phi)^{-1}~,
\end{align}
we can write the operators on the left in \eqref{Weyl1} as
\begin{align}
& \left[ \vec{\sigma} \cdot \vec{\pd} -i \sigma^\phi  \frac{p_\mu}{2} (\epsilon 1 - \cos{\theta}) \pm   \left( x_\mu - \frac{p_\mu}{2 r} \right) \right] = \cr
& \quad = U(\theta,\phi) \Biggl\{ \sigma^3 \left(\pd_r + \frac{1}{r} \right) \pm \left(  x_\mu - \frac{p_\mu}{2 r} \right) + \cr
& \qquad \qquad \qquad \quad + \frac{\sigma^1}{r} \left[ \pd_\theta + \frac{i \sigma^3}{\sin{\theta}} \left( \pd_\phi -  \frac{i \epsilon p_\mu}{2} + \frac{i}{2} ( p_\mu - \sigma^3) \cos{\theta} \right) \right] \Bigg\} U(\theta,\phi)^{-1}~, \qquad \quad
\end{align}
where the upper (lower) sign corresponds to the operator acting on $\psi_{\epsilon}^{(i_\mu)}$ ($\chi_{\epsilon}^{(i_\mu)}$).

This suggests the ansatz
\begin{align}
\psi_{\epsilon}^{(i_\mu)}(\vec{x}) =&~ e^{i \epsilon p_\mu \phi/2} U(\theta,\phi) \tilde{\psi}^{(i_\mu)}(\vec{x})~, \qquad \textrm{with} \quad \tilde{\psi}^{(i_\mu)}(\vec{x}) = \tilde{\psi}^{(i_\mu)}(r,\theta) e^{-im\phi}~,
\end{align}
and similarly for $\chi$.  Note these have the correct relation on the overlap of the northern and southern patches.  Plugging into \eqref{Weyl1} yields the equations
\begin{align}\label{Weyl2}
& \left[ \sigma^3 \left(\pd_r + \frac{1}{r} \right) + \left(  x_\mu - \frac{p_\mu}{2 r} \right) + \frac{1}{r} K \right] \tilde{\psi}^{(i_\mu)} = -i E \tilde{\chi}^{(i_\mu)}~, \cr
& \left[ \sigma^3 \left(\pd_r + \frac{1}{r} \right) - \left(  x_\mu - \frac{p_\mu}{2 r} \right) + \frac{1}{r} K \right] \tilde{\chi}^{(i_\mu)} = -i E \tilde{\psi}^{(i_\mu)}~,
\end{align}
where
\begin{equation}
K := \sigma^1 \left[ \pd_\theta + \frac{\sigma^3}{\sin{\theta}} \left( m - \half ( p_\mu - \sigma^3) \cos{\theta} \right) \right]~.
\end{equation}
One can compute
\begin{align}
K^2 =&~ \pd_{\theta}^2 + \cot{\theta} \pd_{\theta} - \frac{1}{\sin^2{\theta}} \left[ m^2 - m (p_\mu - \sigma^3) \cos{\theta} + \frac{1}{4} (p_\mu - \sigma^3)^2 \right] + \frac{1}{4} (p_{\mu}^2 - 1)~,
\end{align}
and observe that it commutes with the operators in \eqref{Weyl2}.  Therefore we can take $\tilde{\psi},\tilde{\chi}$ to be simultaneous eigenfunctions of $K^2$.  The set of eigenfunctions of $K^2$ are well-known; they may be expressed in terms of Wigner (small) $d$ functions,
\begin{equation}\label{K2eigen}
\tilde{\psi}_{j,m}^{(i_\mu)} = \left( \begingroup \renewcommand{\arraystretch}{1.2} \begin{array}{c} \psi_{1}^{(i_\mu)}(r) d^{j}_{m,\half(p_\mu -1)}(\theta) \\ \psi_{2}^{(i_\mu)}(r) d^{j}_{m,\half(p_\mu+1)}(\theta) \end{array} \endgroup \right)~, \qquad   \tilde{\chi}_{j,m}^{(i_\mu)} = \left( \begingroup \renewcommand{\arraystretch}{1.2} \begin{array}{c} \chi_{1}^{(i_\mu)}(r) d^{j}_{m,\half(p_\mu -1)}(\theta) \\ \chi_{2}^{(i_\mu)}(r) d^{j}_{m,\half(p_\mu+1)}(\theta) \end{array} \endgroup \right)~.
\end{equation}
The associated eigenvalues of $K^2$ are $-k^2 \equiv - j(j+1) + \frac{1}{4}(p_{\mu}^2 - 1)$.

In order to use this decomposition in \eqref{Weyl2} we need to determine the action of $K$ on these eigenfunctions.  We have
\begin{equation}
K = \left( \begin{array}{c c} 0 & L_{m,\half(p_\mu+1)}^- \\ L_{m,\half (p_\mu-1)}^+ & 0 \end{array} \right)~,
\end{equation}
where
\begin{equation}
L_{m,m'}^{\pm} := \pd_\theta \pm \frac{1}{\sin{\theta}} \left(m - m' \cos{\theta} \right)~.
\end{equation}
The action of these operators on the Wigner $d$ functions is
\begin{equation}
L_{m,m'}^{\pm} d_{m,m'}^j = \pm \sqrt{j(j+1) -m' (m' \pm 1)} \ d_{m,m' \pm 1}^j~.
\end{equation}
We then have
\begin{equation}
L_{m,\half (p_\mu - 1)}^+ d^{j}_{m,\half (p_\mu -1)} = k d^{j}_{m,\half (p_\mu+1)}~, \quad L_{m,\half (p_\mu + 1)}^- d^{j}_{m,\half (p_\mu +1)} = -k d^{j}_{m,\half (p_\mu-1)}~,
\end{equation}
where
\begin{equation}
k = \sqrt{ j(j+1) - \frac{1}{4}(p_\mu + 1)(p_\mu -1) } = \sqrt{ \left( j + \half \right)^2 - \frac{p_{\mu}^2}{4} }~.
\end{equation}
Regularity of $d^{j}_{m,m'}$ requires the usual quantization conditions: $j \in \{ 0, \half,1,\ldots \}$ and $m,m'$ can take values from $-j$ to $j$ in integer steps.  These conditions together with the form of the eigenfunctions \eqref{K2eigen} ensure that $k \geq 0$.

We must distinguish between the case $k > 0$ and $k = 0$.  $k > 0$ is the generic case and requires $j \geq \half(|p_\mu| + 1)$.  In this case
\begin{equation}\label{generalK}
K \tilde{\psi}_{j,m}^{(i_\mu)} = \left( \begingroup \renewcommand{\arraystretch}{1.2} \begin{array}{c} -k \psi_{2}^{(i_\mu)}(r) d^{j}_{m,\half(p_\mu -1)}(\theta) \\ k \psi_{1}^{(i_\mu)}(r) d^{j}_{m,\half(p_\mu +1)}(\theta) \end{array}  \endgroup \right)~, \qquad \textrm{for} \quad j \geq \half (|p_\mu|+1)~,
\end{equation}
$k=0$ occurs when $j = j_\mu := \half (|p_\mu| - 1)$. In this situation one of the two components of $\tilde{\psi}$, $\tilde{\chi}$ is zero.  There are two cases depending on the sign of $p_\mu$:
\begin{align}\label{specialK}
& K \tilde{\psi}_{j_\mu,m}^{(i_\mu)} = 0~, \qquad \tilde{\psi}_{j_\mu,m}^{(i_\mu)} = \left( \begin{array}{c} \psi_{1}^{(i_\mu)}(r) d^{j_\mu}_{m,j_\mu}(\theta) \\ 0 \end{array} \right)~, \qquad(p_\mu > 0)~, \cr
& K \tilde{\psi}_{j_\mu,m}^{(i_\mu)} = 0~, \qquad \tilde{\psi}_{j_\mu,m}^{(i_\mu)} = \left( \begin{array}{c} 0 \\ \psi_{2}^{(i_\mu)}(r) d^{j_\mu}_{m,-j_\mu}(\theta) \end{array} \right)~, \qquad  (p_\mu < 0)~.
\end{align}
Analogous remarks apply for $K$ acting on $\tilde{\chi}$.  Note that $p_\mu$ is always an integer.  If $p_\mu = 0$ then the special cases \eqref{specialK} do not exist.

Now we plug the ansatz \eqref{K2eigen} into the equations \eqref{Weyl2}.  With the aid of \eqref{generalK} and \eqref{specialK} and the analogous formulae for $\tilde{\chi}$, we find that the $\theta$ dependence cancels out and we are left with the following radial equations:
\begin{align}\label{Weyl3}
& \left[ \sigma^3 \left( \pd_r + \frac{1}{r} \right) + \left( x_\mu - \frac{p_\mu}{2 r} \right) - i \sigma^2 \frac{k}{r} \right] \hat{\psi}^{(i_\mu)} =  -i E \hat{\chi}^{(i_\mu)}~, \cr
& \left[ \sigma^3 \left( \pd_r + \frac{1}{r} \right) - \left( x_\mu - \frac{p_\mu}{2 r} \right) - i \sigma^2 \frac{k}{r} \right] \hat{\chi}^{(i_\mu)} =  -i E \hat{\psi}^{(i_\mu)}~,
\end{align}
where
\begin{equation}
\hat{\psi}^{(i_\mu)} = \left( \begin{array}{c} \psi_{1}^{(i_\mu)}(r) \\ \psi_{2}^{(i_\mu)}(r) \end{array} \right)~, \qquad  \hat{\chi}^{(i_\mu)} = \left( \begin{array}{c} \chi_{1}^{(i_\mu)}(r) \\ \chi_{2}^{(i_\mu)}(r) \end{array} \right)~.
\end{equation}
These equations hold in the special case $k=0$ as well, provided we remember to set $(\psi_{2}^{(i_\mu)},\chi_{2}^{(i_\mu)}) = 0$ or $(\psi_{1}^{(i_\mu)},\chi_{1}^{(i_\mu)}) = 0$ for $p_\mu > 0$, $p_\mu < 0$ respectively.

The equations \eqref{Weyl3} can be solved in full generality; the solutions involve Coulomb wavefunctions.  However we only require the solutions in two limiting cases and it is more practical to split the discussion accordingly at this point.

\subsection{The spectral measure for $(i \hat{\slashed{D}})_{\rho}^{(0)}$}\label{app:Dirac1}

In this subsection we set $x_\mu = 0$ in \eqref{Weyl3} and determine the spectral representation of the resulting operator.  This is the analysis relevant for subsection \ref{sec:infS2index}, as \eqref{CartanDirac} with $X_\infty \to 0$ becomes $(\hat{\slashed{D}})_{\rho}^{(0)}$, \eqref{DiracDirac}.  Let $\hat{\psi}^{(i_\mu)} = \frac{1}{r} f$ with $f = (f_1,f_2)^T$ and let $\hat{\chi}^{(i_\mu)} = \frac{1}{r} g$ with $g = (g_1,g_2)^T$.  Then we have
\begin{equation}\label{fg1storder}
\left[ \sigma^3 \pd_r + \frac{1}{r} C_- \right] f = -i E g~, \qquad \left[ \sigma^3 \pd_r + \frac{1}{r} C_+ \right] g = - i E f~,
\end{equation}
where
\begin{equation}
C_{\pm} = \left( \begin{array}{c c} \pm p_\mu/2 & -k \\ k & \pm p_\mu/2 \end{array} \right)~.
\end{equation}

From \eqref{fg1storder} we derive the second order equations
\begin{align}\label{fg2ndorder}
& \left[ \pd_{r}^2 + E^2 - \frac{ (j + \half)^2}{r^2} - \frac{1}{r^2} \sigma^3 C_- \right] f = 0~, \qquad \left[ \pd_{r}^2 + E^2 - \frac{ (j + \half)^2}{r^2} - \frac{1}{r^2} \sigma^3 C_+ \right] g = 0~.
\end{align}
The matrices $\sigma^3 C_{\pm}$ need to be diagonalized when $k \neq 0$.  They have the same eigenvalues, $\lambda_{\pm} = \pm (j + \half)$, and we have
\begin{equation}
O_{f}^T (\sigma^3 C_-) O_{f} = \left( \begin{array}{c c} j+ \half & 0 \\ 0 & - (j + \half) \end{array} \right) = O_{g}^T (\sigma^3 C_+) O_g~,
\end{equation}
where $O_f,O_g$ are the orthogonal matrices
\begin{equation}\label{OfOg}
O_f = \frac{1}{\sqrt{2j+1}} \left( \begin{array}{c c} a_- & a_+ \\ - a_+ & a_- \end{array}\right)~, \qquad  O_g = \frac{1}{\sqrt{2j+1}} \left( \begin{array}{c c} a_+ & a_- \\ - a_- & a_+ \end{array} \right)~,
\end{equation}
with $a_{\pm} \equiv \sqrt{(j + \half) \pm \frac{p_\mu}{2}}$.  It follows that the solutions to \eqref{fg2ndorder} are of the form
\begin{equation}\label{fgsol1}
\left( \begin{array}{c} f_1 \\ f_2 \end{array} \right) = O_f \left( \begingroup \renewcommand{\arraystretch}{1.2} \begin{array}{c} c_{1}^f \sqrt{r} Z_{j+1}(|E|r) \\ c_{2}^f \sqrt{r} Z_{j}(|E| r) \end{array} \endgroup \right)~, \qquad \left( \begin{array}{c} g_1 \\ g_2 \end{array} \right) = O_g \left( \begingroup \renewcommand{\arraystretch}{1.2} \begin{array}{c} c_{1}^g \sqrt{r} Z_{j+1}(|E|r) \\ c_{2}^g \sqrt{r} Z_{j}(|E| r) \end{array} \endgroup \right)~,
\end{equation}
where $Z_j$ is a Bessel function with index $j$.

The coefficients $c_{1,2}^f, c_{1,2}^g$ must be determined by plugging these solutions back into the first order equations \eqref{fg1storder}.  This is most easily done by first expressing the equations in terms of $O_{f}^T f$ and $O_{g}^T g$.  Using $O_{g}^T \sigma^3 O_f = O_{f}^T \sigma^3 O_g = \sigma^1$ and $O_{g}^T C_- O_f = O_{f}^T C_+ O_g = -i a \sigma^2$, we find that \eqref{fg1storder} is equivalent to
\begin{align}
& \left( \begin{array}{c c} 0 &  \pd_r - \frac{j+\half}{r} \\ \pd_r + \frac{j+\half}{r} & 0 \end{array}\right) (O_{f}^T f) = -i E (O_{g}^T g)~, \cr
&  \left( \begin{array}{c c} 0 &  \pd_r - \frac{j+\half}{r} \\ \pd_r + \frac{j+\half}{r} & 0 \end{array}\right) (O_{g}^T g) = -i E  (O_{f}^T f)~.
\end{align}
Plugging in \eqref{fgsol1} and making use of the Bessel function identities $Z_{\nu}' - \frac{\nu}{r} Z_\nu = - Z_{\nu+1}$, $Z_{\nu}' + \frac{\nu}{r} Z_\nu = Z_{\nu-1}$, we find that these equations are satisfied if and only if
\begin{equation}
c_{1}^g = -i \sgn(E) c_{2}^f~, \qquad c_{2}^g = i \sgn(E) c_{1}^f~.
\end{equation}
This, together with \eqref{fgsol1} and \eqref{OfOg}, yield the following solutions for $f,g$:
\begin{align}\label{fgsol2}
\left(\begin{array}{c} f_1 \\ f_2 \end{array} \right) =&~ c_{1} \left( \begin{array}{c} a_-  \\ - a_+  \end{array}\right) \sqrt{r} Z_{j+1}(|E| r) + c_{2} \left( \begin{array}{c} a_+  \\  a_-  \end{array}\right) \sqrt{r} Z_{j}(|E| r)~, \cr
\left(\begin{array}{c} g_1 \\ g_2 \end{array}\right) =&~ -i \sgn(E) c_{2} \left( \begin{array}{c} a_+ \\ - a_- \end{array} \right) \sqrt{r} Z_{j+1}(|E| r) + i \sgn(E) c_{1}\left( \begin{array}{c} a_- \\ a_+ \end{array}\right) \sqrt{r} Z_j(|E| r)~, \qquad \quad
\end{align}
where we set $c_{1,2}^f/\sqrt{2j+1} = c_{1,2}$.

This is the generic solution to \eqref{fg1storder} when $k > 0$.  The above analysis can also be applied to the case $k=0$.  Recall that in this case, $j = j_\mu = \half (|p_\mu| - 1)$.  There are two cases to consider, $p_\mu > 0$ or $p_\mu < 0$.  When $p_\mu > 0$ we set $f_2 = g_2 = 0$, $a_- = 0$, and we can read off the solution for $f_1,g_1$ from \eqref{fgsol2}:
\begin{equation}
\left( \begin{array}{c} f_1 \\ f_2 \end{array} \right) =  \left( \begin{array}{c} c_+ \\ 0 \end{array} \right) \sqrt{r} Z_{j_\mu}(|E| r)~, \qquad \left( \begin{array}{c} g_1 \\ g_2 \end{array} \right) = -i \sgn(E) \left( \begin{array}{c} c_+ \\ 0 \end{array} \right) \sqrt{r} Z_{j_\mu+1}(|E| r)~.
\end{equation}
Similarly when $p_\mu < 0$ we set $f_1 = g_1 = 0$, $a_+ = 0$, and we get
\begin{equation}
\left( \begin{array}{c} f_1 \\ f_2 \end{array} \right) =  \left( \begin{array}{c} 0 \\ c_- \end{array} \right) \sqrt{r} Z_{j_\mu}(|E| r)~, \qquad \left( \begin{array}{c} g_1 \\ g_2 \end{array} \right) = i \sgn(E) \left( \begin{array}{c} 0 \\ c_- \end{array} \right) \sqrt{r} Z_{j_\mu+1}(|E| r)~.
\end{equation}

Let us collect the pieces and summarize.  We are solving $(i \hat{\slashed{D}})_{\rho}^{(0)} \Psi = E \Psi$, with $(\hat{\slashed{D}})_{\rho}^{(0)}$ as in \eqref{DiracDirac}, and we have expanded $\Psi_\epsilon = \sum_{\mu} \sum_{i_\mu} \Psi_{\epsilon}^{(i_\mu)} {\bf e}_{i_\mu}$.  Here $\epsilon = \pm$ refers to the northern or southern patch of $S^2$, and $\{ {\bf e}_{i_\mu} \}$ is a basis for $V_{\rho}$ associated with a weight decomposition.  We have found a full set of solutions for $\Psi^{(i_\mu)}$ given as follows.  First we have the two families of solutions
\begin{equation}\label{Psifam1}
(\Psi_{\epsilon}^{(i_\mu)})_{j,m,1} = e^{i\epsilon p_\mu \phi/2} {\bf U}(\theta,\phi) \tilde{\Psi}^{(i_\mu)}_{j,m,1}~, \qquad (\Psi_{\epsilon}^{(i_\mu)})_{j,m,2} = e^{i\epsilon p_\mu \phi/2} {\bf U}(\theta,\phi) \tilde{\Psi}^{(i_\mu)}_{j,m,2}~,
\end{equation}
where ${\bf U}(\theta,\phi) = \mathbbm{1}_2 \otimes U(\theta,\phi) = \mathbbm{1}_2 \otimes e^{-i \phi \sigma^3/2} e^{-i \theta \sigma^2/2}$ and
\begin{align}
\tilde{\Psi}_{j,m,1}^{(i_\mu)}(E;\vec{x}) =&~ \frac{c_1 e^{-i m \phi}}{\sqrt{r}} \left( \begingroup \renewcommand{\arraystretch}{1.2}  \begin{array}{c} a_- Z_{j+1}(|E| r)  d^{j}_{m,\half(p_\mu - 1)}(\theta) \\ -a_+ Z_{j+1}(|E| r)  d^{j}_{m,\half(p_\mu + 1)}(\theta) \\ i \sgn(E) a_- Z_{j}(|E| r)  d^{j}_{m,\half(p_\mu - 1)}(\theta) \\ i \sgn(E) a_+ Z_{j}(|E| r)  d^{j}_{m,\half(p_\mu +1)}(\theta) \end{array} \endgroup \right)~, \cr
\tilde{\Psi}_{j,m,2}^{(i_\mu)}(E;\vec{x}) =&~ \frac{c_2 e^{-i m \phi}}{\sqrt{r}} \left( \begingroup \renewcommand{\arraystretch}{1.2}  \begin{array}{c} a_+ Z_{j}(|E| r)  d^{j}_{m,\half(p_\mu - 1)}(\theta) \\ a_- Z_{j}(|E| r)  d^{j}_{m,\half(p_\mu + 1)}(\theta) \\ -i \sgn(E) a_+ Z_{j+1}(|E| r)  d^{j}_{m,\half(p_\mu - 1)}(\theta) \\ i \sgn(E) a_- Z_{j+1}(|E| r)  d^{j}_{m,\half(p_\mu +1)}(\theta) \end{array} \endgroup \right)~,
\end{align}
with $a_{\pm} \equiv \sqrt{j + \half \pm \frac{p_\mu}{2}}$.  We recall that $p_\mu \equiv -i \langle \mu, P \rangle \in \mathbb{Z}$.  For these solutions the allowed values of $m$ run from $-j$ to $j$ in integer steps and the allowed values of $j$ start at $j = \half (|p_\mu| + 1) = j_\mu + 1$ and increase in integer steps.

In addition to these two families, there is one more set of solutions with a fixed $j = \half (|p_\mu| - 1) = j_\mu$.  This family only exists if $p_\mu \neq 0$ and its form depends on the sign of $p_\mu$.  We will denote the two possibilities with a $+$ or $-$:
\begin{equation}\label{Psifam2ap}
(\Psi_{\epsilon}^{(i_\mu)})_{m,+} = e^{i\epsilon p_\mu \phi/2} {\bf U}(\theta,\phi) \tilde{\Psi}^{(i_\mu)}_{m,+}~, \qquad (\Psi_{\epsilon}^{(i_\mu)})_{m,-} = e^{i\epsilon p_\mu \phi/2} {\bf U}(\theta,\phi) \tilde{\Psi}^{(i_\mu)}_{m,-}~,
\end{equation}
with
\begin{align}\label{Psispecial}
\tilde{\Psi}_{m,+}^{(i_\mu)}(E;\vec{x}) =&~ \frac{c_+ e^{-i m \phi}}{\sqrt{r}} \left(  \begin{array}{c} Z_{j_\mu}(|E| r)  d^{j_\mu}_{m,j_\mu}(\theta) \\ 0 \\ -i \sgn(E) Z_{j_\mu+1}(|E| r)  d^{j_\mu}_{m,j_\mu}(\theta) \\   0 \end{array}\right)~, \qquad (p_\mu > 0)~, \cr
\tilde{\Psi}_{m,-}^{(i_\mu)}(E;\vec{x}) =&~ \frac{c_- e^{-i m \phi}}{\sqrt{r}} \left( \begin{array}{c} 0 \\ Z_{j_\mu}(|E| r)  d^{j_\mu}_{m,-j_\mu}(\theta) \\ 0 \\ i \sgn(E) Z_{j_\mu+1}(|E| r)  d^{j_\mu}_{m,-j_\mu}(\theta) \end{array}\right)~, \qquad ( p_\mu < 0)~.
\end{align}
In this case $m$ runs over $|p_\mu|$ values from $-j_\mu$ to $j_\mu$.

Equations \eqref{Psifam1} through \eqref{Psispecial}, where we take a basis of Bessel functions, constitute a basis of general $(C^\infty)$ solutions to the Dirac equation $(\hat{\slashed{D}})_{\rho}^{(0)} \Psi = - i E \Psi$ on $\UU = \mathbb{R}^3 \setminus \{ 0\}$.  We are interested in determining the spectrum of $(i \hat{\slashed{D}})_{\rho}^{(0)}$ as a self-adjoint operator acting on $\LL^2[\UU,\mathbb{C}^4 \otimes V_\rho]$, where the innerproduct is $(\Psi_1,\Psi_2) = \int \ed^3 x \tr_{\mathbb{C}^4 \otimes V_\rho} ( \Psi_{1}^\ast \Psi_2)$.  It is easy to see that none of the solutions listed above is $\LL^2$.  However, the $Z_\nu = J_\nu$ solutions are plane-wave normalizable; \ie\ they represent scattering states.  Making use of the integrals
\begin{align}
&  \int_{S^2} \vol_{S^2} e^{i (m_1 - m_2) \phi} d^{j_1}_{m_1,m'}(\theta) d^{j_2}_{m_2,m'}(\theta) = \frac{4\pi}{2 j_1 +1} \delta^{j_1 j_2} \delta_{m_1 m_2}~, \cr
& \int_{0}^{\infty} dr r J_{\nu}(\mu_1 r) J_{\nu}(\mu_2 r) = \frac{1}{\mu_1} \delta(\mu_1 - \mu_2)~,
\end{align}
one can check that, when $Z_\nu = J_\nu$,
\begin{align}\label{pworth}
& \int_{\UU} \ed^3 x \overline{\Psi_{j_1,m_1,(1,2)}^{(i_\mu)}(E_1;\vec{x})} \Psi_{j_2,m_2,(1,2)}^{(i_\mu)}(E_2;\vec{x}) =  \delta(E_1-E_2)  \delta^{j_1 j_2} \delta_{m_1 m_2}~, \cr
& \int_{\UU} \ed^3 x \overline{\Psi_{m_1,\pm}^{(i_\mu)}(E_1;\vec{x})} \Psi_{m_2,\pm}^{(i_\mu)}(E_2;\vec{x}) = \delta(E_1 - E_2) \delta_{m_1 m_2}~,
\end{align}
provided one chooses the normalization constants
\begin{equation}
c_{1,2} = \frac{\sqrt{|E|}}{2\sqrt{2\pi}}~, \qquad c_{\pm} = \frac{\sqrt{|E| |p_\mu|}}{2\sqrt{2\pi}}~.
\end{equation}
Meanwhile the innerproduct between wavefunctions from different families, \eqref{Psifam1}, \eqref{Psifam2ap}, vanishes.

It follows that the spectrum of $(i \hat{\slashed{D}})_{\rho}^{(0)}$ is purely continuous, consisting of the whole real line.  We can use the above $Z_\nu = J_\nu$ solutions to construct the (integral kernel for) the spectral measure associated with $(i \hat{\slashed{D}})_{\rho}^{(0)}$.  Denoting this spectral measure as $\ed (i \hat{\slashed{D}})_{\rho}^{(0)}$, we have
\begin{align}\label{spectralmeasure}
\ed (i \hat{\slashed{D}})_{\rho}^{(0)}(\vec{x},\vec{y}) =&~ \sum_{\mu \in \Delta_{\rho}} \sum_{i_\mu =1}^{n_\rho(\mu)} {\bf e}_{i_\mu} \overline{{\bf e}}_{i_\mu} \otimes \displaystyle\biggl\{ \bigoplus_{j = j_\mu+1}^{\infty} \ \bigoplus_{m = -j}^j \ \bigoplus_{s=1}^2 \Psi^{(i_\mu)}_{j,m,s}(E;\vec{x}) \overline{\Psi^{(i_\mu)}_{j,m,s}(E;\vec{y})} \ \oplus \cr
& \qquad \qquad  \bigoplus_{m= -j_\mu}^{j_\mu} \Psi^{(i_\mu)}_{m,\sgn(p_\mu)}(E;\vec{x}) \overline{\Psi^{(i_\mu)}_{m,\sgn(p_\mu)}(E;\vec{y})} \displaystyle\biggr\} \ed E~. \quad \raisetag{26pt}
\end{align}
The last summand only exists when $p_\mu \neq 0$, so that $j_\mu \geq 0$.  The spectral measure can be used to evaluate functions of the operator $(i \hat{\slashed{D}})_{\rho}^{(0)}$ via
\begin{equation}
F\left[ (i \hat{\slashed{D}})_{\rho}^{(0)} \right] = \int \ed (i \hat{\slashed{D}})_{\rho}^{(0)} F[E]~.
\end{equation}
A similar relation holds at the level of integral kernels, $F[ (i \hat{\slashed{D}})_{\rho}^{(0)} ](\vec{x},\vec{y})$, using \eqref{spectralmeasure}.

As a check let us consider the resolution of the identity operator, $F[ (i \hat{\slashed{D}})_{\rho}^{(0)} ] = \mathbbm{1}$ on $\LL^2[\UU,\mathbb{C}^4 \otimes V_\rho]$, whose kernel is $\mathbbm{1}_{ \mathbb{C}^4 \otimes V_\rho} \cdot \delta^{(3)}(\vec{x} - \vec{y})$.  We should obtain this kernel from evaluating
\begin{align}\label{idkernel1}
\mathbbm{1}(\vec{x},\vec{y}) =&~ \sum_{\mu \in \Delta_{\rho}} \sum_{i_\mu =1}^{n_\rho(\mu)} {\bf e}_{i_\mu} \overline{{\bf e}}_{i_\mu}  \otimes \displaystyle\biggl\{ \sum_{j = j_\mu+1}^{\infty}  \sum_{m = -j}^j  \sum_{s=1}^2 \int_{-\infty}^{\infty} \ed E  \Psi^{(i_\mu)}_{j,m,s}(E;\vec{x}) \overline{\Psi^{(i_\mu)}_{j,m,s}(E;\vec{y})} + \cr
& \qquad \qquad \qquad  + \sum_{m= -j_\mu}^{j_\mu} \int_{-\infty}^{\infty} \ed E \Psi^{(i_\mu)}_{m,\sgn(p_\mu)}(E;\vec{x}) \overline{\Psi^{(i_\mu)}_{m,\sgn(p_\mu)}(E;\vec{y})} \displaystyle\biggr\}~. \qquad \raisetag{24pt}
\end{align}
Let $(r,\theta,\phi)$ be spherical coordinates for $\vec{x}$ and $(r',\theta',\phi')$ be spherical coordinates for $\vec{y}$.  Working from the inside out we have schematically $\Psi \overline{\Psi'} = e^{i \epsilon p_\mu(\phi - \phi')/2} {\bf U} \tilde{\Psi} \overline{\tilde{\Psi}'} \overline{{\bf U}'}$, using \eqref{Psifam1}.  The unitary matrix ${\bf U}$ is independent of $E,j,m,s$ so we can carry out the integral and sums on the $4 \times 4$ matrices $\tilde{\Psi} \overline{\tilde{\Psi}'}$.  The sum on $s$ together with the integral over $E$ diagonalizes this matrix, so that we have
\begin{align}\label{ssum}
& \int_{-\infty}^{\infty} \ed E \sum_{s=1}^2 \tilde{\Psi}^{(i_\mu)}_{j,m,s}(E;\vec{x}) \overline{\tilde{\Psi}^{(i_\mu)}_{j,m,s}(E;\vec{y})} = \diag(A_-,A_+,A_-,A_+)~, \qquad \textrm{with} \cr
& A_{\pm} = \frac{(2j + 1) e^{i m(\phi'-\phi)}}{4\pi r^2} \delta(r - r') d_{m,\half(p_\mu \pm 1)}^j(\theta) d_{m,\half(p_\mu \pm1)}^j(\theta')~.
\end{align}
and, when $p_\mu \neq 0$,
\begin{align}\label{extrapiece}
& \int_{-\infty}^{\infty} \ed E  \tilde{\Psi}^{(i_\mu)}_{m,\sgn(p_\mu)}(E;\vec{x}) \overline{\tilde{\Psi}^{(i_\mu)}_{m,\sgn(p_\mu)}(E;\vec{y})} = \left\{ \begin{array}{l l} \diag(A_{-}^0,0,A_{-}^0,0)~,~ & p_\mu > 0 \\ \diag(0,A_{+}^0,0,A_{+}^0)~,~ & p_\mu < 0 \end{array} \right. \quad  \textrm{with} \qquad \cr
& A_{\pm}^0 = \frac{|p_\mu| e^{i m(\phi'-\phi)}}{4\pi r^2} \delta(r - r') d_{m,\pm j_\mu}^{j_\mu}(\theta) d_{m,\pm j_\mu}^{j_\mu}(\theta')~. \raisetag{24pt}
\end{align}
Observe that \eqref{extrapiece} adds to \eqref{ssum} so that the sum over $j$ of $d^{j}_{m,m'}(\theta) d^{j}_{m,m'}(\theta')$ in \eqref{idkernel1} starts at $j = |m'|$ in both cases $m' = \pm j_\mu$.  When $p_\mu = 0$ \eqref{extrapiece} doesn't exist and \eqref{ssum} already satisfies this property.  Now the Wigner $d$ functions appear in the Wigner $D$ functions via $D^{j}_{m,m'}(\vec{\theta}) = e^{-i m \phi} d^{j}_{m,m'}(\theta) e^{-i m' \psi}$ and the $D$'s satisfy the following completeness relation on the $SU(2)$ group manifold parameterized by Euler angles $\vec{\theta} \equiv (\phi,\theta,\psi)$, with $\psi \sim \psi + 4\pi$, $\phi \sim \phi + 2\pi$:
\begin{equation}
\sum_{j = 0}^{\infty} \frac{2j+1}{16 \pi^2} \sum_{k,m = -j}^j D^{j}_{m,k}(\vec{\theta})^\ast  D^{j}_{m,k}(\vec{\theta}')= \delta(\psi - \psi') \delta(\cos{\theta} - \cos{\theta'}) \delta(\phi - \phi')~, \quad
\end{equation}
where $j$ increases in half-integer steps.  Multiplying both sides of this relation by $e^{i m' \psi'}$ for any integer or half-integer $m'$, and integrating over $\psi'$ leads to the completeness relation
\begin{equation}\label{S2completeness}
\sum_{j = |m'|}^{\infty} \frac{2j+1}{4\pi} \sum_{m=-j}^j e^{i m (\phi - \phi')} d^{j}_{m,m'}(\theta) d^{j}_{m,m'}(\theta') = \delta(\cos{\theta} - \cos{\theta'}) \delta(\phi - \phi')~,
\end{equation}
where now the $j$-sum starts at $|m'|$ and increases in integer steps.  The completeness relation \eqref{S2completeness} is exactly what is required to evaluate \eqref{idkernel1}.  We thus find
\begin{align}
\mathbbm{1}(\vec{x},\vec{y}) =& \sum_{\mu \in \Delta_{\rho}} \sum_{i_\mu = 1}^{n_\rho(\mu)} {\bf e}_{i_\mu} \overline{{\bf e}}_{i_\mu} \otimes e^{i \epsilon p_\mu (\phi - \phi')/2} {\bf U} \left[ \frac{1}{r^2} \delta(r-r')\delta(\cos{\theta} - \cos{\theta'}) \delta(\phi - \phi') \mathbbm{1}_{\mathbb{C}^4} \right] \overline{{\bf U}'} \cr
=&~ \mathbbm{1}_{\mathbb{C}^4 \otimes V_\rho} \cdot \delta^{(3)}(\vec{x} - \vec{y})~, \raisetag{24pt}
\end{align}
as required.

\subsection{Zero-modes}\label{app:zeromodes}

In this subsection we return to \eqref{Weyl3} and look for $\LL^2$-normalizable zero-energy solutions.  This clearly requires keeping the asymptotic Higgs vev $x_\mu \neq 0$ since the analysis of the previous section shows that there are no zero-energy bound states when $x_\mu = 0$.  Thus our task is to find the $\LL^2$-normalizable solutions of
\begin{align}\label{Weyl4}
& \left[ \sigma^3 \left( \pd_r + \frac{1}{r} \right) + \left( x_\mu - \frac{p_\mu}{2 r} \right) - i \sigma^2 \frac{k}{r} \right] \hat{\psi}^{(i_\mu)} =  0~, \cr
&  \left[ \sigma^3 \left( \pd_r + \frac{1}{r} \right) - \left( x_\mu - \frac{p_\mu}{2 r} \right) - i \sigma^2 \frac{k}{r} \right] \hat{\chi}^{(i_\mu)} =  0~.
\end{align}
Recall that $k = \sqrt{ (j+\half)^2 - \frac{p_{\mu}^2}{4}}$.  In particular, $k$ is invariant under $p_\mu \to -p_\mu$. Therefore it is sufficient to solve the top equation for $\hat{\psi}^{(i_\mu)}$ since we can obtain the bottom equation by sending $(x_\mu,p_\mu) \to (-x_\mu,-p_\mu)$.  We again write $\hat{\psi}^{(i_\mu)} = \frac{1}{r} f$, $f = (f_1,f_2)^T$.  The first of \eqref{Weyl4} is equivalent to
\begin{equation}\label{1storderbound}
\left[ r \pd_r + x_\mu r - \frac{p_\mu}{2} \right] f_1 = k f_2~, \qquad \left[ r \pd_r - x_\mu r + \frac{p_\mu}{2} \right] f_2 = k f_1~.
\end{equation}

Suppose first that $k \neq 0$.  Then we find the second order equations
\begin{equation}\label{2ndorderbound}
\left[ r^2 \pd_{r}^2 + r \pd_r + (p_\mu \pm 1) x_\mu r - x_{\mu}^2 r^2 - \left( j + \half \right)^2 \right] f_{1,2} = 0~,
\end{equation}
where the $+(-)$ is for $f_1(f_2)$.  This equation has a regular singularity at $r = 0$ and an irregular one at $r = \infty$.  The possible behaviors around each of these points are $r^{\pm (j + 1/2)}$ and $e^{\pm |x_\mu| r}$ respectively.  $\LL^2$-normalizability for $\Psi$ on $\UU$ requires that we choose the decaying exponential and that $f \propto r^s$ with $s > -1/2$ as $r \to 0$, which rules out the $r^{-(j+1/2)}$ solutions.  These two conditions turn out to be incompatible with each other.  Defining $f_{1,2} = r^{j+1/2} e^{-|x_\mu| r} \tilde{f}_{1,2}$ and letting $\xi \equiv 2 |x_\mu| r$ one finds that \eqref{2ndorderbound} can be put in the form of the associated Laguerre differential equation:
\begin{equation}\label{Laguerre}
\left[ \xi \pd_{\xi}^2 + (\alpha + 1 - \xi) \pd_\xi + \nu_{1,2} \right] \tilde{f}_{1,2} = 0~,
\end{equation}
where
\begin{equation}
\alpha = 2j+1~, \qquad \nu_{1,2} = \half \sgn(x_\mu) (p_\mu \pm 1) - (j+1)~,
\end{equation}
with the $+(-)$ for $\nu_1(\nu_2)$.  The general solution to \eqref{Laguerre} is $\tilde{f} = b L_{\nu}^\alpha(\xi) + c U(-\nu;1+\alpha;\xi)$, where $L$ is an associated Laguerre function and $U$ is a confluent hypergeometric function.  The $U$ solution must be discarded as it reintroduces the $r^{-(j+1/2)}$ behavior for $f$ as $r \to 0$.  The associated Laguerre function will reintroduce the $e^{+|x_\mu| r}$ behavior at infinity unless the power series truncates so that we get an associated Laguerre polynomial.  It will truncate if and only if $\nu_{1,2}$ is a non-negative integer.  However we have
\begin{equation}
\nu_{1,2} \leq \half (|p_\mu| + 1) - (j+1) = j_\mu - j < 0~,
\end{equation}
where the last inequality is strict because $k > 0$ implies $j > j_\mu$.  Hence there are no $\LL^2$ solutions to \eqref{Weyl4} when $k > 0$.

Now we focus on the $k=0$ case.  The first comment is that the angular analysis above only allows for $k = 0$ when $p_\mu \neq 0$.  ($k = p_\mu = 0$ would imply a negative value for the angular momentum quantum number $j$).  Hence \eqref{Weyl4} has no $\LL^2$ solutions when $p_\mu = 0$.  The general solutions to \eqref{1storderbound} are
\begin{equation}
f_1 = c_1 r^{p_\mu/2} e^{- x_\mu r}~, \qquad f_2 = c_2 r^{-p_\mu/2} e^{x_\mu r}~.
\end{equation}
The second comment is that the angular analysis requires $f_2 = 0$ when $p_\mu > 0$ and $f_1 = 0$ when $p_\mu < 0$.  We see that normalizable solutions for $f_1$ exist when $p_\mu > 0$ provided $x_\mu > 0$, and normalizable solutions for $f_2$ exist when $p_\mu < 0$ provided $x_\mu < 0$.  In particular, normalizable solutions for $\hat{\psi}^{(i_\mu)}$ exist if and only if $k = 0$ and $p_\mu$, $x_\mu$ have the same sign.

Consider, however, the solutions for $\hat{\chi}$ when $k =0$.  We set $\hat{\chi}^{(i_\mu)} = \frac{1}{r} g$ with $g = (g_1,g_2)^T$ as before.  Then, since the equations for $g$ are obtained from those for $f$ by sending $(x_\mu,p_\mu) \to (-x_\mu,-p_\mu)$, we get
\begin{equation}
g_1 = c_1 r^{-p_\mu/2} e^{x_\mu r}~, \qquad g_2 = c_2 r^{p_\mu/2} e^{-x_\mu r}~.
\end{equation}
Now the conditions for $\LL^2$-normalizability are incompatible with the requirements from the angular analysis.  When $p_\mu > 0$ we must set $g_2 = 0$, but then $g_1$ is not normalizable.  This is because $p_\mu$ is always an integer and thus $p_\mu > 0$ implies $p_\mu \geq 1$.  This leads to a non-normalizable behavior: $\hat{\chi}$ of the form $r^s$ with $s \leq -3/2$ near $r = 0$.  Similarly, when $p_\mu < 0$ we must set $g_1 = 0$, but then $g_2$ is not normalizable.  We conclude that $\hat{\chi}^{(\mu_i)}$ has no $\LL^2$-normalizable zero-mode solutions.  Note that any such solution would have implied a nontrivial kernel for the operator $L_{\rho}^\dag$, and this would have been in contradiction with the general argument that $\ker{L_{\rho}^\dag} = \{ 0\}$.  (See discussion around \eqref{positiveL}.)


\bibliographystyle{utphys}
\bibliography{MRVdimP1}

\end{document}